\documentclass[]{jkas}

\usepackage[utf8]{inputenc}
\usepackage{array}
\DeclareUnicodeCharacter{2500}{--}

\newcommand{\subsubsubsection}[1]{\paragraph{#1}}
\titleformat{\paragraph}[display]{\normalfont\normalsize\bfseries\color{jkasSectnColor}}{}{0pt}{}
\titlespacing*{\paragraph}{0pt}{1.8ex plus .4ex minus .2ex}{0.6ex plus .2ex}
\titleformat{\subparagraph}[display]{\normalfont\normalsize\itshape\bfseries}{}{0pt}{}
\titlespacing*{\subparagraph}{0pt}{1.4ex plus .4ex minus .2ex}{0.5ex plus .2ex}

\def\year{2026}

\def\volume{}
\def\issue{}
\def\beginpage{1}
\def\received{---}
\def\accepted{---}
\def\published{---}
\date{Received \received; Accepted \accepted; Published \published}

\title{The Korean Space Telescope (KST): A Dedicated Space Observatory for the Dark Universe}

\author[1,$\star$]{M. James Jee}{0000-0002-5751-3697}
\author[2,3,4]{Ho Seong Hwang}{0000-0003-3428-7612}
\author[1]{Dongwook Lim}{0000-0001-7277-7175}
\author[1]{Minjin Kim}{0000-0002-3560-0781}
\author[2,3]{Myungshin Im}{0000-0002-8537-6714}
\author[5,6]{Sungwook E. Hong}{0000-0003-4923-8485}
\author[5]{Ho-Gyu Lee}{0000-0002-3808-7143}
\author[5]{Soung-Chul Yang}{0000-0001-9842-639X}
\author[5,6]{David Parkinson}{0000-0002-7464-2351}
\author[5,6]{Arman Shafieloo}{0000-0001-6815-0337}
\author[5]{Jun-Qian Jiang}{0000-0003-0957-3633}
\author[7]{Haeng Pal Heo}{0000-0001-9139-4583}
\author[7]{Dongok Ryu}{0000-0003-3767-8934}
\author[8]{Hagyong Kihm}{0000-0002-3992-0758}
\author[7]{Hyung Yun Noh}{}
\author[7]{Wooyong Kang}{}
\author[7]{Youngsun Kim}{}
\affil[1]{Department of Astronomy, Yonsei University, 50 Yonsei-ro, Seodaemun-gu, Seoul 03722, Republic of Korea}
\affil[2]{Department of Physics and Astronomy, Seoul National University, 1 Gwanak-ro, Gwanak-gu, Seoul 08826, Republic of Korea}
\affil[3]{SNU Astronomy Research Center, Seoul National University, Gwanak-gu, Seoul 08826, Republic of Korea}
\affil[4]{Institute for Data Innovation in Science, Seoul National University, Seoul 08826, Republic of Korea}
\affil[5]{Korea Astronomy and Space Science Institute, 776 Daedeok-daero, Yuseong-gu, Daejeon 34055, Republic of Korea}
\affil[6]{Astronomy Campus, University of Science and Technology, 776 Daedeok-daero, Yuseong-gu, Daejeon 34055, Republic of Korea}
\affil[7]{Korea Aerospace Research Institute, 169-84 Gwahak-ro, Yuseong-gu, Daejeon 34133, Republic of Korea}
\affil[8]{Korea Research Institute of Standards and Science, 267 Gaejeong-ro, Yuseong-gu, Daejeon 34113, Republic of Korea}

\def\corrauthor{%
M. James Jee, \email{mkjee@yonsei.ac.kr}
}

\def\runningauthor{Jee et al.}
\def\runningtitle{KST: Probing the Dark Sector}
\def\keywords{%
cosmology: dark energy --- dark matter --- gravitational waves --- surveys --- space vehicles: instruments
}

\def\abstracttext{%
The Korean Space Telescope (KST) is a next-generation 1.4-meter wide-field space observatory, a dedicated probe of the \textbf{dark sector}. Its single instrument provides diffraction-limited visible imaging ($\sim0.15^{\prime\prime}$ in the optical) together with simultaneous near-infrared ($R\sim100$) spectrophotometry. \textit{Every detected source} is thus observed in the NIR by the survey itself, with spectral information delivered at a resolution set by its brightness (full $R\sim100$ typing at the bright end, synthetic medium-band or broadband photometry below it) and with no separate follow-up, a capability decisive for time-domain science. With it, KST addresses the three outstanding questions of dark physics in concert: the \textbf{dynamics of dark energy}, the \textbf{nature of dark matter}, and gravitational-wave \textbf{dark sirens} as independent cosmological rulers. KST combines measurements of expansion and structure growth to test dark-energy models. The two cosmology tiers of a ``wedding-cake'' survey ($2{,}000~\text{deg}^2$ Wide plus $10~\text{deg}^2$ cluster-targeted Deep Survey) combine weak lensing, baryon acoustic oscillations, and a near-infrared-standardized Type Ia supernova Hubble diagram to constrain dark energy at its pivot redshift ($\sigma(w_p)\approx0.03$). KST also maps dark matter from cluster cores to galactic substructure to test cold dark matter against warm and self-interacting alternatives, and builds a deep near-infrared host catalog for a distance-ladder-independent $H_0$ from dark sirens. The same data yield a \textbf{Legacy Science} portfolio (galaxy evolution, active galactic nuclei, exoplanets, and the Milky Way), at no additional observing cost apart from a dedicated bulge microlensing campaign. As Korea's first meter-class space telescope dedicated to astrophysics, KST occupies a niche complementary to the Vera C. Rubin Observatory, \textit{Euclid}, and Roman.
}

\begin{document}
\jkashead

\section{Introduction}

\begin{figure}[t]
\centering
    \includegraphics[width=\linewidth]{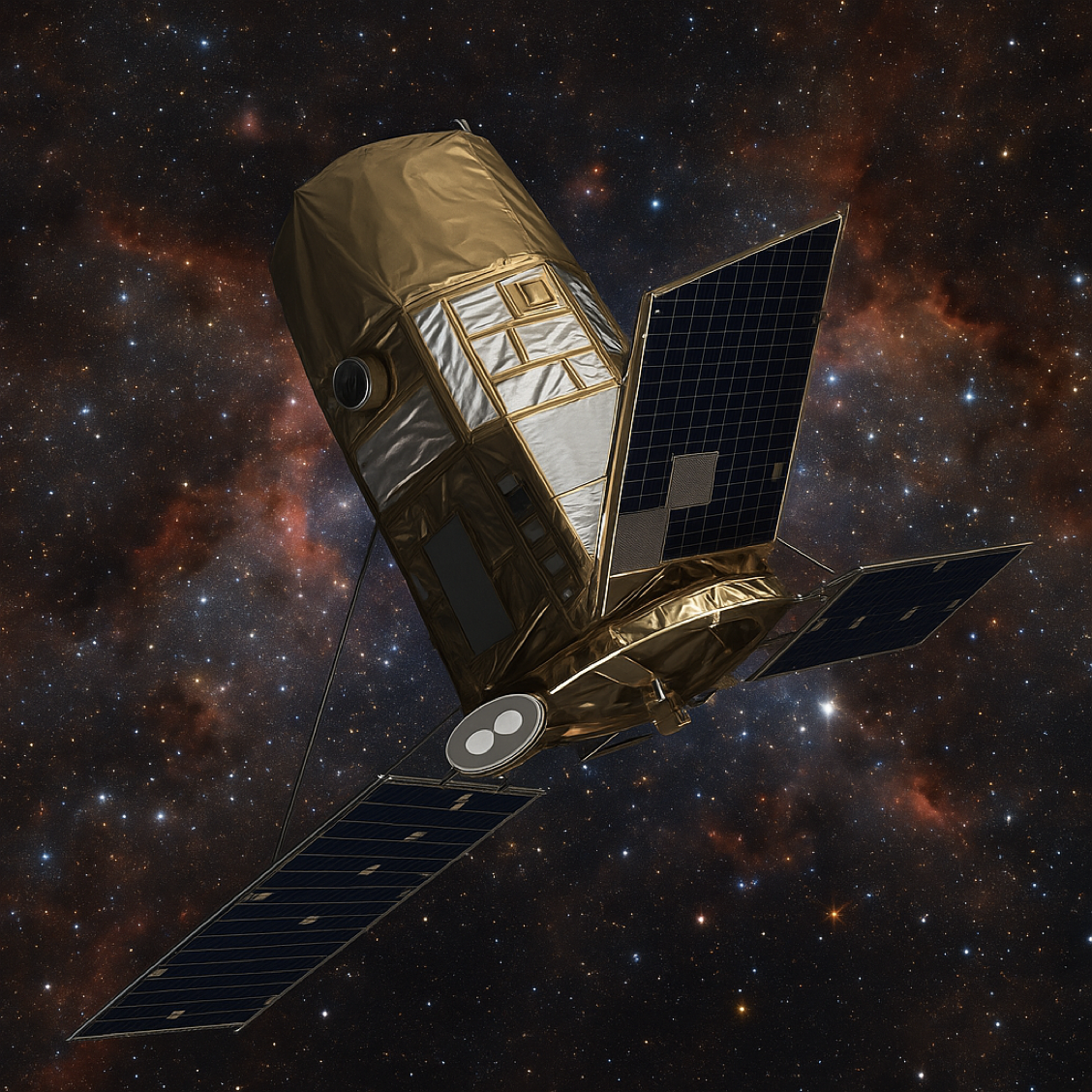}
\caption{Conceptual rendering of the Korean Space Telescope (KST) in its operational orbit. The spacecraft design leverages the high-resolution imaging and orbital stability heritage of the Korean Multi-Purpose Satellite (KOMPSAT) series \citep{Lee2021KOMPSAT,Seo2020KOMPSATIQ} to support a 1.4-meter wide-field astronomical surveyor.}
\label{fig:kst}
\end{figure}

The current paradigm of physical cosmology, the spatially flat $\Lambda$ cold dark matter ($\Lambda$CDM) model, rests on a startling premise. Roughly $95\%$ of the present energy content of the Universe is invisible. About $68\%$ is attributed to \textit{dark energy}, a smooth component with negative pressure that drives the accelerating cosmic expansion discovered through Type~Ia supernovae (SNe~Ia) in the late 1990s \citep{Riess1998, Perlmutter1999}, while a further $\sim27\%$ is non-baryonic \textit{dark matter}, whose gravity assembles the cosmic web and binds galaxies and clusters \citep{Planck2020}. With only six free parameters, $\Lambda$CDM reproduces the cosmic microwave background (CMB), large-scale structure, and the abundance of clusters with remarkable economy. Yet its two dominant ingredients remain physically unidentified, and over the past few years a series of high-precision measurements has begun to strain the model itself. Three questions now stand out as the principal scientific drivers of the Korean Space Telescope (KST, Figure \ref{fig:kst}): \textit{Is dark energy dynamical? What is the particle nature of dark matter? And can gravitational-wave ``standard sirens'' deliver an independent verdict on the expansion of the Universe?}

\paragraph{The dynamics of dark energy.}
In $\Lambda$CDM, dark energy is a cosmological constant with an equation-of-state parameter fixed at $w \equiv p/\rho c^2 = -1$, even though the value expected from vacuum energy differs from the observed one by many orders of magnitude \citep{Weinberg1989}. The minimal extension promotes $w$ to a function of time, conventionally written in the Chevallier--Polarski--Linder form $w(a) = w_0 + w_a(1-a)$ \citep{Chevallier2001, Linder2003}. Deciding whether the data demand $w_0 \neq -1$ or $w_a \neq 0$ has become the central question of observational cosmology. The urgency is twofold. First, the ``Hubble tension'' has grown into a $\sim5\sigma$ discrepancy between the early-Universe value inferred from the \textit{Planck} CMB under $\Lambda$CDM ($H_0 = 67.4 \pm 0.5~\mathrm{km\,s^{-1}\,Mpc^{-1}}$; \citealt{Planck2020}) and the local distance-ladder measurement of the SH0ES (Supernovae and $H_0$ for the Equation of State of dark energy) program ($H_0 = 73.17 \pm 0.86~\mathrm{km\,s^{-1}\,Mpc^{-1}}$; \citealt{Breuval2024,Riess2025JWST}). A milder, survey-dependent ``$S_8$'' discrepancy in the inferred amplitude of structure persists at lower significance \citep{DESY6_2026,Wright2025}, and late-time modifications to dark energy are among the proposed resolutions \citep{DiValentino2021}. Second, and most strikingly, baryon acoustic oscillation (BAO) measurements from the Dark Energy Spectroscopic Instrument (DESI) now favor the dynamical-dark-energy quadrant ($w_0 > -1$, $w_a < 0$), implying a component that crossed $w=-1$ at low redshift \citep{DESI2024,DESIDR2Cosmo}. Model-agnostic reconstructions reach similar conclusions \citep{Calderon2024,Lodha2025}, and, if interpreted as a minimally coupled fluid within general relativity, the $w<-1$ phase violates the null energy condition and is nontrivial to realize in consistent single-field models \citep{Vikman2005,Hu2005,Fang2008}. In the DESI DR2 analysis the preference for evolving dark energy over $\Lambda$CDM reaches $3.1\sigma$ from BAO$+$CMB alone, rising to $2.8$--$4.2\sigma$ when SNe~Ia from Pantheon$+$ \citep{Brout2022}, Union3 \citep{Rubin2023}, and the Dark Energy Survey five-year sample \citep{DESSN5YR2024} are successively added \citep{DESIDR2Cosmo}. This significance depends sensitively on the SNe sample and on the assumed parametrization, however, and could in principle be mimicked by unrecognized supernova systematics. Notably, the combined ``Supernovae Unite'' sample retains the preference for evolving dark energy \citep{Camilleri2026}. Settling the question demands independent, high-redshift probes that measure cosmic \textit{geometry} and the \textit{growth} of structure \textit{together}: weak-lensing shear and BAO pushed into the optically inaccessible ``redshift desert'' ($1.5 < z < 3$), combined with SNe~Ia standardized in the near-infrared (NIR), where dust systematics are strongly suppressed, coverage that also enables a direct test of the phantom crossing \citep{LinderHuterer2003}.

\paragraph{The nature of dark matter.}
On large scales the cold, collisionless dark matter (CDM) of $\Lambda$CDM is extraordinarily successful, and on cluster scales the spatial offset between the lensing mass and the X-ray gas in merging systems such as the Bullet Cluster is a direct, model-independent demonstration that the mass is dark and nearly collisionless \citep{Clowe2006,cha2025}. Its particle identity, however, is unknown. The long-favored weakly interacting massive particle has eluded direct, indirect, and collider searches, broadening interest in alternatives whose signatures lie in the \textit{small-scale} structure of the Universe rather than in the laboratory. Warm dark matter free-streams and suppresses the matter power spectrum below a cutoff scale \citep{Irsic2017}. Fuzzy (ultralight axion) dark matter imprints a de~Broglie-scale suppression and solitonic cores \citep{Hui2017}. Self-interacting dark matter reshapes halo interiors into cores set by the cross-section $\sigma/m$ \citep{TulinYu2018}. Primordial black holes survive as dark matter only in narrow mass windows \citep{CarrKuhnel2020}. The classic galaxy-scale ``small-scale challenges'' to CDM are now largely attributable to baryonic feedback and are no longer clean discriminants \citep{Bullock2017}. Decisive tests have therefore moved to the lowest-mass, essentially starless halos (e.g. \citealt{LeeG2024,KwonM2025}), where dark-matter physics alone sets the structure. These are reached through gravitational lensing (flux-ratio anomalies in multiply imaged sources \citep{DalalKochanek2002} and the direct gravitational imaging of dark subhalos that perturb Einstein rings \citep{Vegetti2012}), and through perturbations in cold stellar streams \citep{Bonaca2019} and satellite-galaxy counts \citep{Nadler2021}. Across all of these methods the limiting factor is statistical. Only a few hundred suitable strong lenses are known, and resolving the relevant sub-arcsecond image structure and low-surface-brightness substructure demands space-based, diffraction-limited imaging over wide fields \citep{Vegetti2024}. Near-infrared coverage is integral to each of these measurements. It supplies the photometric redshifts that push weak-lensing source samples, and hence convergence and cluster/self-interaction mass maps, to $z\sim2$--$3$, where the lensing signal is largest. It recovers dust-reddened lensed arcs and lens galaxies and enables the achromatic optical$+$NIR flux-ratio test that separates dark substructure from dust and stellar microlensing. And it resolves the old, red-giant populations of dust-obscured, low-surface-brightness satellites. Simultaneous visible imaging and NIR spectrophotometry are thus integral to the dark-matter program, not incidental to it.

\paragraph{Dark sirens and the cosmic expansion.}
Gravitational-wave (GW) sources are ``standard sirens'': the inspiral waveform encodes the luminosity distance directly, calibrated only by general relativity and independent of the cosmic distance ladder \citep{Schutz1986}. Since the first detection of a binary black-hole merger \citep{Abbott2016}, GW astronomy has opened a route to $H_0$ whose systematics are orthogonal to both the CMB and the distance ladder, \textit{provided} that a redshift can be attached to each event. The binary neutron-star merger GW170817, with its kilonova counterpart, gave the first such ``bright-siren'' measurement, $H_0 = 70.0^{+12.0}_{-8.0}~\mathrm{km\,s^{-1}\,Mpc^{-1}}$ \citep{AbbottMM2017, AbbottSiren2017}, but electromagnetic counterparts have proven rare and no comparable event has been secured since. The ``dark-siren'' technique instead assigns a statistical redshift by weighting the candidate host galaxies within the GW localization volume using a redshift catalog \citep{DelPozzo2012, Gray2020}. Combining the O1--O3 events with galaxy catalogs yields $H_0 \approx 68^{+13}_{-7}~\mathrm{km\,s^{-1}\,Mpc^{-1}}$ \citep{AbbottGWTC3cosmo2023}, a precision that will sharpen toward the percent level as the event count grows into the next-generation-detector era \citep{ChenHolz2018, Maggiore2020}. The dominant limitation is no longer the GW data but the galaxy catalog. Dark sirens require host catalogs that are deep and \textit{complete} within the localization volume, with accurate redshifts, while kilonovae, which peak and persist longest in the NIR, demand rapid, wide-field NIR follow-up. A deep, wide NIR photometric-redshift survey is precisely the enabling dataset, and precisely where current optical catalogs thin out.

These three frontiers, distinct in their physics, converge on a single observational prescription: a wide-field, space-based facility that measures expansion and growth \textit{together}, resolves faint structure at the diffraction limit, reaches deep NIR photometric redshifts over thousands of square degrees, and responds rapidly to transients. No existing or planned mission provides this combination in a single instrument. The Vera~C. Rubin Observatory lacks space-based resolution and NIR coverage. \textit{Euclid}~\citep{Euclid} exposes its single-band visible imager in parallel with grism spectroscopy from its Near Infrared Spectrometer and Photometer (NISP) but acquires its NIR imaging separately, and its Wide Survey is single-epoch by design, with repeated visits confined to the Deep and calibration fields. The Nancy Grace Roman Space Telescope observes its filters and dispersers sequentially through a single optical path. KST is designed to fill exactly this gap.

KST is a 1.4-meter wide-field observatory optimized around the ``aperture--etendue'' balance (Section~\ref{sec:rationale}). With a $1^{\circ}\times0.25^{\circ}$ field of view it delivers diffraction-limited visible imaging ($\sim0.15^{\prime\prime}$ in the optical) together with simultaneous NIR spectrophotometry, spectrophotometrically characterizing every detected source within the survey itself at brightness-matched resolution (Section~\ref{sec:simultaneity}) while bypassing the survey ``efficiency cliff'' that constrains larger space telescopes. Its cosmology program is realized through the two cosmology tiers of a ``wedding-cake'' survey strategy: a $2{,}000~\text{deg}^2$ \textbf{Wide Survey} to map the growth of cosmic structure, and a $10~\text{deg}^2$ \textbf{Deep Survey} centered on massive galaxy clusters to probe the time-domain and strongly-lensed Universe. Together, these two surveys measure the geometry of cosmic expansion and the growth of structure, the combination required to confront all three dark-sector questions, and each pillar is developed in detail in Section~\ref{sec:darksector}.

In this sense KST occupies a distinctive niche. Simultaneity alone is not the claim: \textit{Euclid} already exposes its visible imager in parallel with grism spectroscopy. What no existing or planned wide-field mission offers is KST's combination of \textit{four-band} optical imaging acquired in parallel, contiguous $R\sim100$ linear-variable-filter (LVF) spectral mapping of \textit{every} source rather than slitless dispersion optimized for emission-line objects, repeated time-domain coverage of the same fields, and the internal cross-calibration these enable, the specific combination that each dark-sector pillar requires (deep NIR photometric redshifts for structure growth and for dark-siren host catalogs, and simultaneous optical imaging and NIR spectrophotometry for lensing substructure and NIR-standardized supernovae). 
 The same survey data simultaneously underpin a broad \textbf{Legacy Science} portfolio (galaxy evolution, active galactic nuclei, exoplanets, and Milky Way astronomy), produced as a natural by-product. KST will complement these surveys through multiband optical imaging, NIR spectral mapping, and repeated observations of selected fields.

The remainder of this paper is organized as follows. Section~\ref{sec:design} describes the telescope design and instrumentation, and Section~\ref{sec:darksector} develops the three core dark-sector science pillars. Section~\ref{sec:legacy} presents the Legacy Science portfolio, and Section~\ref{sec:survey} details the survey strategy and mission time budget. Section~\ref{sec:synergy} sets KST in the context of contemporary international facilities. We summarize in Section~\ref{sec:summary}.

\section{Telescope Design and Instrumentation}
\label{sec:design}

KST utilizes a Korsch-type three-mirror anastigmat (TMA) optical design, selected for its excellent wide-field imaging performance and compact folded geometry.

\subsection{Design Philosophy: The Aperture--Etendue Optimization}
\label{sec:rationale}

A common misconception in observational astronomy is the direct equivalence of aperture size to scientific merit. While ground-based facilities have reached the 8-meter to 30-meter class, KST's 1.4-meter aperture represents a strategic ``sweet spot'' for space-based surveys that balances sensitivity, wavefront stability, and survey efficiency.

\begin{figure}[t]
\centering
    \includegraphics[width=\linewidth]{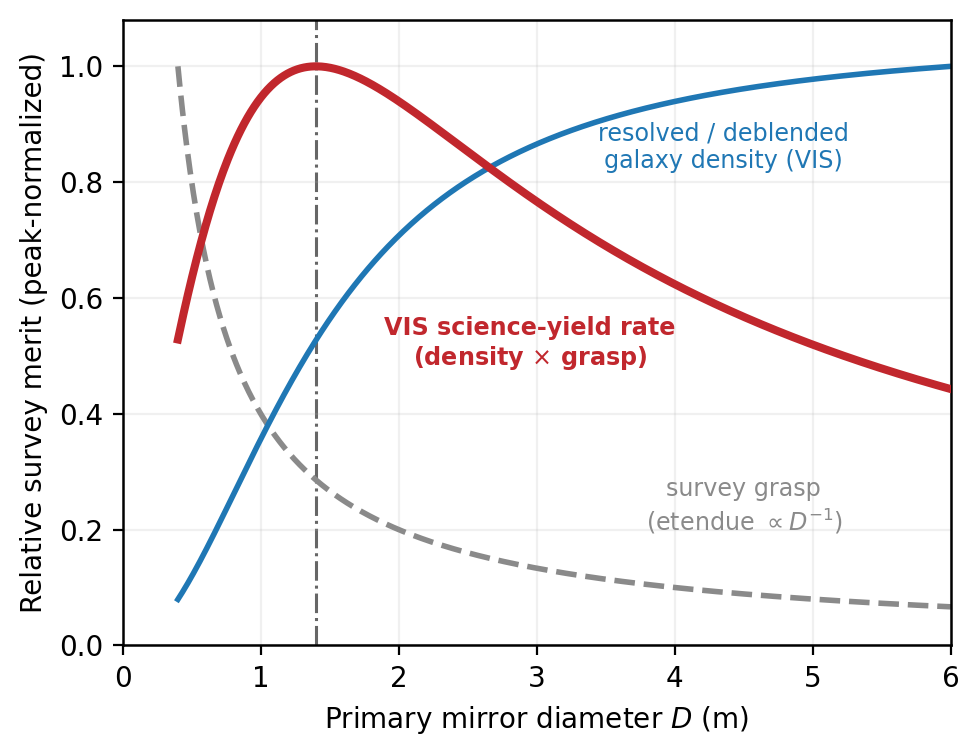}
\caption{\textbf{Aperture optimization for the imaging channel.} The survey grasp (etendue $\propto D^{-1}$; gray dashed) \textit{decreases} with aperture, while the density of cleanly resolved and deblended galaxies in the visible (VIS) imaging channel (blue) \textit{increases} as the point spread function (PSF) shrinks below the faint-galaxy size and then saturates. Their product, the rate at which science-usable galaxies are surveyed (red), has a genuine interior optimum at $D\approx1.4$~m: the aperture whose $\sim0.15''$ visible PSF matches the $\theta_g\sim0.15''$ scale that must be resolved for faint-galaxy shape measurement and deblending. Below this the PSF blends the faint population and degrades shapes, while above it survey grasp is spent resolving already-resolved sources. Curves are peak-normalized. The construction is an illustrative design rationale, with the model and its inputs given in Section~\ref{sec:etendue}. The near-infrared channel provides $R\sim100$ spectrophotometry (photometric redshifts and spectral energy distributions, SEDs), not imaging, and is not what sets the imaging optimum.}
\label{fig:etendue_scaling}
\end{figure}

\label{sec:etendue}

To quantify the merit of the 1.4-meter aperture, we analyze the \textbf{Etendue} ($E$), defined as the product of the telescope's collecting area ($A$) and its instantaneous field of view ($\Omega$). In high-precision optical systems such as the Korsch-type TMA used by KST, maintaining a diffraction-limited wavefront error ($\sigma$) across the field imposes a strict constraint. Following third-order aberration theory \citep[e.g.,][]{Schroeder2000}, the RMS wavefront error  scales with aperture diameter ($D$) and field angle ($\theta$) as $\sigma \propto D^3 \cdot \theta^2$. Here $\sigma$ is the \textit{physical} wavefront error (in length units), fixed by the optical prescription, aperture, and field angle, and is itself independent of wavelength. The \textit{diffraction-limited} condition, however, is wavelength-relative. The Mar\'echal criterion $\sigma\lesssim\lambda/14$ is easier to satisfy at longer $\lambda$. The binding constraint on the usable diffraction-limited field is therefore set by the bluest (optical) band, and the near-infrared falls comfortably within it. The angular resolution itself does scale as $\lambda/D$ ($\sim0.15''$ in the optical, $\sim0.2''$ in the NIR). Since weak-lensing shapes are measured in the visible imaging channel while the NIR provides $R\sim100$ spectrophotometry, the coarser NIR PSF does not degrade the high-redshift shear measurement.

To maintain a constant system-level PSF (i.e., keeping $\sigma$ constant), the square of the usable field angle must scale inversely with the cube of the aperture: $\theta^2 \propto D^{-3}$. Since the solid angle of the field of view is proportional to the square of the field angle ($\Omega \propto \theta^2$), it follows that $\Omega \propto D^{-3}$. Consequently, the survey power scales as:
\begin{equation}
    E = A \times \Omega \propto D^2 \times D^{-3} = \frac{1}{D}.
\end{equation}
\noindent
We note that this scaling applies strictly to diffraction-limited space systems. Ground-based surveyors like the Vera C. Rubin Observatory (8.4 m) achieve significantly larger fields because they operate in the \textit{seeing-limited} regime, where the allowable wavefront error is relaxed by a factor of $\sim$5 (to a seeing-limited PSF of $\sim 0.7''$). We emphasize that this is a \textit{fixed-complexity} scaling. It assumes a common optical prescription and wavefront-error budget, and dedicated corrector optics or additional aspheric surfaces can extend the diffraction-limited field at the cost of mass, volume, and alignment tolerances. The 1.4~m value should therefore be read as the sweet spot for KST's specific TMA architecture and cost class rather than a universal limit.

Etendue therefore \textit{decreases} monotonically with aperture, producing a ``survey efficiency cliff'': a larger mirror surveys a fixed area to a fixed depth more slowly, not faster. Grasp alone thus favors a \textit{small} aperture and does not, by itself, define an optimum. That optimum appears only when grasp is weighed against angular resolution. In the visible imaging channel, the surface density of galaxies that are cleanly \textit{resolved and deblended}, and hence usable for weak-lensing shapes, morphology, and cluster/large-scale-structure (LSS) science, rises with aperture through the resolution factor $n_{\rm res}\propto\theta_g^2/(\theta_g^2+\theta_{\rm PSF}^2)$, saturating once the PSF falls well below the characteristic scale $\theta_g\sim0.15''$ that must be resolved for reliable shape measurement and deblending. The resulting \textit{science-yield rate}, $n_{\rm res}(D)\times E(D)$ (resolved galaxies surveyed per unit time), is an increasing, saturating function times a decreasing one, and therefore \textit{peaks} at a well-defined aperture, namely the one whose imaging PSF matches $\theta_g$. Taking the visible PSF to scale as $\theta_{\rm PSF}(D)=0.15''\,(1.4~{\rm m}/D)$ (normalized to KST's delivered value), together with $\theta_g\simeq0.15''$ and $E\propto D^{-1}$, this yield rate peaks at $D\approx1.4$~m (Figure~\ref{fig:etendue_scaling}). A smaller aperture blends the faint population and degrades shapes, while a larger one only spends survey grasp resolving sources that are already resolved. Because $\theta_g$ is set to the delivered PSF scale, the figure is a design rationale, showing that matching the visible PSF to the faint-galaxy scale maximizes yield, rather than an independent survey optimization. The 1.4~m design therefore maximizes science yield per unit time (at $E\approx0.385\,\text{m}^2\text{deg}^2$), enabling:
\begin{itemize}
    \item \textbf{Cosmological Statistics:} Rapid mapping of massive galaxy samples to minimize shape noise in weak lensing measurements.
    \item \textbf{Microlensing Cadence:} Continuous monitoring of millions of stars to detect rare, short-duration planetary deviations.
    \item \textbf{GW Counterpart Localization:} Efficient tiling of large gravitational-wave error contours to capture fast-fading kilonovae.
\end{itemize}

\subsection{Wavefront Stability and Wide-Field Precision}
Beyond survey speed, the 1.4 m aperture ensures superior image stability. Larger mirrors are more susceptible to thermal gradients and mechanical jitter.
\begin{itemize}
    \item \textbf{Aberration Control:} The 1.4 m aperture allows for the correction of third-order aberrations (astigmatism, coma) using achievable aspheric tolerances, ensuring a stable PSF across the full $1^{\circ}\times0.25^{\circ}$ field.
    \item \textbf{System Stability:} The KOMPSAT series provides directly relevant heritage in high-resolution optical payload design, geometric calibration, and long-term on-orbit image-quality stability \citep{Seo2016KOMPSAT3A,Jeon2016KOMPSAT3A,Seo2020KOMPSATIQ}. The shear-calibration requirement adopted here, $m < 10^{-3}$, is a weak-lensing specification flowed down to the KST platform rather than a performance demonstrated by those Earth-observation missions. Verifying that the thermal and mechanical design meets it is a mission-definition-phase deliverable.
\end{itemize}

\subsection{Near-Infrared Sensitivity and Duty Cycle}
The space environment amplifies the capabilities of the 1.4 m aperture:
\begin{itemize}
    \item \textbf{Dark Sky Advantage:} Free of atmospheric OH airglow and thermal emission, the $1.1\text{--}1.8~\mu {\rm m}$ sky seen by KST is so faint that its $R\sim100$ channel depths are set by detector noise rather than by the sky (Table~\ref{tab:lvf_modes}), while its visible imaging reaches $r\approx27$ (point source, 1~hr; Table~\ref{tab:kst_specs}).
    \item \textbf{Photometric Stability:} Freedom from atmospheric extinction allows for $<1\%$ photometric precision, vital for standardizing Type Ia Supernovae.
    \item \textbf{Uninterrupted Duty Cycle:} Near-continuous ($24/7$) coverage enables the high-cadence monitoring of transients that is impossible from ground-based sites due to the diurnal cycle and weather.
\end{itemize}

These space-based advantages are decisive for Korea in particular. The domestic ground-based sky, limited by humidity, cloud cover, atmospheric turbulence, and the light pollution of a densely populated peninsula, cannot support diffraction-limited imaging or deep-field surveys reaching $m_{\rm AB}>26$. For high-precision wide-field astronomy a space platform is therefore not merely advantageous but necessary, positioning Korea as a leader in space-based wide-field surveys.

\subsection{Optical Configuration}
The telescope features a 1.4-meter primary mirror with an effective focal length of 19 meters ($f/13.57$). While the theoretical diffraction limit for a 1.4~m aperture at $r$-band is $\sim0.11''$, KST is designed to deliver a system-level PSF FWHM of $\sim0.15''$, providing a robust engineering margin for wavefront errors, jitter, and detector effects. The field of view is $1.0^{\circ} \times 0.25^{\circ}$, enabling efficient wide-area surveys.

\begin{figure*}[t]
\centering
\includegraphics[width=0.92\textwidth]{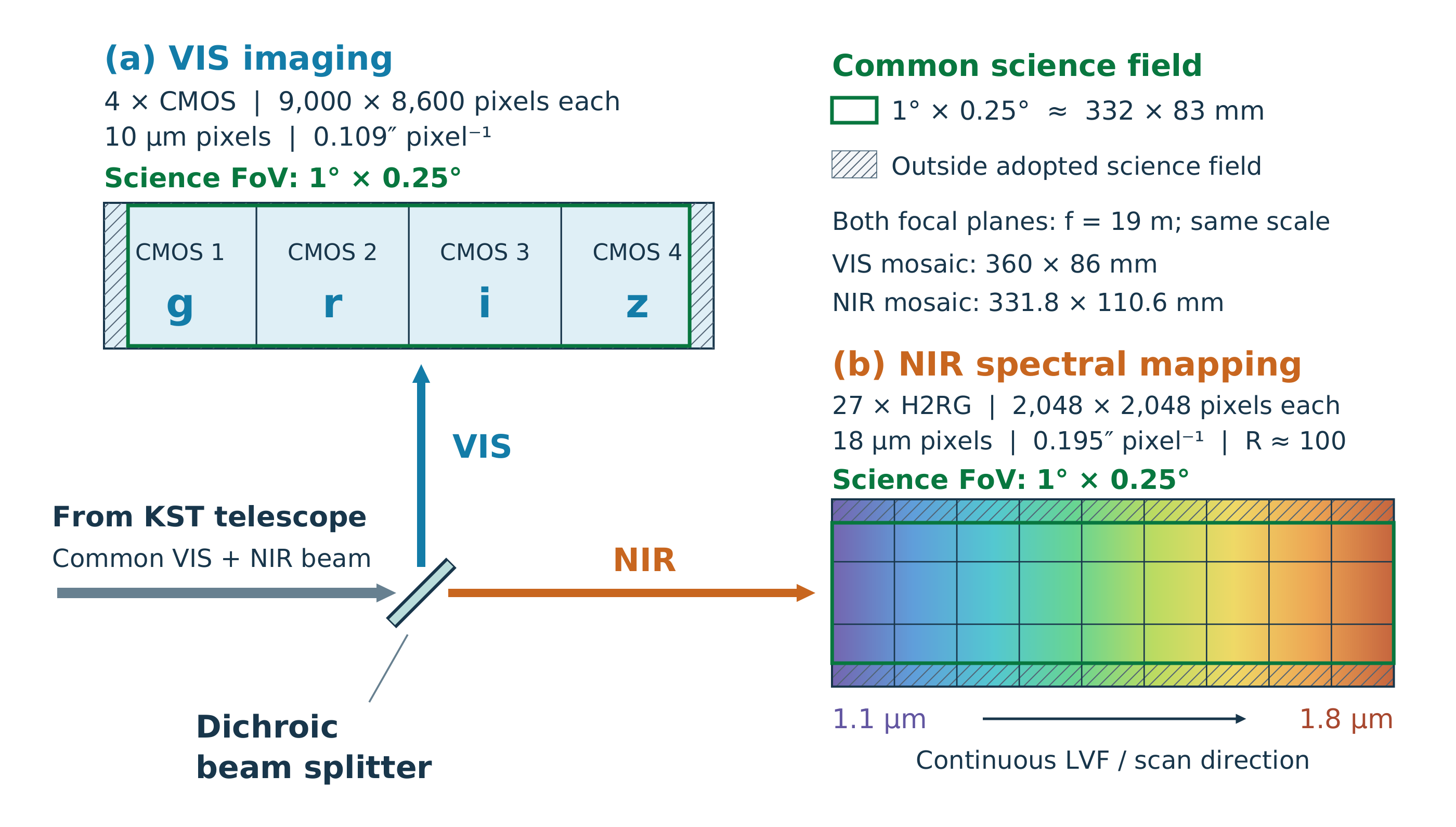}
\caption{\textbf{Schematic of the KST dichroic beam splitter and detector mosaics.} The common incident beam is separated into simultaneous VIS and NIR channels. \textbf{(a)} Four CMOS detectors form a $4\times1$ mosaic, with one fixed $g$, $r$, $i$, or $z$ filter per detector. \textbf{(b)} Twenty-seven square H2RG detectors form a $9\times3$ mosaic beneath an LVF whose transmission wavelength varies continuously from 1.1 to 1.8~$\mu$m along the scan direction, providing $R\approx100$ spectral mapping. Green outlines mark the identical $1^{\circ}\times0.25^{\circ}$ science fields. Hatched regions indicate detector area outside the adopted field. Both mosaics are drawn on a common scale using the 19~m effective focal length and the specified pixel pitches. Detector gaps and H2RG reference-pixel borders are omitted. Beam routing and channel orientation are schematic, and relay optics are not shown.}
\label{fig:focal_plane}
\end{figure*}

\subsection{Simultaneous Dual-Channel Observation}
A dichroic beam splitter directs light into two independent channels—Visible (VIS) and Near-Infrared (NIR)—simultaneously (Figure~\ref{fig:focal_plane}).
\begin{itemize}
    \item \textbf{Visible (VIS):} Four $9,000 \times 8,600$ pixel CMOS devices ($4 \times 1$ layout), each directly coated with a fixed broadband filter ($g$, $r$, $i$, or $z$). The $1^{\circ}\times0.25^{\circ}$ field thus comprises four $0.25^{\circ}\times0.25^{\circ}$ single-band sub-fields. Full $g,r,i,z$ coverage of a sky position is built up as the survey scan steps it across the four sub-fields. The $0.109''/\text{pixel}$ scale is matched to the diffraction limit.
    \item \textbf{Near-Infrared (NIR):} Twenty-seven H2RG detectors ($9 \times 3$ mosaic) with a $0.195''/\text{pixel}$ scale, optimized for the $1.1$--$1.8~\mu {\rm m}$ range.
\end{itemize}

Because the broadband filters are coated directly onto the VIS sensors and the LVF is bonded to the NIR mosaic, the payload contains no filter wheel or other moving spectral element. Every mechanism eliminated from the optical path removes a single-point failure mode and a source of repeatability error, a significant reliability and calibration advantage for a small dedicated mission.

\subsection{Spectrophotometric Mapping and Strategic Niche}
The NIR channel incorporates a linear variable filter (LVF) system, providing spectral mapping ($R \sim 100$) across the full focal plane. This creates $\sim48$ spectral channels per scan, bridging the gap between broadband imaging and high-resolution spectroscopy. The LVF approach is already flight-proven. NASA's SPHEREx mission employs LVFs of the same type for its all-sky spectral survey \citep{spherex}.

\subsubsection{Rationale for $R \sim 100$ Resolution}
The $R \sim 100$ resolution balances sensitivity with diagnostic power. It is sufficient to identify the 4000~\AA\ break and major emission lines (H$\alpha$, [O\,\textsc{iii}]), supporting a target photometric-redshift accuracy of $\sigma_z/(1+z) \lesssim 0.01$ (adopted as a design goal throughout this paper) and the classification of high-redshift SNe~Ia. The realized accuracy will depend on galaxy type, brightness, and redshift, with detailed photometric-redshift simulations deferred to future work.

\subsubsection{Comparison with NASA's SPHEREx}

\begin{figure}[t]
\centering
    \includegraphics[width=\columnwidth]{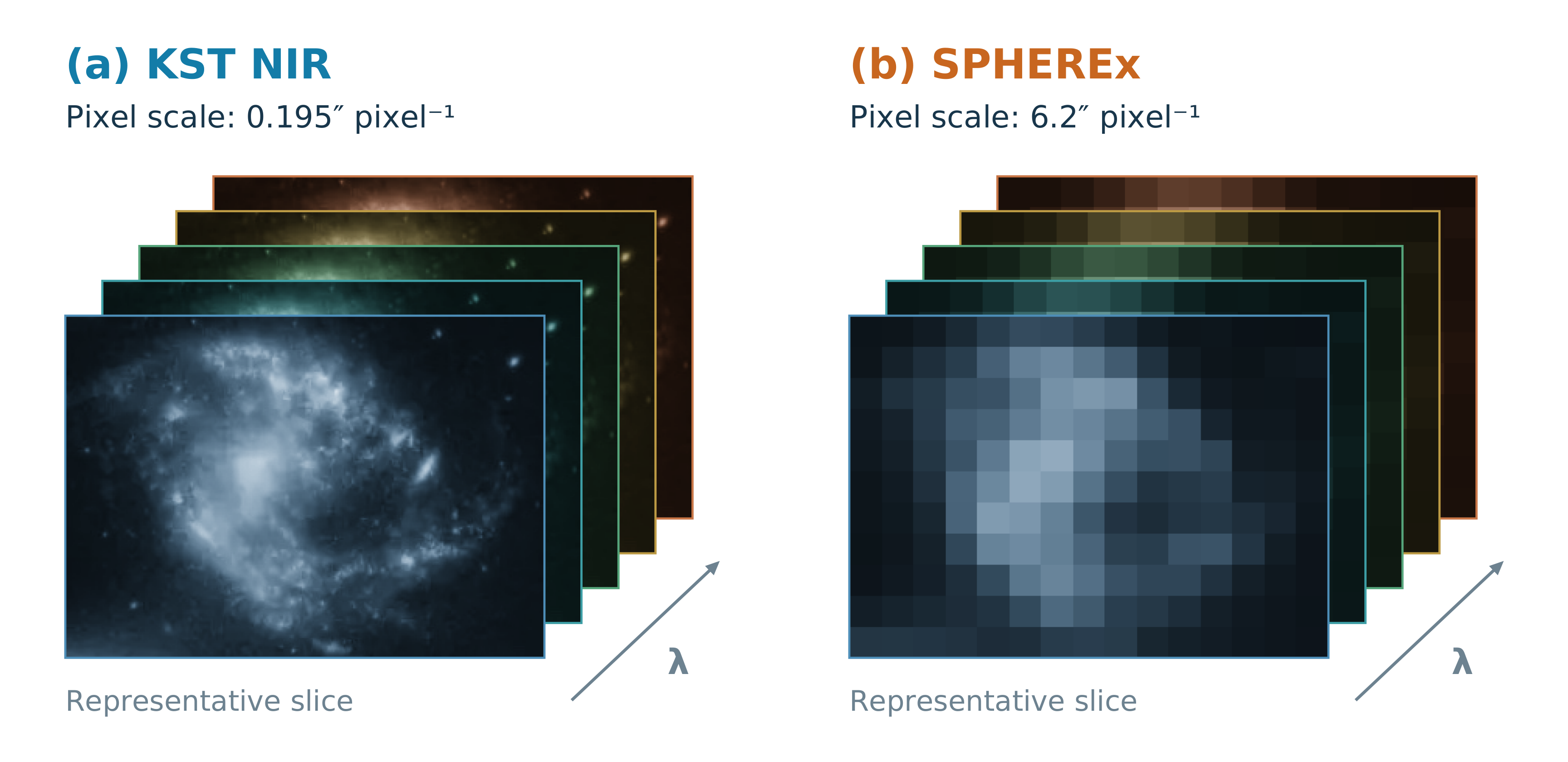}
\caption{Illustrative comparison of spatial sampling in KST and SPHEREx spectral cubes. The unobscured front planes show the same illustrative galaxy with fine and coarse sampling, while the offset planes and arrows indicate the wavelength dimension. The labels give the nominal detector pixel scales: $0.195''$/pixel for KST's NIR channel and $6.2''$/pixel for SPHEREx. Colors distinguish schematic wavelength slices, with the same underlying morphology used throughout each stack. The images provide a qualitative comparison of spatial sampling and do not incorporate instrument-specific PSF convolution, noise, or wavelength-dependent galaxy morphology.}
\label{fig:kst_vs_spherex}
\end{figure}

While both missions use LVFs, KST occupies a distinct niche:
\begin{itemize}
    \item \textbf{Pixel Resolution:} KST ($\sim0.195''$) vs. SPHEREx ($6.2''$, \citealt{spherex}). KST resolves galaxy morphology (Figure~\ref{fig:kst_vs_spherex}).
    \item \textbf{Wavelength Synergy:} KST offers VIS imaging simultaneous with its NIR spectrophotometry, while SPHEREx is NIR-only.
    \item \textbf{Depth:} KST performs deep targeted surveys ($m_{AB} \sim 26$), whereas SPHEREx is a shallower all-sky survey.
\end{itemize}

\begin{table*}[t]
\centering
\caption{Key Parameters of the Korean Space Telescope (KST)}
\label{tab:kst_specs}
\begin{tabular}{ll}
\hline
\textbf{Parameter} & \textbf{Value} \\ \hline
\multicolumn{2}{l}{\textit{Optical System}} \\
Optical Design & Korsch-type three-mirror anastigmat \\
Primary Mirror & 1.4 m \\
Focal Length & 19 m ($f/13.57$) \\
Field of View & $1^{\circ} \times 0.25^{\circ}$ \\ \hline
\multicolumn{2}{l}{\textit{VIS Channel}} \\
Detectors & CMOS (Teledyne CIS301/302) \\
Pixel Scale & 0.109 arcsec/pixel \\
Filters & g, r, i, z \\
PSF FWHM & $\sim 0.15''$ ($r$-band) \\
5$\sigma$ 1-hr Depth ($d=0.3''$ galaxy) & g: 26.1, r: 25.7, i: 25.3, z: 24.9 \\
~~~~~~~~~~~~~~~~~~~~~(point source)$^{\dagger}$ & g: 27.6, r: 27.0, i: 26.5, z: 26.0 \\ \hline
\multicolumn{2}{l}{\textit{Near-Infrared (NIR) Channel}} \\
Detectors & 27 H2RG ($9 \times 3$ mosaic) \\
Pixel Scale & 0.195 arcsec/pixel \\
Spectroscopy & LVF ($1.1$--$1.8~\mu$m, $R \sim 100$) \\
5$\sigma$ Depth (1hr) & $\sim 22.5$ AB mag per $R\sim100$ channel (conservative; Table~\ref{tab:lvf_modes}) \\ \hline
\end{tabular}
\par\smallskip
\parbox{0.92\textwidth}{\footnotesize $^{\dagger}$Point-source depths assume near-optimal PSF extraction ($\sim$50\% encircled-energy aperture), a conservative sky model (zodiacal light $+$ Earthshine, bounding the low-Earth-orbit case; an L2 orbit would be darker), an end-to-end throughput of $\approx$0.45, and the $0.15''$ PSF. The $0.3''$-galaxy depths include additional conservative margins adopted as weak-lensing requirements; the tabulated point-to-extended offset therefore exceeds the purely photometric offset ($\approx$0.85 mag).}
\end{table*}

\begin{figure*}[t]
\centering
    \includegraphics[width=0.95\textwidth]{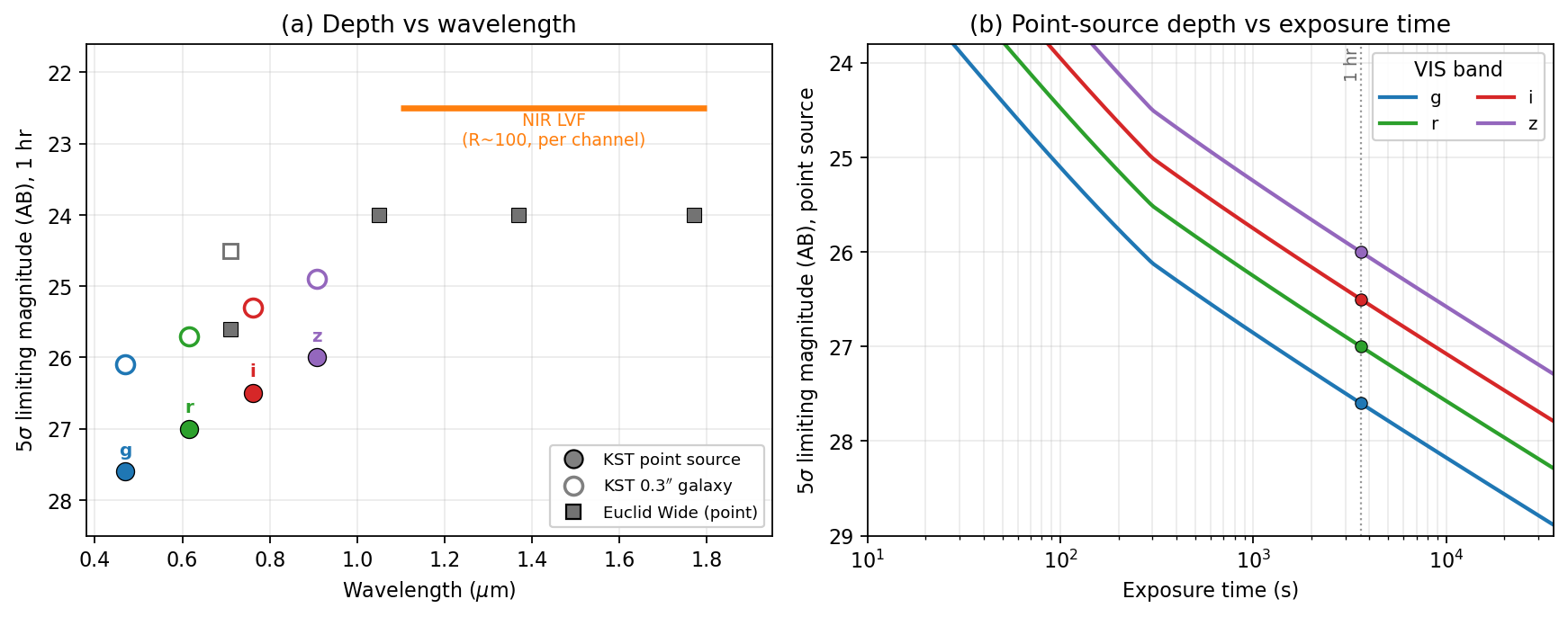}
\caption{\textbf{KST sensitivity.} \textbf{(a)} $5\sigma$ limiting AB magnitude versus wavelength for a 1-hr exposure. Filled circles show KST point-source depths in $g,r,i,z$ from a first-principles exposure-time calculator (ETC) calibrated to reproduce the tabulated depths (Table~\ref{tab:kst_specs}). Open circles show the corresponding $0.3''$-galaxy depths. The orange bar marks the NIR LVF per-resolution-element ($R\sim100$) continuum depth across $1.1$--$1.8~\mu$m, which is naturally shallower than broadband imaging. Gray squares show the \textit{Euclid} Wide survey for comparison \citep[filled: point source; open: extended VIS;][]{Euclid}. \textbf{(b)} Point-source $5\sigma$ depth versus integration time for the four VIS bands. The dotted line and filled points mark the 1-hr survey depth. The ETC assumes a 1.4-m aperture with central obscuration, correlated-double-sampling (CDS) read noise $<5\,e^-$, dark current, and a conservative zodiacal$+$Earthshine sky model (bounding the low-Earth-orbit case). The effective per-band throughput is calibrated to the point-source depths of Table~\ref{tab:kst_specs}.}
\label{fig:sensitivity}
\end{figure*}

\subsection{Spectrophotometry at the Moment of Detection}
\label{sec:simultaneity}
A defining operational advantage of KST follows from its simultaneous dual-channel design: deep visible imaging and NIR LVF spectrophotometry of the \textit{entire} field are acquired in the \textit{same} exposure. Every detected source is measured in one optical band (set by the sub-field on which it falls) and in one $R\sim100$ spectral channel simultaneously with its discovery image. Both the optical bands and the NIR wavelengths are encoded as focal-plane position (fixed per-detector filters in the VIS channel; the LVF in the NIR). The full $g,r,i,z$ set and the complete $1.1$--$1.8~\mu$m spectrum are therefore assembled by the \textit{same} survey scan, which steps each source across the four optical sub-fields and the $\sim$48 spectral channels (Section~\ref{sec:lvf_scan}), the principle employed by SPHEREx. Spectral information thus begins accumulating at the instant of detection, and a full low-resolution spectrum follows within the same scan pass (Section~\ref{sec:lvf_scan}), from the same data, with no external follow-up. The fidelity of the resulting classification and redshift is signal-to-noise dependent, and the delivered per-channel depth depends on the observing mode (Table~\ref{tab:lvf_modes}). In the Wide Survey, sources brighter than $\approx$20.8~AB receive full $R\sim100$ spectral classification from a single passage, and intermediate sources are characterized through synthetic medium-band ($R\sim20$) spectrophotometry formed by binning adjacent channels. Fainter sources retain synthesized broadband NIR photometry to $22.9$~AB in a single Wide passage ($24.0$~AB in the Deep mission stack), below which the NIR data enter the spectral-energy-distribution fit as upper limits (see Figure~\ref{fig:sensitivity}). Spectrophotometric typing and LVF redshifts therefore apply to the brighter fraction of the optical sample, while the four optical bands together with NIR detections or upper limits yield photometric redshifts for the full catalog. Deeper per-channel spectra accumulate in the Deep fields and in dedicated target-of-opportunity (ToO) stares.

\begin{table}[ht]
\centering
\caption{NIR LVF spectrophotometric depth by observing mode}
\label{tab:lvf_modes}
\small
\setlength{\tabcolsep}{4pt}
\begin{tabular}{@{}lcccc@{}}
\hline
\textbf{Mode} & $t_{\rm ch}$ & \multicolumn{3}{c}{$5\sigma$ depth (AB)} \\
 & & $R\sim100$ & $R\sim20$ & broadband \\ \hline
Benchmark (Table~\ref{tab:kst_specs}) & 1 hr & 22.5 & 23.4 & 24.6 \\
Wide Survey passage & 255 s & 20.8 & 21.7 & 22.9 \\
Deep Survey epoch & 26 s & 18.5 & 19.4 & 20.6 \\
Deep mission stack & 2.1 hr & 21.9 & 22.7 & 24.0 \\ \hline
\end{tabular}\\[3pt]
\parbox{0.97\columnwidth}{\footnotesize Per-channel exposure $t_{\rm ch}$ assumes uniform sampling of the $\sim$48 channels within each mode's total NIR integration. Depths are computed with the exposure-time calculator of Figure~\ref{fig:sensitivity}, which charges the $5\,e^-$ correlated-double-sampling read noise once per exposure in addition to dark current ($0.05\,e^-\,\mathrm{s^{-1}}$) and the zodiacal$+$Earthshine sky. Because a single $R\sim100$ channel collects little sky ($\approx$$0.006\,e^-\,\mathrm{s^{-1}\,pix^{-1}}$), the channel depths are detector-noise limited, and short exposures and stacks of short exposures fall below the background-limited $1.25\log_{10}t_{\rm ch}$ scaling. \textit{The depths are deliberately conservative:} the $22.5$~AB benchmark is adopted as a requirement-level anchor, and reproducing it requires an effective per-channel throughput of only $\approx$0.08, well below the $\approx$0.45 end-to-end throughput assumed for the visible channel (Table~\ref{tab:kst_specs}). An LVF throughput of $0.3$ would deepen every row by $\approx$1.3--1.4~mag. All rows are referenced to this anchor and should be read as lower limits pending the LVF throughput budget of the mission-definition phase. Binning $N$ adjacent channels into a synthetic band gains $1.25\log_{10}N$ mag ($R\sim20$: $N=5$; broadband: $N=48$). The $\sim$48 channels thus yield $\approx$10 independent $R\sim20$ bands. Sliding bins give more measurements but no additional independent information. Dedicated ToO stares integrate a selectable channel set and reach the benchmark row for 1-hr dwells. Claims of spectral classification and spectrophotometric redshifts throughout the science sections refer to these mode-dependent depths.}
\end{table}

\begin{figure*}[t]
\centering
\includegraphics[width=0.82\textwidth]{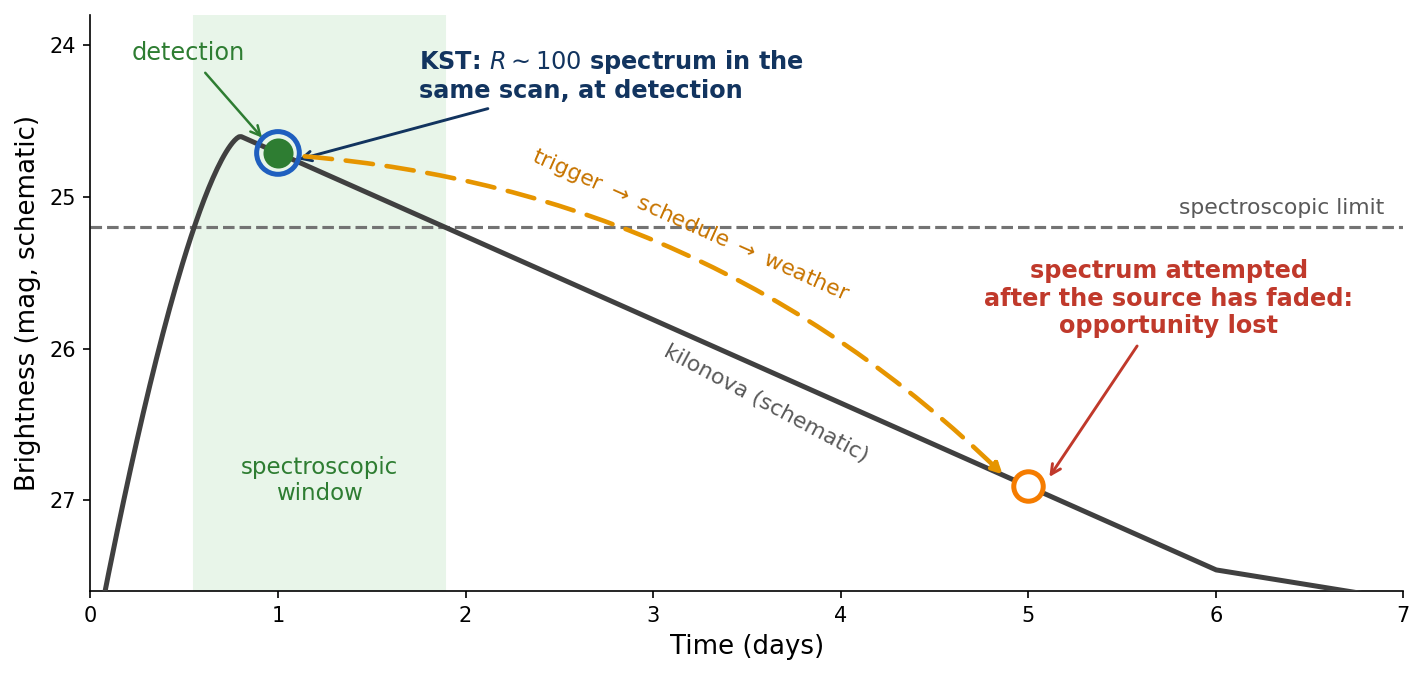}
\caption{\textbf{Spectroscopy without external follow-up.} Schematic kilonova light curve illustrating the finite spectroscopic window (green band: the source is brighter than the spectroscopic limit, dashed line). The survey detects the transient near peak (green point), and its $R\sim100$ spectrum accumulates during the same KST scan (blue circle). An externally triggered spectrum (orange) arrives only after trigger, scheduling, and weather latencies of days, by which time the kilonova has faded far below the spectroscopic limit and the opportunity is lost. The brightness scale is schematic.}
\label{fig:simultaneity}
\end{figure*}

This collapses the detection $\rightarrow$ trigger $\rightarrow$ spectroscopic-follow-up loop that bottlenecks time-domain astronomy (Figure~\ref{fig:simultaneity}). Wide-field imaging surveys such as the Vera C. Rubin Observatory detect transients photometrically but must refer candidates to external spectrographs, incurring scheduling, weather, and target-of-opportunity latencies of days, an interval over which fast-evolving sources such as kilonovae and young supernovae fade or change character. A spectrum, if obtained at all, then no longer corresponds to the discovery epoch. Because KST's spectroscopy is part of the survey itself, beginning at discovery and completing within the scan, its advantage is greatest precisely where speed is decisive: confirming and typing a kilonova counterpart to a gravitational-wave event (Section~\ref{sec:dark_sirens}), assigning a host redshift and, for the brighter events, a spectrophotometric type to a Type~Ia supernova from the same scan that detects it (Section~\ref{sec:dark_energy}), and flagging microlensing anomalies and other transients in real time.

The benefit is also statistical and calibration-level. The spectrophotometric measurement is obtained with a selection function set by brightness alone (Table~\ref{tab:lvf_modes}) rather than by follow-up scheduling. The spectroscopic subsample is therefore defined by a known, uniform flux limit instead of the heterogeneous biases that contaminate externally-followed samples, while the light curve and the spectral time series, recorded together, are intrinsically synchronized and share a single photometric calibration. Among contemporary facilities the combination is distinctive: \textit{Euclid} exposes its NISP grism in parallel with its single-band visible imager, and the Chinese Space Station Telescope (CSST) obtains slitless spectra on the same focal plane as its multi-band imaging, but both rely on grisms whose redshift yield is dominated by emission-line sources, whereas the LVF records contiguous $R\sim100$ continuum spectrophotometry at a fidelity set by source brightness. The Nancy Grace Roman Space Telescope observes its filters and dispersers sequentially.

\subsection{LVF Scan Strategy: Dispersion Orientation}
\label{sec:lvf_scan}
\begin{figure}[t]
\centering
\includegraphics[width=\columnwidth]{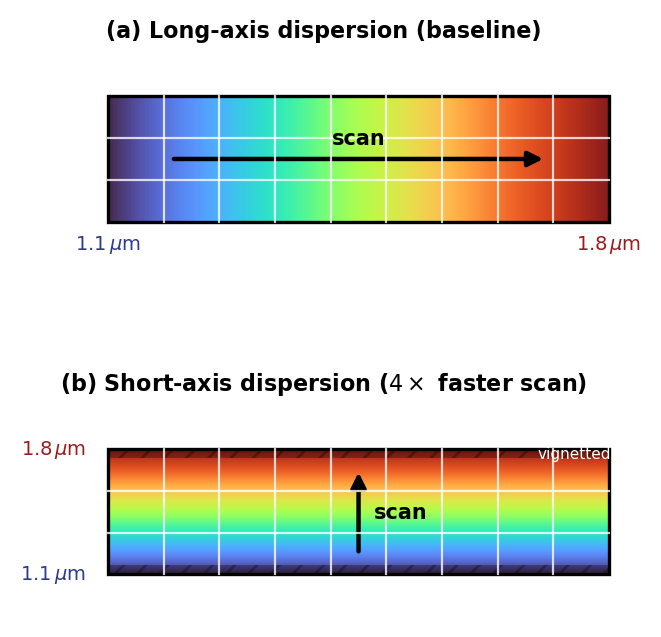}
\caption{\textbf{LVF dispersion orientation.} Two alternative design orientations, fixed at integration, for the LVF's wavelength gradient on the $1^\circ\times0.25^\circ$ focal plane (color shows the $1.1$--$1.8~\mu$m mapping; white lines, the $9\times3$ NIR detector mosaic; arrow, the scan direction). \textbf{(a)} Dispersing along the long axis (baseline) gives a gentle gradient and keeps the vignetted edges in the cross-scan direction, at the cost of a long scan to complete each spectrum. \textbf{(b)} Dispersing along the short axis completes a spectrum in a $4\times$ shorter scan and crosses fewer detector gaps, but steepens the gradient and places the vignetted edges on the extreme wavelength channels.}
\label{fig:lvf_scan}
\end{figure}

Because the LVF encodes wavelength as a function of focal-plane position, a complete $R\sim100$ spectrum is assembled by scanning a source along the dispersion direction. The orientation of that direction on the focal plane is a survey-strategy choice that leaves the first-order spectroscopic survey speed unchanged but strongly affects data quality and cadence (Figure~\ref{fig:lvf_scan}). The two options disperse the $1.1$--$1.8~\mu$m range either along the long ($1^\circ$, nine-detector) axis or along the short ($0.25^\circ$, three-detector) axis.

Dispersing along the \textbf{long axis} (the baseline) produces a gentle wavelength gradient ($\sim0.7~\mu$m\,deg$^{-1}$; a resolution element subtends $\sim75''$), which minimizes spatial--spectral mixing across resolved sources and keeps all $\sim48$ channels clear of the NIR array's vignetted rows (detector area outside the adopted science field, hatched in Figure~\ref{fig:focal_plane}), which then fall in the cross-scan direction and cost only field. Its drawbacks are a long ($1^\circ$) incomplete-spectrum ramp at the scan ends, narrow $0.25^\circ$ cross-scan strips, eight detector-gap crossings in wavelength, and a spectrum that is completed only after a full $1^\circ$ scan. Dispersing along the \textbf{short axis} compresses the same range into $0.25^\circ$, a $\sim$4$\times$ steeper gradient ($\sim19''$ per element). A spectrum is then completed in a $4\times$ shorter scan, with smaller edge losses, wider $1^\circ$ strips, and only two gap crossings. The costs are stronger spatial--spectral mixing for extended sources and vignetting of the extreme (bluest and reddest) channels.

The trade therefore pits spectrophotometric fidelity, which favors the long axis and the wide galaxy-evolution survey, against spectral cadence, which favors the short axis and the time-domain programs that exploit the rapid spectroscopy of Section~\ref{sec:simultaneity}. Because the VIS filters are coated onto the sensors and the LVF orientation relative to the detector mosaic is likewise fixed at integration, the two configurations are alternative designs rather than selectable observing modes, and the choice applies mission-wide. We adopt the long-axis orientation as the baseline. The short-axis alternative remains under consideration in the design trade should the time-domain programs (the cluster Deep fields and the high-cadence Galactic-bulge monitoring, where spectral cadence dominates) be weighted more heavily in the final mission definition.

In the baseline long-axis configuration, the observing sequence involves three timescales. A single exposure (minutes) records one optical band and one $\sim$15~nm LVF channel for every source in the field. As the scan advances ($\sim$0.30$^\circ$/hr at the Wide-survey rate), each sky position moves through the four optical sub-fields, with the per-band integration times of Section~\ref{sec:survey}, and across the $\sim$48 spectral channels. A complete $g,r,i,z$ set and a complete spectrum are thus both assembled within one scan pass, with $\sim$3.4 hours elapsing between the first and the last wavelength measurement ($\sim$4$\times$ shorter in the short-axis configuration). The step pitch, per-channel dwell, exposure sub-structure, and overlap strategy are science-case dependent and remain mission-definition-phase deliverables. The values quoted here follow directly from the survey rate and the LVF gradient.

\subsection{Science-to-Engineering Requirements Traceability}

To realize the scientific potential of KST, the mission design is governed by a strict flow-down of requirements from science goals to engineering specifications. The 1.4-meter aperture is only the starting point. Achieving a diffraction-limited survey requires rigorous control over spacecraft dynamics and environmental stability. The key engineering drivers derived from the science objectives are detailed below and summarized in Table \ref{tab:req_matrix}.

\subsubsection{Pointing, Tracking, and Jitter Control}
The most critical engineering constraint is driven by the Weak Gravitational Lensing science case, which requires precise shape measurements of galaxies. Any uncorrected motion of the spacecraft during an exposure blurs the Point Spread Function (PSF), introducing systematic errors that can mimic cosmic shear.

\begin{itemize}
    \item \textbf{Spacecraft Jitter:} High-frequency vibrations (jitter) from reaction wheels or cryocoolers must be minimized. To maintain a system-level PSF FWHM of $\sim0.15''$, the RMS pointing jitter is constrained to be less than $10\%$ of the PSF width, requiring $\sigma_{\rm jitter} \lesssim 0.015''$ ($1\sigma$) over the integration time.
    \item \textbf{Tracking Stability:} For deep-field exposures necessary to reach $m_{AB} \sim 26$, the attitude control system (ACS) must maintain a rigid lock on the field. This requires fine guidance sensors (FGS) located directly on the focal plane to provide closed-loop feedback with a tracking accuracy better than $0.02''$.
\end{itemize}

\subsubsection{Thermal and Wavefront Stability}
\label{sec:thermal_stability}
To distinguish between competing dark energy models, the multiplicative shear calibration bias ($m$) must be controlled to $m < 10^{-3}$. This imposes strict stability requirements on the optical bench:
\begin{itemize}
    \item \textbf{Athermal Design:} The optical structural assembly must utilize low-expansion materials (e.g., Silicon Carbide or Carbon Fiber Reinforced Polymer) to prevent focus drift or the induction of coma due to orbital thermal variations.
    \item \textbf{Wavefront Error (WFE) Drift:} The time-dependent variation of the WFE is limited to $< 5$ nm RMS between wavefront sensing calibration epochs.
\end{itemize}

\begin{table*}[t]
\centering
\caption{Expanded Science-to-Engineering Requirements Traceability Matrix}
\label{tab:req_matrix}
\resizebox{\textwidth}{!}{%
\begin{tabular}{p{3.5cm} p{5.0cm} p{7.0cm}}
\hline
\textbf{Science Driver} & \textbf{Measurement Requirement} & \textbf{Engineering Constraint} \\ \hline
\textbf{Weak Lensing \newline (Dark Energy $+$ Dark Matter)} & Minimize shape noise \& shear bias & $\bullet$ \textbf{Jitter:} $< 0.015''$ RMS ($1\sigma$) \newline $\bullet$ \textbf{Tracking:} $< 0.02''$ stability (300~s) \newline $\bullet$ \textbf{WFE Drift:} $< 5$ nm RMS (thermal) \\ \hline
\textbf{Dark Matter \newline (substructure lensing)} & Resolve lensed-image structure \& low-surface-brightness arcs & $\bullet$ \textbf{Resolution:} diffraction-limited $\sim0.15''$ PSF \newline $\bullet$ \textbf{PSF Knowledge:} stable, well-sampled, modeled to $m<10^{-3}$ \\ \hline
\textbf{Dark Energy (SNe Ia)} & Standardize luminosity distances & $\bullet$ \textbf{Photometric Stability:} $< 1\%$ (relative) \newline $\bullet$ \textbf{Calibration:} Onboard Flat Field Source ($0.1\%$ stability) \\ \hline
\textbf{Dark Sirens \newline (GW follow-up)} & Rapid counterpart ID \& complete host redshifts & $\bullet$ \textbf{Slew and Settle:} Settling $< 2$ min for ToO \newline $\bullet$ \textbf{Data Latency:} Priority downlink $< 15$ min \newline $\bullet$ \textbf{Host Catalog:} deep NIR photo-$z$ \\ \hline
\textbf{Legacy \newline (deep fields, microlensing)} & Background-limited detection \& bright-star dynamic range & $\bullet$ \textbf{Read Noise:} $< 5\,e^-$ (CDS) \newline $\bullet$ \textbf{Dark Current:} $< 0.05\,e^-\,\text{s}^{-1}$ ($T_{\rm det} < 100$~K) \newline $\bullet$ \textbf{Full Well:} $> 100{,}000\,e^-$, non-lin.\ $< 1\%$ \newline $\bullet$ \textbf{Orbit:} Sun-synchronous low Earth orbit (LEO) or Sun--Earth L2 (continuous-viewing capability; trade study in the mission-definition phase) \\ \hline
\end{tabular}%
}
\end{table*}

\subsubsection{Detector Electronics and Dynamic Range}
The simultaneous observation requirement imposes conflicting demands on the readout electronics, which must be reconciled to satisfy both faint and bright science cases.
\begin{itemize}
    \item \textbf{Low-Noise High-z Sensitivity:} To reach the background-limited depth of $m_{AB} \sim 26$ for galaxy evolution studies, the read noise must be minimized. The CMOS (VIS) and H2RG (NIR) readout electronics are required to achieve a Correlated Double Sampling (CDS) read noise of $< 5\,e^-$ RMS. 
    \item \textbf{High Dynamic Range for Legacy Fields:} The crowded, low-latitude fields underpinning the Legacy Science programs (the Galactic bulge and inner disk, used for both Milky Way bright-star science and exoplanet microlensing) contain bright and faint stars in the same frame. To monitor the brightest source stars without saturation while detecting faint time-domain anomalies, the detectors must support a high dynamic range. The engineering requirement is a full-well capacity of $> 100{,}000\,e^-$ with non-linearity corrected to $< 1\%$ up to $90\%$ of saturation.
\end{itemize}

\subsubsection{Data Volume and Onboard Processing}
The KST focal plane represents a massive data generator. With a $\sim423$ Megapixel aggregate focal plane (approx.\ 310 MP VIS $+$ 113 MP NIR) and high-cadence microlensing observations, the raw data rate exceeds standard downlink capacities.
\begin{itemize}
    \item \textbf{Onboard Compression:} To fit within the K/Ka-band radio downlink budget ($\sim 1.5$ Gbps), the mission requires lossless onboard data compression (e.g., Rice algorithm) with a compression ratio of $\sim 2.5:1$.
    \item \textbf{Transient Alert Latency:} For the Dark Sirens program (Section~\ref{sec:dark_sirens}), the latency between photon arrival and ground alert must be minimized. The spacecraft avionics must support an autonomous ``burst mode'' downlink trigger upon receiving a target-of-opportunity (ToO) command, ensuring GW counterpart data is prioritized in the downlink queue within $< 15$ minutes of observation.
\end{itemize}

\subsubsection{Calibration Hardware and Metrology}
Achieving the $< 1\%$ photometric stability for Type Ia Supernovae and the precise shape measurement for Weak Lensing requires dedicated onboard calibration hardware.
\begin{itemize}
    \item \textbf{Relative Flux Calibration:} A stable onboard illumination source (e.g., LED-based flat field stimulator) is required to monitor pixel-to-pixel sensitivity variations to a precision of $0.1\%$ on daily timescales.
    \item \textbf{Wavefront Sensing:} To track the PSF stability requirements ($\Delta \text{WFE} < 5$ nm), the focal plane will include Phase Retrieval capabilities using out-of-focus imagery of stars, performed during slew operations to update the active optics control loop without interrupting science operations.
\end{itemize}

\section{Core Science: Probing the Dark Sector}
\label{sec:darksector}
KST is conceived around a single mission identity. It is the only dedicated space observatory built to weigh all three components of the dark sector with one instrument. Its three flagship science pillars---\textbf{Dynamical Dark Energy}, the \textbf{Nature of Dark Matter}, and \textbf{Dark Sirens}---are not independent programs but facets of one strategy. By measuring \textit{both} the geometry of cosmic expansion and the growth of cosmic structure from the same simultaneous VIS+NIR survey, KST addresses the dark sector along the very axes whose combination separates evolving dark energy from modified gravity and pins down the microphysics of dark matter. The following three subsections develop each pillar in turn.

\subsection{Pillar I: Dynamical Dark Energy}
\label{sec:dark_energy}

The first science pillar of the Korean Space Telescope (KST) targets the physics of dark energy, the $\sim 68\%$ of the universe's energy budget driving cosmic acceleration. For over two decades, the standard $\Lambda$CDM model—postulating a cosmological constant ($\Lambda$) and cold dark matter (CDM)—has successfully described the universe's large-scale properties \citep{Planck2020}. However, as measurement precision has improved, significant cracks have emerged in this foundation.

The most prominent of these is the ``Hubble Tension'', a statistically significant $5\sigma$ discrepancy between the Hubble constant ($H_0$) predicted by early-universe physics and the value measured locally \citep{Riess2022}. Additionally, the ``$S_8$ Tension'' suggests the universe is smoother than predicted by the Cosmic Microwave Background (CMB).
Dynamical dark energy may be a missing piece needed to bridge this gap.

Most critically, \textbf{recent results from the Dark Energy Spectroscopic Instrument (DESI)} have provided some of the strongest observational hints that dark energy may not be a cosmological constant ($\Lambda$), but instead evolves with time \citep{DESI2024,DESIDR2Cosmo}. When combined with CMB and SN data, DESI BAO measurements favor dynamical dark energy over $\Lambda$CDM at a significance level of $\sim 3\sigma$, with the preferred Chevallier--Polarski--Linder (CPL; \citealt{Chevallier2001,Linder2003}) equation of state exhibiting phantom crossing \citep{DESI2024,DESIDR2Cosmo}. Model-agnostic and non-parametric reconstructions have yielded similar results \citep{Calderon2024,Lodha2025}. Confirming this dynamical dark energy scenario requires datasets with independent and complementary systematics, while robustly establishing phantom crossing requires broader and higher-redshift coverage \citep{LinderHuterer2003}.

By combining wide-field optical imaging with contiguous NIR LVF spectrophotometry ($R \sim 100$), KST will construct a multi-probe dataset free from atmospheric systematics. Within this pillar, KST constrains the dark energy equation of state, $w(z)$, through four orthogonal probes: Weak Gravitational Lensing (WL), Strong Lensing (SL), Type Ia Supernovae (SNe Ia), and Large-Scale Structure (LSS). The decisive strength of KST is that the \textit{same} dataset measures both the geometry of cosmic expansion (SNe, BAO and the Alcock--Paczy\'nski (AP) test, time delays) and the growth of structure (WL, cluster abundance), the combination required to distinguish evolving dark energy from a modification of gravity. The microphysical nature of dark matter probed by these same lensing data is developed separately in Pillar~II (Section~\ref{sec:dark_matter}).

\subsubsection{Weak Gravitational Lensing}
Weak gravitational lensing (WL)—the subtle, coherent distortion of background galaxy shapes by foreground matter—is widely regarded as the most powerful technique for mapping the dark universe. Unlike galaxy clustering, which traces light (baryons) and relies on uncertain ``bias'' factors to infer the underlying mass, weak lensing relies solely on General Relativity. Because gravity couples to all energy densities, weak lensing provides a direct, unbiased map of the total matter distribution, dominated by dark matter \citep{Bartelmann2001}.

KST will use background-galaxy shapes to reconstruct the projected matter distribution in tomographic redshift bins. This map allows us to measure the history of cosmic expansion and the growth rate of structure simultaneously, providing a stringent test of general relativity and the nature of dark energy.

\subsubsubsection{Theoretical Framework and Observables}
The fundamental observable in weak lensing is the ``Cosmic Shear'' ($\gamma$). As light from a distant source traverses the inhomogeneous universe, the differential gravitational potential deflects the light rays, mapping a circular source into an ellipse. This distortion is described by the Jacobian matrix of the lens mapping:
\begin{equation}
    A = \left( \begin{array}{cc} 1-\kappa-\gamma_1 & -\gamma_2 \\ -\gamma_2 & 1-\kappa+\gamma_1 \end{array} \right)
\end{equation}
where $\kappa$ is the convergence (isotropic magnification related to the projected surface mass density) and $\gamma = \gamma_1 + i\gamma_2$ is the complex shear (anisotropic stretching).

\subparagraph{The Shear Power Spectrum and $S_8$}
In the weak lensing regime ($\kappa, \gamma \ll 1$), we measure the statistical correlation of galaxy ellipticities. The two-point correlation functions of the shear field, $\xi_{\pm}(\theta)$, are Fourier-transformed into the shear power spectrum, $C_{\ell}^{\gamma\gamma}$. This power spectrum is a line-of-sight projection of the 3D matter power spectrum $P_\delta(k,z)$, weighted by the geometric ``lensing efficiency'' kernel:
\begin{equation}
    C_{\ell}^{\gamma\gamma} \propto \int_0^{z_{\text{source}}} dz \frac{W(z)^2}{H(z)\, D_M^2(z)}\, P_\delta\!\left(k=\frac{\ell+1/2}{D_M(z)},\, z\right)
\end{equation}
where $D_M(z)$ is the transverse comoving distance (equal to the comoving distance $\chi$ in a flat universe), $W(z)$ is the lensing efficiency of the source redshift distribution, and $k=(\ell+1/2)/D_M$ is the Limber approximation \citep{LoVerde2008}. This equation encapsulates the dual sensitivity of weak lensing:
\begin{itemize}
    \item \textbf{Geometry:} The kernel $W(z)$ depends on ratios of lens and source distances, making it sensitive to the expansion history $H(z)$ and dark energy ($w_0, w_a$).
    \item \textbf{Growth:} The matter power spectrum $P_\delta(k,z)$ traces the clustering of matter, which is sensitive to the total matter density $\Omega_m$ and the amplitude of fluctuations $\sigma_8$.
\end{itemize}
Current ground-based surveys suffer from a degeneracy between these parameters, typically constrained as the aggregate parameter $S_8 \equiv \sigma_8 (\Omega_m / 0.3)^{0.5}$. The ``$\sigma_8$ Tension''—where low-redshift lensing surveys measure a lower clustering amplitude than predicted by the high-redshift CMB—is one of the most significant anomalies in current cosmology. KST's precision is designed to resolve this tension by reaching higher redshifts and removing the atmospheric systematics that plague ground-based measurements. Scaling from current Stage~III surveys to KST's source density ($n_{\rm eff}\approx57~\text{arcmin}^{-2}$) and its NIR tomography to $z\sim3$ over the $2{,}000~\text{deg}^2$ footprint, we forecast an $\approx$1\% measurement of the clustering amplitude, $\sigma(S_8)\approx0.008$.

\paragraph{Systematic Control: The Space Advantage}
The cosmological signal from weak lensing is minute. The typical shear induced by large-scale structure is of order $\gamma \approx 0.01$ (a 1\% change in axis ratio). In contrast, the intrinsic ellipticity of galaxies is $\sigma_e \approx 0.3$, and instrumental distortions can induce spurious ellipticities of $\sim$5--10\%. Success in weak lensing is therefore defined not by raw photon count, but by the control of systematic errors.

\subparagraph{The Point Spread Function (PSF) Challenge}
The primary enemy of weak lensing is the PSF. If the telescope and atmosphere smear the galaxy image anisotropically, the galaxy will appear elongated in the direction of the smear, mimicking a gravitational lensing signal.
\begin{itemize}
    \item \textbf{Ground-Based Limitations:} Ground-based telescopes view the universe through the turbulent atmosphere. The PSF is stochastic, varying on millisecond timescales and changing shape across the field of view. Furthermore, the atmosphere acts as a chromatic prism (differential refraction), causing the PSF size and ellipticity to depend on the color of the galaxy (which is often unknown to high precision). Removing these effects requires complex modeling that leaves residual biases.
    \item \textbf{KST's Diffraction-Limited Stability:} KST operates in the vacuum of space, free from atmospheric turbulence. Its PSF is determined solely by the telescope optics, which are thermally stabilized. This yields a PSF that is diffraction-limited ($\text{FWHM} \approx 0.15''$) and temporally stable.
    \item \textbf{Chromatic PSF Control:} A diffraction-limited PSF scales with wavelength. The effective PSF of any source is therefore weighted by its spectral energy distribution across the filter. \textit{Euclid} measures shapes in a single ultra-broad visible band ($\sim$550--900~nm), across which the PSF size varies by tens of percent. Because the stars used to model the PSF have different SEDs from the galaxies being measured, the PSF must be reconstructed per-object from external multi-band photometry, and residual spectral-energy-distribution (SED) and color-gradient errors propagate directly into shear bias \citep{Cypriano2010, EriksenHoekstra2018}. KST instead measures shapes in four separate $g,r,i,z$ bands of a few times narrower fractional bandwidth, shrinking the in-band chromatic PSF variation accordingly, while its own $griz$$+$LVF SEDs enable per-object PSF interpolation and a null test unavailable to a single-band imager—shear measured independently in four bands must agree.
    \item \textbf{Multiplicative Bias Control:} The stability of the KST PSF allows for precise deconvolution. We forecast that KST will control the multiplicative shear calibration bias $m$ (where $\gamma_{\text{obs}} = (1+m)\gamma_{\text{true}}$) to levels below $m < 10^{-3}$. This represents an order-of-magnitude improvement over current ground-based limits, ensuring that the inferred dark-energy parameters are not limited by measurement error.
\end{itemize}

\subparagraph{High-Resolution Shape Measurement}
The second limiting factor is ``shape noise''—the random intrinsic shapes of galaxies. The only way to beat this noise is to average over a larger number of galaxies.
\begin{itemize}
    \item \textbf{Resolving the Faint Universe:} Ground-based surveys are limited by ``seeing'' (typically $\sim 0.7''$). Small, faint galaxies—which are the most numerous—are blurred into unresolved blobs, making their shapes unmeasurable. KST's $0.15''$ resolution resolves these faint galaxies ($r \sim 26$), effectively unlocking a new population of background sources.
    \item \textbf{Effective Source Density:} This resolution gain roughly quadruples the effective source density to $n_{\text{eff}} \approx 57~\text{gal/arcmin}^{2}$. High source density allows for finer tomographic binning and higher angular resolution in the reconstructed mass maps.
\end{itemize}

\paragraph{Tomographic Lensing: A 3D Map of the Universe}
Lensing is inherently a cumulative 2D effect (integrated along the line of sight). However, by using photometric redshifts (photo-$z$), we can divide the source galaxies into redshift ``shells.'' This technique, known as tomography, recovers the 3D evolution of the dark matter distribution.

\subparagraph{Mapping the ``Redshift Desert''}
Current weak lensing surveys (e.g., the Dark Energy Survey, DES, and the Kilo-Degree Survey, KiDS) are limited to $z < 1.0$--$1.5$ due to the lack of deep NIR data. This leaves the ``Redshift Desert'' ($1.5 < z < 3.0$), the matter-dominated epoch that anchors the expansion history \textit{before} dark energy becomes dynamically dominant at $z \lesssim 0.7$, largely unmapped.
\begin{itemize}
    \item \textbf{NIR Synergy:} KST's synthetic NIR photometry, formed by binning the LVF channels (Table~\ref{tab:lvf_modes}), combined with deep optical data enables precise photo-z estimates out to $z \sim 3$ for NIR-detected sources.
    \item \textbf{Evolution of Dark Energy:} By measuring the growth of structure $D(z)$ across these high-redshift bins, KST breaks the degeneracy between dark energy and modified gravity. If dark energy is dynamical, its signature will be most prominent in the growth rate differences at these intermediate redshifts.
\end{itemize}

\paragraph{Mitigating Intrinsic Alignments}
The most pernicious astrophysical systematic in weak lensing is Intrinsic Alignment (IA). Galaxies that form in the same tidal field tend to align their shapes physically, creating a false correlation signal that mimics lensing.
\begin{itemize}
    \item \textbf{The ``Gravitational-Intrinsic (GI)'' Term:} A background galaxy is lensed by a foreground structure, while a foreground galaxy is physically aligned with that same structure. This creates a negative correlation (Gravitational-Intrinsic) that suppresses the signal.
    \item \textbf{Self-Calibration with KST:} The IA signal depends strongly on the galaxy type (red passive galaxies align strongly, while blue star-forming galaxies do not) and redshift. KST's high-quality photo-z and LVF-based spectral classification allow for the rigorous separation of the sample into ``red'' and ``blue'' tomographic bins. By jointly modeling cosmic shear, galaxy--galaxy lensing, and galaxy clustering across these bins (the ``$3 \times 2$pt'' analysis), KST can self-calibrate the IA amplitude, turning a major systematic source into a valuable probe of galaxy formation physics.
\end{itemize}

\paragraph{Beyond the Power Spectrum: Non-Gaussian Statistics}
The matter density field is Gaussian at early times, but gravitational collapse introduces significant non-Gaussianity at late times (structure formation). The standard power spectrum analysis discards this non-Gaussian information.
\begin{itemize}
    \item \textbf{Peak Counts:} KST's high source density allows for the construction of high-resolution ``mass maps.'' Counting the number of high-significance peaks in these maps provides a direct census of massive galaxy clusters. Peak counts are highly sensitive to $\Omega_m$ and break the degeneracies inherent in 2-point statistics.
    \item \textbf{Minkowski Functionals:} Topological descriptors of the density field (area, perimeter, genus) provide a powerful test of primordial non-Gaussianity and Modified Gravity. KST's wide-field, high-resolution maps are ideal for these higher-order statistical tests, extracting information that is completely invisible to standard power spectrum analysis.
\end{itemize}

\paragraph{Cluster Abundance Cosmology}
The same convergence maps yield a catalog of galaxy clusters detected as high-significance weak-lensing peaks and confirmed by optical richness. Scaling the cluster surface density of current Stage~III imaging surveys to KST's depth over the $2{,}000~\text{deg}^2$ footprint, we expect of order $10^{4}$ clusters out to $z\sim1.5$. Crucially, KST measures each cluster's mass \textit{directly} from its diffraction-limited weak-lensing signal, with the multiplicative shear bias controlled to $m<10^{-3}$. The abundance as a function of mass and redshift is thus calibrated without the mass--observable scaling uncertainties that limit optical and X-ray cluster cosmology. This delivers an independent constraint on $\sigma_8$ and $\Omega_m$ and a further, growth-based handle on dark energy that is complementary to, and jointly analyzable with, the cosmic-shear and BAO probes above.

\subsubsection{Strong Lensing: Time-Delay Cosmography for $H_0$}
Strong gravitational lensing occurs when a massive foreground object—such as a massive elliptical galaxy or a galaxy cluster—lies directly along the line of sight to a background source. The immense gravitational potential of the lens warps the surrounding spacetime, creating multiple, highly distorted, and magnified images of the background object.

Strong lensing offers a purely geometric probe of the universe's expansion history and a gravitational test of the nature of dark matter on sub-galactic scales. KST, combining wide-field monitoring, high angular resolution, and cluster-targeted deep fields, is well placed to advance this field substantially.

\paragraph{Time-Delay Cosmography: A Late-Universe Anchor for $H_0$}
Resolving the $5\sigma$ Hubble tension introduced above requires a probe that is independent of both the Planck CMB systematics and the Cepheid-based distance ladder. Strong lensing time delays provide exactly this anchor.

\subparagraph{The Physics of the Time Delay}
When a source is multiply imaged, the light rays for each image traverse different paths through the universe. Consequently, they arrive at the observer at different times. The time delay $\Delta t_{ij}$ between any two images $i$ and $j$ is given by:
\begin{equation}
    \Delta t_{ij} = \frac{D_{\Delta t}}{c} \left[ \frac{1}{2} (\theta_i - \beta)^2 - \frac{1}{2} (\theta_j - \beta)^2 - (\psi(\theta_i) - \psi(\theta_j)) \right]
\end{equation}
where $\theta$ are the image positions, $\beta$ is the unlensed source position, and $\psi$ is the lensing potential. The critical cosmological information is contained in the Time-Delay Distance, $D_{\Delta t}$, which is a combination of angular diameter distances to the lens ($D_d$), the source ($D_s$), and between them ($D_{ds}$):
\begin{equation}
    D_{\Delta t} \propto \frac{D_d D_s}{D_{ds}} \propto \frac{1}{H_0}
\end{equation}
Because $D_{\Delta t}$ is inversely proportional to the Hubble Constant, measuring the time delay provides a direct measurement of $H_0$. Crucially, this measurement is purely geometric and independent of the standard candle calibration used in the local distance ladder.

\subparagraph{The Mass-Sheet Degeneracy: The Quasar Limit}
Traditionally, time-delay cosmography has relied on lensed quasars \citep[e.g.,][]{Wong2020}. While successful, quasar lensing suffers from a fundamental systematic limitation known as the Mass-Sheet Degeneracy (MSD). The MSD arises because the mass distribution of the lens can be mathematically transformed (by adding a uniform sheet of mass density $\kappa_\lambda$) in a way that leaves the image positions and flux ratios unchanged, but rescales the time delay—and thus the inferred $H_0$—by a factor of $(1-\lambda)$.

This degeneracy is the dominant source of uncertainty in quasar-based measurements. Breaking it requires independent constraints on the lens mass, typically obtained via stellar kinematics, which introduces its own complex systematics regarding anisotropy and the Initial Mass Function (IMF).

\begin{figure}[t]
\centering
\includegraphics[width=\columnwidth]{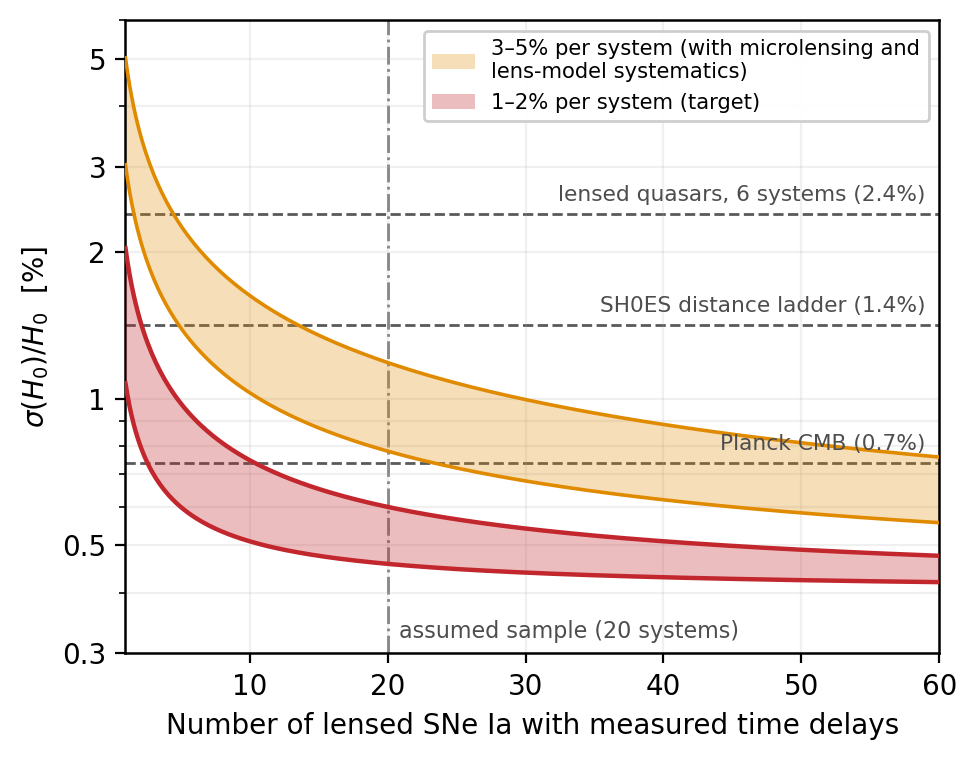}
\caption{\textbf{Time-delay $H_0$ from cluster-lensed supernovae.} Projected fractional precision on $H_0$ versus the number of multiply imaged SNe~Ia with measured time delays, for the target per-system precision of $1$--$2\%$ enabled by the standard-candle breaking of the mass-sheet degeneracy (red) and for a degraded $3$--$5\%$ per system that allows for microlensing of the images and cluster lens-model systematics (orange). Both include an illustrative $0.4\%$ systematic floor in quadrature. Horizontal lines mark current published precisions: the six-system lensed-quasar measurement \citep{Wong2020}, the SH0ES distance ladder \citep{Riess2022}, and the \textit{Planck} CMB inference \citep{Planck2020}. For an assumed sample of 20 systems (vertical line), the two cases give $\sigma(H_0)/H_0\approx0.5$--$0.6\%$ and $\approx0.8$--$1.2\%$. The sample size itself is a planning assumption.}
\label{fig:h0lens}
\end{figure}

\subparagraph{Lensed Type Ia Supernovae and the Mass-Sheet Degeneracy}
KST complements this approach by targeting lensed Type Ia supernovae (SNe Ia) \citep{OguriMarshall2010,Goldstein2019,Birrer2022}. Unlike stochastic quasars, SNe Ia are standard candles with a known intrinsic luminosity.
\begin{itemize}
    \item \textbf{Breaking the MSD:} The standard candle nature of SNe Ia directly constrains the absolute magnification $\mu$ of the lensed images. Since the Mass-Sheet transformation rescales the magnification ($\mu \to \mu / (1-\lambda)^2$), knowing the intrinsic luminosity fixes $\lambda$. This breaks the Mass-Sheet Degeneracy directly from the photometric data, without requiring external velocity dispersion data.
    \item \textbf{Precision Target and Limitations:} Breaking the MSD in this way sets a target precision of $\sim$1--2\% per system \citep{Birrer2022}. In practice, microlensing of the individual images by stars and intracluster light perturbs their magnifications by $\sim$0.1--0.3~mag, and cluster lens-model systematics and the calibration of the SN absolute magnitude add further uncertainty. We therefore also consider a degraded case of $\sim$3--5\% per system. The cluster-lensed supernovae analyzed to date have yielded $H_0$ to $\sim$7\% (SN Refsdal; \citealt{Kelly2023}) and $\sim$9\% (SN H0pe; \citealt{Pascale2025}). For an assumed sample of $\sim$20 systems, whose yield in the Deep fields remains to be established by simulation, the target and degraded cases give $\sigma(H_0)/H_0\approx0.5\%$ and $\approx1\%$, respectively (Figure~\ref{fig:h0lens}), a percent-level cross-check on the Hubble tension that depends on the distance ladder only through the SN absolute-magnitude calibration. At the lower end of the $10$--$20$-system yield estimated in Section~\ref{sec:survey}, these figures degrade by a factor of $\sqrt{2}$.
\end{itemize}

\paragraph{KST Implementation: The Cluster-Targeted Strategy}
Lensed supernovae are rare transient events. Catching them requires monitoring a massive amount of lensing mass with high cadence. KST's survey strategy is optimized for this:
\begin{itemize}
    \item \textbf{Cluster-Centric Fields:} By centering the $10~\text{deg}^2$ Deep Survey on massive galaxy clusters (the most efficient lenses in the universe), KST maximizes the lensing cross-section. This effectively turns the galaxy clusters into ``cosmic telescopes,'' boosting the discovery rate of background high-z supernovae.
    \item \textbf{High-Cadence Monitoring:} Measuring the time delay requires constructing a precise light curve for each lensed image to determine the time-shift. KST's 5-day continuous cadence is ideal for sampling the rise and fall of SNe Ia, allowing for delay measurements with uncertainties of just a few days.
    \item \textbf{NIR Advantage:} KST's NIR sensitivity ($1.1$--$1.8~\mu$m) reduces contamination from the lens galaxy's light and from any residual dust, and favors detection of highly magnified, high-redshift images near the lens core.
\end{itemize}

\subsubsection{Type Ia Supernovae: Standard Candles in an Accelerating Universe}
While Type Ia Supernovae (SNe Ia) provided the initial discovery of cosmic acceleration, they remain the most direct and potent probe of the expansion history $H(z)$ and the dark energy equation of state parameter $w$. However, the next generation of dark energy experiments is no longer limited by statistics but by systematic uncertainties. Current optical surveys are hitting a ``systematics floor'' driven by dust extinction correction, intrinsic color dispersion, and population drift.

KST is designed to break through this floor. By extending the observation window into the NIR (the rest-frame NIR for $z\lesssim0.6$ and the rest-frame red optical beyond) and targeting massive galaxy clusters, KST constructs a Hubble Diagram that is fundamentally different—and cleaner—than any predecessor.

\paragraph{Near-Infrared Standardization of Type Ia Supernovae}
The standard practice in supernova cosmology relies on the empirical ``width-luminosity'' relation (the Phillips relation) to standardize SN Ia peak magnitudes. In the optical rest-frame ($B$-band), SNe Ia are not standard candles. They exhibit significant intrinsic scatter ($\sim 0.15$ mag) and a strong correlation between decline rate and luminosity.

In the NIR, the physics changes in our favor.
\begin{itemize}
    \item \textbf{True Standard Candles:} Theoretical models and empirical data show that the width-luminosity relation flattens significantly in the NIR ($Y$, $J$, $H$ bands). The secondary maximum in the IR light curve, driven by the recombination of Fe III to Fe II, renders the peak luminosity less sensitive to the $^{56}$Ni mass and opacity variations that drive optical scatter. As a result, the intrinsic scatter of SNe Ia drops to $\sigma_{\text{int}} \approx 0.08$--$0.10$ mag in the NIR. This effectively doubles the statistical weight of each KST-discovered supernova compared to an optical counterpart.
    \item \textbf{Breaking the Dust Degeneracy:} The correction for host galaxy dust extinction is among the largest systematic errors in supernova cosmology. The extinction law $A_\lambda$ scales roughly as $\lambda^{-1}$. KST's $1.1$--$1.8~\mu$m channel samples the rest-frame $Y$ band for $z\lesssim0.6$ and the $J$ band for $z\lesssim0.45$, where the extinction dimming is a factor of $\sim$3--5 smaller than in the rest-frame $B$ band ($0.44~\mu\text{m}$), and the rest-frame red optical at higher redshift, where the gain is smaller. Furthermore, this reduces the reliance on the highly uncertain total-to-selective extinction ratio $R_V$, which is known to vary between host galaxies.
\end{itemize}

\paragraph{The Cluster Advantage: A Dual-Population Strategy}
A defining strategic choice in KST's design is the placement of its Deep Survey fields centered on massive galaxy clusters. This is not merely to utilize gravitational lensing but to secure a clean, systematics-controlled sample of supernovae. KST will detect two distinct populations of supernovae in the same fields:

\subparagraph{The ``Clean'' Sample: Cluster Member Supernovae}
The majority of SNe Ia detected in current surveys occur in star-forming field galaxies (spirals/irregulars). These environments are fraught with dust lanes and complex stellar populations. By monitoring massive galaxy clusters, KST accesses the population of SNe Ia occurring in ``massive elliptical cluster members''.
\begin{itemize}
    \item \textbf{Dust-Free Environments:} Massive elliptical galaxies in cluster cores are gas-poor and essentially dust-free. SNe Ia occurring in these hosts require negligible extinction corrections, removing the primary systematic uncertainty entirely.
    \item \textbf{Progenitor Systematics Control:} Recent studies hint at an environmental dependence of SN Ia luminosity (the ``mass step''), where SNe in massive, passive galaxies appear intrinsically brighter than those in active, low-mass hosts. By comparing the Hubble residuals of the passive Cluster Member sample against the active Field Galaxy sample (detected in the background/foreground of the same fields), KST can rigorously calibrate and remove this environmental bias, which would otherwise mimic a cosmological signal.
\end{itemize}

\subparagraph{The Lensed Sample.}
The cluster cores of the Deep Survey act as strong lenses for supernovae exploding in background galaxies. The multiply imaged SNe~Ia among them supply the time-delay $H_0$ measurement described above (Figure~\ref{fig:h0lens}), with the precision caveats stated there.

\paragraph{The Progenitor-Age Bias: A Direct Test of a Dark-Energy Systematic}
A specific and currently unresolved instance of this environmental dependence has become one of the most consequential systematics debates in supernova cosmology. Several analyses argue that the standardized SN~Ia luminosity correlates with the \textit{stellar-population age} of the host, with a slope of order $-0.04~\text{mag\,Gyr}^{-1}$ \citep{Kang2020, Lee2022, Son2025}. Because the mean progenitor age declines with redshift, such a correlation would impose a redshift-dependent luminosity drift of up to $\sim$0.2~mag over $0<z<1$, comparable to the dimming attributed to dark energy, and its proponents argue that correcting for it weakens the evidence for cosmic acceleration. The mainstream view holds that the effect is largely degenerate with host mass and dust \citep{BroutScolnic2021}, is absorbed by standard bias corrections, and shows no significant redshift evolution in the data \citep{Rose2020, Vincenzi2024, Wiseman2026}. The debate is sharpened by the DESI hint of evolving dark energy, whose significance depends on the supernova sample adopted \citep{DESI2024,DESIDR2Cosmo}. Distinguishing a genuine $w(z)$ from a progenitor-age drift has therefore become central to cosmology.

KST is specifically configured to address the \textit{observational} question, because it breaks the underlying age--mass--dust degeneracy in three complementary ways. First, its rest-frame NIR photometry standardizes SNe~Ia where the intrinsic scatter is smallest ($\sim$0.10~mag) and extinction, and hence the dust-driven host step, is suppressed by $\sim$3--5$\times$ relative to the rest-frame $B$ band \citep{Avelino2019, BaroneNugent2012}. Second, the simultaneous $R\sim100$ spectrophotometry measures the host stellar-population age \textit{directly}, for every supernova and across redshift, allowing the age--luminosity slope to be \textit{fitted} rather than assumed and to be disentangled from host mass and star-formation rate \citep{Rigault2020}. Third, the cluster-targeted Deep Survey supplies a control sample of SNe~Ia in old, passive, dust-poor elliptical hosts. Their stellar populations do age with cosmic time (they are $\sim$5~Gyr younger at $z\approx1$ than today), but along a well-modeled passive track that the same spectrophotometry constrains. The luminosity drift predicted by the progenitor-age hypothesis can then be tested against a sample with minimal dust and star-formation variation. We note that KST's rest-frame NIR coverage is limited to $z\lesssim0.6$ by the $1.8~\mu$m cutoff. The high-redshift leverage is instead carried by the host-age measurements and the passive cluster sample rather than by NIR distances alone. Together these make KST one of the few facilities able to confirm or refute a progenitor-age origin for the supernova Hubble diagram, and thereby to test the robustness of cosmic acceleration itself.

\paragraph{Spectroscopic Purity via the Linear Variable Filter}
A major bottleneck for future ``Stage IV'' dark energy missions (e.g., Rubin/LSST) is the lack of spectroscopic confirmation. Relying solely on photometric classification introduces contamination from Core-Collapse SNe (Type Ibc/II), which can bias the dark energy equation of state parameter $w$ by several percent.
KST mitigates this bottleneck for the brighter events using its LVF.
\begin{itemize}
    \item \textbf{The Silicon Fingerprint:} The defining spectral feature of a Type Ia supernova is the deep absorption trough of singly ionized silicon (Si~II $\lambda6355$, blueshifted to $\approx$6150~\AA).
    \item \textbf{In-Survey Confirmation:} Every supernova position in the Deep Survey accumulates an $R \sim 100$ LVF spectrum within the survey itself, with no external follow-up (Section~\ref{sec:simultaneity}). Near-peak epoch stacks provide direct spectral classification of the brighter events, and the mission-co-added spectrum ($\approx$22 AB per channel; Table~\ref{tab:lvf_modes}) delivers a spectrophotometric redshift and stellar-population diagnostics for hosts above this limit, with adjacent-channel binning ($R\sim20$, $\approx$0.9~mag deeper) extending the coverage to fainter hosts. The Si~II feature enters the LVF bandpass only at $z\gtrsim0.8$, where SNe~Ia are fainter than the per-channel limits. There, classification relies on the optical light curves and colors together with the host redshift.
    \item \textbf{Yield:} This capability allows KST to build a sample of $\sim 1,000$ SNe Ia at $z > 0.7$ with LVF host redshifts and observer-frame NIR light curves, a redshift regime where ground-based spectroscopy is prohibitively expensive. This well-characterized sample provides the high-redshift anchor for the Hubble diagram, maximizing the lever arm for constraining the time evolution of dark energy ($w_a$).
\end{itemize}

\paragraph{Projected Dark-Energy Constraints}
By combining this supernova Hubble diagram (with the systematic control afforded by NIR observations and the ``Clean'' cluster sample) with weak-lensing tomography and a \textit{Planck} CMB prior, KST will constrain the dark-energy equation of state at its best-constrained pivot redshift ($z_p\approx0.35$) to $\sigma(w_p) \approx 0.03$ and its time evolution to $\sigma(w_a) \approx 0.25$--$0.36$, a Dark Energy Figure of Merit (FoM) $[\sigma(w_p)\,\sigma(w_a)]^{-1} \approx 80$--$180$. The BAO/AP and cluster probes are not included in this computation. Adding them would move the realized value from the lower toward the upper end of the range. The weak-lensing analysis independently measures the growth amplitude to $\approx$1\% precision, $\sigma(S_8) \approx 0.008$, and the lensed-SN time-delay $H_0$ provides an independent cross-check on the local expansion rate.

\begin{figure}[t]
\centering
\includegraphics[width=\columnwidth]{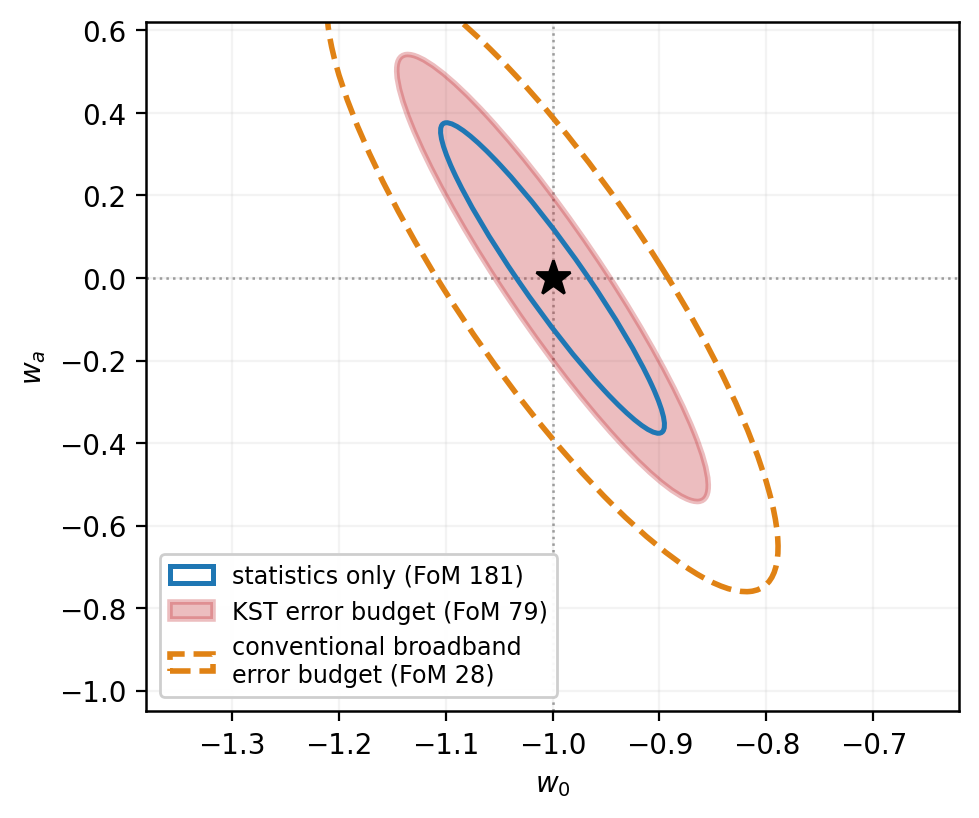}
\caption{\textbf{The error budget, not the statistics, sets the constraint.} Forecast $68\%$ confidence contours in the $w_0$--$w_a$ plane for the \textit{same} KST survey under three error budgets, from the systematics-marginalized Fisher analysis described in the text: statistical precision only (blue); the KST error budget, with its internally measured shear calibration, mean-redshift, and NIR supernova systematics (red); and an otherwise identical survey carrying a conventional broadband systematics chain (orange dashed). Contours are centered on the $\Lambda$CDM fiducial (star). The prior widths defining the two chains are stated in the text. Marginalizing over systematics costs the conventional chain a factor $\sim$3 in figure of merit relative to KST's. The KST-error-budget contour corresponds to $\sigma(w_0)=0.096$, $\sigma(w_a)=0.36$, FoM $=79$ ($\sigma(w_p)\approx0.035$ at the pivot), computed for the current survey ($5{,}000$ SNe to $z\approx1$, $n_{\rm eff}=57$, $(1+z)$-scaled redshift-bias priors). The bridge to the full-probe headline forecasts is given in the text.}
\label{fig:w0wa}
\end{figure}

What distinguishes this measurement is less the size of the ellipse than the provenance of its error budget. Each leading systematic of contemporary dark-energy experiments is addressed \textit{structurally}, within the KST dataset itself: (i) supernova standardization is performed in the NIR, where dust extinction and color-calibration systematics, the dominant terms in current Hubble-diagram budgets, are suppressed at the source; (ii) shear is measured with a thermally stable, diffraction-limited space PSF, controlling the multiplicative bias to $m<10^{-3}$ without the stochastic atmospheric corrections that limit ground-based lensing; (iii) shapes are measured in four separate $g,r,i,z$ bands, which reduces the chromatic PSF errors inherent to single-band imagers \citep{Cypriano2010, EriksenHoekstra2018} at the design level and cross-checks them with a four-band shear null test; and (iv) the $R\sim100$ spectrophotometry obtained for every source internally calibrates the tomographic mean redshifts, the leading lensing systematic that broadband surveys must constrain with external spectroscopy. KST thus delivers a constraint approaching Stage~IV precision whose dominant systematics are internally measured rather than externally imported, an error budget orthogonal to those of DESI, \textit{Euclid}, and Rubin. A confirmation (or refutation) of evolving dark energy from such an independent systematics chain carries evidentiary weight beyond its formal figure of merit, and in joint analyses the orthogonality is what breaks the degeneracies that no single experiment resolves alone.

An illustrative systematics-marginalized Fisher analysis makes the stakes quantitative (Figure~\ref{fig:w0wa}). We augment a ten-bin tomographic cosmic-shear Fisher matrix for the $2{,}000$~deg$^2$, $n_{\rm eff}=57$~arcmin$^{-2}$ survey with twenty nuisance parameters (a multiplicative shear bias $m_i$ and a tomographic mean-redshift shift $\Delta z_i$ per bin) marginalized with Gaussian priors that encode the systematic control of each error-budget chain, and combine the result with the supernova Hubble diagram (including its systematic magnitude floor) and a \textit{Planck} prior.

The prior widths are stated assumptions, chosen to bracket published requirements and achieved values. For KST-like control we adopt $\sigma(m)=10^{-3}$ (the stability budget of Section~\ref{sec:thermal_stability}), $\sigma(\Delta z)=0.002(1+z)$ (LVF self-calibration from bright-end $R\sim100$ spectra and the $\approx$10 independent synthetic $R\sim20$ bands, an internal calibration sample that is flux-limited but spans the full survey footprint, serving the role that broadband missions must fill with external spectroscopy; \citealt{EuclidVII2020}), and a $0.010$~mag NIR supernova floor. For a conventional broadband chain we adopt $\sigma(m)=5\times10^{-3}$ (chromatic single-band PSF residuals; \citealt{Cypriano2010, EriksenHoekstra2018}), $\sigma(\Delta z)=0.005(1+z)$ (external photometric-redshift calibration, intermediate between the Stage-IV requirement and current achieved values), and a $0.025$~mag optical supernova floor. The KST mean-redshift prior is a requirement rather than a demonstrated capability. The depth, size, and representativeness of the internal calibration sample needed to reach $\sigma(\Delta z)=0.002(1+z)$ at the flux limits of Table~\ref{tab:lvf_modes} remain to be validated by end-to-end simulation in the mission-definition phase. The latter chain is a stylized composite of the broadband approach rather than a forecast for any specific mission. Its prior widths sit between formal Stage-IV requirements and values achieved by current surveys, and a mission meeting its stated requirements would fall between the two scenarios shown.

Under identical survey statistics, the weak-lensing figure of merit retains $74\%$ of its statistical value under KST-like control but only $34\%$ under the conventional chain, and in the full probe combination the marginalized figure of merit differs by a factor of $\sim$$3$.

For traceability, with the current survey inputs ($5{,}000$ SNe to $z\approx1$ including $1{,}000$ at $z>0.7$, $n_{\rm eff}=57$~arcmin$^{-2}$, and redshift-bias priors scaling as $(1+z)$), the KST-error-budget case of Figure~\ref{fig:w0wa} corresponds to $\sigma(w_0)=0.096$, $\sigma(w_a)=0.36$, and FoM $=79$ for this WL$+$SN$+$\textit{Planck} combination. The strong $w_0$--$w_a$ correlation ($\rho=-0.93$) implies a pivot uncertainty $\sigma(w_p)=\sigma(w_0)\sqrt{1-\rho^2}\approx0.035$ at $z_p\approx0.33$ and hence $\mathrm{FoM}\equiv[\sigma(w_p)\,\sigma(w_a)]^{-1}=79$, consistent by construction with $1/\sqrt{\det\mathbf{C}_{w_0w_a}}$. Switching off the systematic priors in the \textit{same} calculation gives $\sigma(w_0)=0.069$, $\sigma(w_a)=0.248$, $\sigma(w_p)=0.022$ at $z_p\approx0.35$, and FoM $=181$. Including the BAO/AP and cluster probes, which are largely immune to the nuisances marginalized here, moves the realized constraint from the former floor toward the latter ceiling. The headline range quoted in Section~\ref{sec:forecast}, $\sigma(w_p)\approx0.03$ with FoM $\approx80$--$180$, is bounded by these two computed cases rather than extrapolated beyond them. Consistent with the transparent, order-of-capability approach of Section~\ref{sec:forecast}, these numbers illustrate scaling rather than a full likelihood analysis. The structural conclusion, however, is robust—at Stage-IV statistical power the delivered constraint is set less by the raw statistics than by the systematics chain, and KST's chain is internally measured.

\subsubsection{Large-Scale Structure: Verifying the Dynamical Dark Energy Hypothesis}
The distribution of galaxies in the universe is not random but traces the underlying ``Cosmic Web'' of dark matter, shaped by the physics of the early universe and the subsequent expansion history. By mapping the 3D positions of millions of galaxies, KST will construct a ``standard ruler'' and a ``standard shape'' to measure the geometry of the universe with precision sufficient to distinguish between a cosmological constant ($\Lambda$) and dynamical dark energy.

This probe has gained urgent priority. Recent results from ground-based spectroscopic surveys (specifically DESI) have hinted at a potential deviation from the standard $\Lambda$CDM model, favoring a dynamical equation of state. KST is well positioned to test this claim by providing an independent, orthogonal dataset in a redshift regime that ground-based instruments struggle to probe.

\paragraph{Confronting Dynamical Dark Energy and Phantom Crossing}

The standard model assumes that the dark energy density is constant in time ($w = -1$). However, many physically motivated models allow the dark energy density to evolve with time. As noted above, recent DESI BAO analyses combined with CMB and SN data show a $\sim 3\sigma$ preference for dynamical dark energy whose preferred equation of state crosses the phantom divide at $w=-1$, a behavior that is nontrivial to realize for a minimally coupled fluid.

Confirming phantom crossing requires precise measurements of the expansion history over a broad redshift range. The existing evidence rests on spectroscopic tracers available in the optical window, namely emission-line galaxies, quasars, and the Lyman-$\alpha$ forest. Extending the measurement to higher redshift with independent systematics is precisely where KST contributes.
\begin{itemize}
    \item \textbf{Orthogonal Tracers:} The DESI result is derived from specific galaxy populations selected in optical bands. These selections are subject to complex bias functions and redshift failure rates that could mimic a cosmological signal. KST provides a completely independent check using NIR selected galaxies (both $H\alpha$ emitters and continuum-selected passive galaxies). These populations trace the same underlying dark matter structures but have different bias factors and selection systematics, allowing for a powerful cross-correlation test.
    \item \textbf{Mapping the ``Redshift Desert'':} The ``Redshift Desert'' ($1.5 < z < 3.0$) provides a critical high-redshift lever arm for testing dynamical dark energy and phantom crossing. In this range, the key spectral features used by ground-based optical spectrographs (like the 4000 \AA\ break) shift into the NIR, where the Earth's atmosphere is bright (OH lines) and opaque. Ground-based surveys thus suffer from a dramatic drop in sampling density in this regime. KST, operating above the atmosphere with deep NIR sensitivity, will map this era with high fidelity. Measuring the distance scale at $z \sim 2$ anchors the time-evolution parameter $w_a$ and directly tests the phantom crossing.
\end{itemize}

\paragraph{Baryon Acoustic Oscillations (BAO): The Standard Ruler}
Acoustic oscillations in the pre-recombination photon--baryon plasma froze out at recombination ($z \sim 1100$), imprinting a characteristic comoving scale, the sound horizon $r_s \approx 150$ Mpc, on the late-time galaxy distribution. KST will measure this standard ruler in the galaxy correlation function across multiple tomographic redshift bins.
\begin{itemize}
    \item \textbf{Transverse and Radial Modes:} In a 3D galaxy map, the BAO scale manifests in two projections. Perpendicular to the line of sight, the angular separation $\Delta\theta$ of the comoving sound horizon measures the comoving angular distance,
    \begin{equation}
        D_M(z) = \frac{r_s}{\Delta \theta}, \qquad D_A(z) = \frac{D_M(z)}{1+z},
    \end{equation}
    where $D_A$ is the angular-diameter distance. Along the line of sight, the redshift interval $\Delta z = r_s H(z)/c$ would measure the expansion rate directly, but resolving it requires spectroscopic precision. Photometric redshifts with $\sigma_z/(1+z)\approx0.01$ smear radial separations by $\sigma_\chi = c\,\sigma_z/H(z) \approx 20$--$35\,h^{-1}$ Mpc, a substantial fraction of the $\approx$$105\,h^{-1}$ Mpc sound-horizon scale. KST's primary BAO observable is therefore the \textit{angular} (transverse) scale $D_M(z)$ measured in tomographic shells. Residual radial information enters only through the broad-bin anisotropy of the projected clustering, and its delivered precision is a mission-definition-phase deliverable to be established with end-to-end simulations.
    \item \textbf{Precision Forecast:} By detecting the BAO peak in the $1.5 < z < 3.0$ range, KST effectively ``anchors'' the expansion history curve at high redshift. This constrains the early behavior of dark energy, preventing the degeneracy between $w_0$ and $w_a$ that plagues low-redshift-only surveys.
\end{itemize}

\paragraph{The Alcock--Paczyński (AP) Test: The Standard Shape}
While BAO uses a standard scale, the Alcock--Paczyński (AP) test uses a ``standard shape'', the statistical isotropy of large-scale clustering. If the reference cosmology assumed when converting observed angles and redshifts into comoving coordinates differs from the true one, the recovered clustering appears squashed or stretched along the line of sight.
\begin{itemize}
    \item \textbf{The AP Parameter:} This geometric distortion is quantified by the AP parameter $F_{\rm AP}(z)$ \citep{Alcock1979}:
    \begin{equation}
        F_{\rm AP}(z) = \frac{(1+z)\,D_A(z)\,H(z)}{c} = \frac{D_M(z)\,H(z)}{c},
    \end{equation}
    with $D_M = (1+z)D_A$ the comoving angular distance. If the recovered clustering signal exhibits residual anisotropy, the reference cosmology is incorrect, and the parameters are iterated until isotropy is restored (see also \citealt{ap2010,ap2023} for the extended AP test, which does not require an isotropic assumption). With photometric redshifts the test is applied to the anisotropy of clustering projected in broad tomographic bins. Its delivered precision depends on the same radial smearing quantified above and is likewise a mission-definition-phase deliverable.
    \item \textbf{Curvature Sensitivity:} The AP test is uniquely sensitive to the product $H(z)\,D_M(z)$, which depends strongly on the cosmic curvature $\Omega_k$. Breaking the degeneracy between dark energy and curvature is essential, because a universe with non-zero curvature can mimic the expansion history of a dynamical dark energy model. KST's AP measurement at high redshift therefore contributes a constraint on cosmic flatness ($\Omega_k = 0$) complementary to those from the CMB and spectroscopic BAO.
\end{itemize}

\paragraph{Growth versus Expansion: Testing Gravity}
Peculiar velocities sourced by large-scale structure shift galaxy positions in a spectroscopic redshift survey, and the resulting ``Redshift-Space Distortions'' (RSD) trace the logarithmic growth rate of structure,
    \begin{equation}
        f(z) = \frac{d \ln D}{d \ln a} \approx \Omega_m(z)^\gamma .
    \end{equation}
Resolving this anisotropy requires spectroscopic redshift precision well beyond KST's photometric $\sigma_z/(1+z) \approx 0.01$. KST therefore does not measure $f\sigma_8$ directly, and its role in the gravity test is complementary.
\begin{itemize}
    \item \textbf{Modified Gravity vs. Dark Energy:} General Relativity (GR) predicts a growth index of $\gamma \approx 0.55$ \citep{Linder2005}, whereas modified-gravity theories (such as $f(R)$ gravity or Dvali-Gabadadze-Porrati braneworld models) predict a different growth rate ($\gamma \neq 0.55$) for the \textit{same} expansion history. KST supplies the two ingredients of this test that photometric data measure best: a precise expansion history from angular BAO and the AP test across the redshift desert, and an independent, lensing-based measurement of structure growth from cosmic-shear tomography (described above). Confronting these with the spectroscopic $f\sigma_8(z)$ measured by DESI and other ground-based surveys over the same redshift range yields a stringent consistency test of General Relativity on cosmological scales. A mismatch would indicate that cosmic acceleration reflects a breakdown of Einstein's theory of gravity rather than a new energy component.
\end{itemize}

\paragraph{Higher-Order Statistics: Breaking the Bias Degeneracy}
The standard cosmological analysis relies on the 2-point correlation function (or Power Spectrum), which captures all available information only if the density field is Gaussian. However, non-linear gravitational collapse induces significant non-Gaussianity at late times.
\begin{itemize}
    \item \textbf{The Bispectrum:} KST's high galaxy density allows for the precise measurement of the 3-point correlation function (Bispectrum). The Bispectrum is sensitive to the filamentary nature of the cosmic web. Crucially, the dependence of the Bispectrum on galaxy bias $b$ is different from that of the Power Spectrum. By analyzing both simultaneously, KST can break the degeneracy between galaxy bias and the amplitude of fluctuations $\sigma_8$, sharpening the cosmological constraints by a factor of $\sim 2$.
\end{itemize}

\subsubsection{Synergy: Breaking the Dark Sector Degeneracies}
The true power of KST lies in the combination of these four probes.
\begin{itemize}
    \item \textbf{Geometric Probes (SNe, BAO/AP):} Constrain expansion history but suffer from $w$-$\Omega_k$ degeneracy.
    \item \textbf{Growth Probes (WL, Clusters; RSD from external spectroscopic surveys):} Constrain matter content but suffer from $\Omega_m$-$\sigma_8$ degeneracy.
\end{itemize}
By conducting a joint analysis (Table~\ref{tab:cosmo_probes}), KST breaks these degeneracies, delivering dark-sector constraints whose precision is set by an internally measured systematics budget rather than by external calibration.

\begin{table*}[t]
\centering
\caption{Summary of KST Cosmological Probes and Target Parameters}
\label{tab:cosmo_probes}
\begin{tabular}{p{3.5cm} p{2.5cm} p{4.5cm} p{4.5cm}}
\hline
\textbf{Target Physics} & \textbf{Key Parameters} & \textbf{Primary KST Probe} & \textbf{Primary Observable} \\ \hline
\textbf{Cosmic Expansion} & $H_0$, $H(z)$ & Strong Lensing (SL) \newline Type Ia SNe (SN Ia) \newline BAO & Time-delay of lensed images \newline Luminosity distance $d_L(z)$ \newline Galaxy clustering scale ($r_s$) \\ \hline
\textbf{Dark Energy Nature} & $w_0$, $w_a$, $\Omega_\Lambda$ & SN Ia + WL + BAO \newline (Checking DESI hints) & Joint constraint on expansion history vs. growth of structure \\ \hline
\textbf{Dark Matter Properties} & $\Omega_m$, $\sigma_8$ \newline Subhalo Mass & Weak Lensing (WL) \newline Strong Lensing (SL) & Cosmic shear tomography \newline Flux-ratio anomalies in lensed arcs \\ \hline
\textbf{Testing Gravity (GR)} & $\gamma$; $f\sigma_8$ (external) & WL Tomography \newline Angular BAO $+$ AP \newline \textit{RSD: external-survey synergy} & Cosmic-shear growth of structure \newline Expansion history \newline \textit{Anisotropic clustering (DESI et al.)} \\ \hline
\textbf{Cosmic Geometry} & $\Omega_k$, $D_M(z)$ & Angular BAO \newline Alcock--Paczyński (AP) Test & Transverse (angular) sound-horizon scale \newline Projected clustering anisotropy in broad tomographic bins \\ \hline
\end{tabular}
\end{table*}

\subsection{Pillar II: The Nature of Dark Matter}
\label{sec:dark_matter}

Dark matter constitutes $\sim27\%$ of the cosmic energy budget, yet its particle identity remains
unknown. On large scales, the cold, collisionless dark matter (CDM) of $\Lambda$CDM passes every
test yet devised, and merging clusters supply the cleanest single demonstration of its nature. In
systems such as the Bullet Cluster, the lensing mass is spatially offset from the collisional
X-ray gas, demonstrating that the dominant mass is dark and nearly collisionless \citep{Clowe2006,cha2025}. The
once-favored weakly interacting massive particle has nonetheless eluded direct, indirect, and
collider searches, broadening interest in alternatives whose signatures lie not in the laboratory
but in the \textit{small-scale} structure of the Universe. Warm dark matter (WDM)
free-streams and suppresses structure below a cutoff scale \citep{Irsic2017}, fuzzy (ultralight
axion) dark matter (FDM) imprints a de~Broglie-scale suppression and solitonic cores
\citep{Hui2017}, self-interacting dark matter (SIDM) reshapes halo interiors through a
cross-section $\sigma/m$ \citep{TulinYu2018}, and primordial black holes remain viable only in narrow
mass windows \citep{CarrKuhnel2020}.

The classic galaxy-scale ``small-scale challenges'' to CDM (the missing-satellites, core--cusp,
and too-big-to-fail problems) are now largely attributable to baryonic feedback and are no longer
clean discriminants of dark-matter microphysics \citep{Bullock2017}. Decisive tests have therefore
shifted to the lowest-mass, essentially \textit{starless} halos, where dark-matter physics rather
than galaxy formation sets the structure. These are accessible only through gravity. Because
gravitational lensing responds to all mass regardless of whether it shines, the same
diffraction-limited weak- and strong-lensing data that anchor Pillar~I
(Section~\ref{sec:dark_energy}) give KST a direct, luminosity-independent probe of dark matter
across three decades of scale: the suppression of the small-scale matter power spectrum, the
abundance of galaxy-scale subhalos in strong lenses, and the collisional behavior of dark matter
in the cores of massive clusters. The limiting factor in every one of these measurements is
\textit{statistical and angular}. Only a few hundred suitable strong lenses are currently known,
and resolving the relevant sub-arcsecond image structure demands space-based, diffraction-limited
imaging over wide fields \citep{Vegetti2024}, which is precisely KST's design point. A complementary,
non-lensing avenue, high-cadence gravitational microlensing, extends KST's reach to the
compact-object end of the mass spectrum, where primordial black holes may reside
(Section~\ref{sec:dm_pbh}). Simultaneous near-infrared coverage is integral to this program, not incidental to it. It delivers the photometric redshifts that place lensed sources and faint background populations at the $z\sim1$--$3$ where lensing efficiency peaks, and it recovers the dust-reddened arcs and lens galaxies that optical data alone would miss, as the individual probes below make explicit.

\subsubsection{The Small-Scale Matter Power Spectrum}
While the large-scale power spectrum constrains cosmology, the small-scale power spectrum (high
wavenumbers $k$) carries the imprint of dark-matter microphysics. In CDM, structure forms
hierarchically down to sub-solar scales. The alternatives instead predict a cutoff. Thermal-relic
WDM washes out structure below a free-streaming scale, suppressing power for $k \gtrsim
10~h\,\text{Mpc}^{-1}$ (current Lyman-$\alpha$ forest data bound the thermal-relic mass to
$m_{\rm WDM} \gtrsim 5.3$~keV; \citealt{Irsic2017}), while FDM produces an analogous suppression
set by the de~Broglie wavelength of an $m \sim 10^{-22}$~eV particle \citep{Hui2017}.

Measuring this suppression is difficult because baryonic processes (feedback from active galactic nuclei, AGN, and supernova winds)
also redistribute matter and damp power on these scales. KST breaks this degeneracy in two ways.
Its diffraction-limited cosmic-shear maps (Pillar~I) extend the weak-lensing power spectrum to
smaller angular scales than seeing-limited surveys can reach, isolating the regime where
WDM/FDM and baryonic effects diverge. Cross-correlation with external X-ray and
Sunyaev--Zel'dovich cluster catalogs, which trace the hot gas directly, allows the baryonic
contribution to be marginalized, isolating the pure dark-matter signal.

\subsubsection{Dark-Matter Substructure from Strong Lensing}
Strong gravitational lensing is a direct laboratory for the nature of dark matter. CDM predicts
that galaxy-sized halos are self-similar, teeming with thousands of bound subhalos extending to
sub-galactic masses, whereas free-streaming (WDM/FDM) truncates this subhalo mass function below a
characteristic scale. Counting the lowest-mass, starless subhalos is therefore a direct test of
the dark-matter particle, and strong lensing detects them purely through their gravity.

\paragraph{Flux-ratio anomalies}
In a smooth lens potential, the relative brightness (flux ratio) of the multiple images of a
compact background source follows strict relations set by the macro-model, particularly for
fold and cusp configurations. A subhalo projected near one of the images perturbs its
magnification, producing a ``flux-ratio anomaly'', a direct gravitational signature of
substructure sensitive to all mass regardless of luminosity \citep{DalalKochanek2002}. The
amplitude and frequency of anomalies across a lens sample constrain subhalos down to
$10^{7}$--$10^{8}\,M_\odot$.

\paragraph{Gravitational imaging of Einstein rings}
For extended sources lensed into arcs and Einstein rings, a subhalo perturbs the surface-brightness
distribution in a spatially localized way. Modeling these residuals ``images'' the perturber
directly, yielding its mass and position, the technique that achieved the first detection of an
individual dark substructure at cosmological distance \citep{Vegetti2012}. It is sensitive both to
subhalos bound to the main lens and to isolated halos anywhere along the line of sight, providing a
tomographic census of small-scale structure.

\paragraph{Discriminating among dark-matter models}
The abundance of low-mass (sub)halos is the discriminator: CDM predicts a power-law mass function
($dN/dM \propto M^{-1.9}$) extending to Earth-mass scales, WDM a cutoff set by the particle mass,
and SIDM modified inner profiles and abundances through core formation and gravothermal collapse.
Measuring the mass function and detecting (or excluding) a truncation near $10^{7}\,M_\odot$ thus
maps directly onto the particle physics.

\paragraph{Why KST}
Both techniques require resolving image structure on sub-arcsecond scales, which ground-based
seeing blends and washes out. KST's diffraction-limited $\sim0.15''$ PSF cleanly separates close
images and resolves arc perturbations, while its wide-field, NIR-sensitive survey transforms the
sample size from the few hundred lenses known today toward the $\sim10^{5}$ galaxy-scale lenses
expected from the coming generation of wide surveys, of which KST can characterize the subset
bright and well-configured enough for substructure analysis at the resolution required. The near-infrared is decisive here for two reasons. Many lens galaxies and their reddened, dusty arcs are detected only in the NIR, and comparing image flux ratios \textit{between} the optical and NIR isolates the achromatic gravitational perturbation of a subhalo from the chromatic contamination of dust extinction and stellar microlensing that would otherwise mimic it. This
population-level subhalo-mass-function measurement, rather than a handful of individual systems, is
what turns substructure lensing into a competitive constraint on the dark-matter particle.

\subsubsection{Dwarf Galaxies, Satellite Counts, and Stellar Streams}
\label{sec:dm_dwarfs}
The substructure detected gravitationally in distant strong lenses has a directly countable counterpart in the Local Group, where the lowest-mass dark-matter halos can be enumerated one by one. Two complementary populations bracket the subhalo mass function across the threshold below which halos no longer form stars. \textbf{Satellite galaxies}, down to the ultra-faint regime, occupy the subhalos that \textit{do} host stars. Their abundance as a function of luminosity and distance constrains the halo mass function and hence the warm-dark-matter free-streaming cutoff, with current Milky Way satellite censuses bounding the thermal-relic mass to $m_{\rm th}\gtrsim6$~keV \citep{Nadler2021}. \textbf{Cold stellar streams}, the tidal debris of disrupted clusters and dwarfs, probe the subhalos \textit{below} that threshold that host no stars at all. A passing dark subhalo opens a characteristic density gap in an otherwise smooth stream, as established for GD-1 \citep{Bonaca2019}.

KST contributes to both with a niche distinct from optical all-sky surveys. Its NIR sensitivity pierces the dust of the Galactic plane and bulge, opening the low-latitude ``zone of avoidance'' to satellite searches that optical surveys cannot penetrate, while its diffraction-limited resolution resolves member stars in distant and compact systems and traces low-surface-brightness streams to faint limits. The cluster and wide fields yield a deep census of dwarf and ultra-diffuse galaxies (Section~\ref{sec:galaxy_evolution}). KST will thus extend the satellite census into obscured and crowded regions and detect (or bound) dark-subhalo perturbations in cold streams, delivering a Local-Group measurement of the subhalo mass function that independently cross-checks the strong-lensing substructure result above and contributes to tightening the free-streaming bound toward $m_{\rm th}\sim10$~keV. Finally, because dwarf galaxies probe self-interactions at the low velocities ($\lesssim50~\text{km\,s}^{-1}$) inaccessible to clusters, the structural data on KST-discovered and KST-resolved dwarfs, with ground-based spectroscopic follow-up, anchor the low-velocity end of the velocity-dependent self-interaction cross-section, complementing the high-velocity cluster constraint (Section~\ref{sec:dm_cluster}).

\subsubsection{Cluster-Scale Dark Matter and Self-Interactions}
\label{sec:dm_cluster}
The cluster-targeted Deep Survey (Section~\ref{sec:survey}) provides KST's most direct
\textit{dynamical} test of dark-matter interactions. Self-interactions transfer energy and momentum
in the dense cores of massive halos, producing two observable signatures: a spatial offset between
the collisionless galaxies/dark matter and the collisional gas (or between substructure dark matter
and its galaxies) in ongoing mergers, and constant-density cores in relaxed clusters in place of
the cuspy profiles predicted by collisionless CDM. Both are sensitive to $\sigma/m$, which
cluster-scale measurements currently bound to $\sigma/m \lesssim 1~\text{cm}^2\,\text{g}^{-1}$
\citep{TulinYu2018}.

By centering its ten Deep Survey fields on massive lensing clusters, KST will build joint
diffraction-limited strong- and weak-lensing mass maps \citep[reconstructed with methods such as the
maximum-entropy-regularized code MARS;][]{chajee2022} and compare them, system by system, against
the positions of the cluster galaxies and the diffuse intracluster light (ICL), a collisionless
stellar tracer of the potential \citep{joo2025} (the ICL analysis is developed as Legacy science in
Section~\ref{sec:galaxy_evolution}). Because the weak-lensing convergence is an integral over the source-redshift distribution, the accuracy of these cluster mass maps hinges on the near-infrared. NIR photometry pins the photometric redshifts of the faint $z\sim1$--$3$ background galaxies, calibrating the cluster masses and extending the convergence reconstruction to the high source redshifts where the lensing signal is strongest. In merging ``Bullet-Cluster analogues'' these comparisons
measure dark-matter--baryon and dark-matter--galaxy offsets \citep{Clowe2006,cha2025}. In relaxed systems
they measure inner density slopes. Crucially, KST scales this approach from the samples of a few tens of merging systems that
define current constraints \citep{finner2025, Jee2026} to the $\gtrsim$1000 merging clusters of
its Deep and Wide Surveys, converting the scatter of
centroid offsets and core radii into a competitive, population-level limit on $\sigma/m$,
interpreted against cosmological simulations of cluster mergers under the self-interacting
dark-matter paradigm \citep{lee2026}.

\begin{figure}[t]
\centering
\includegraphics[width=\columnwidth]{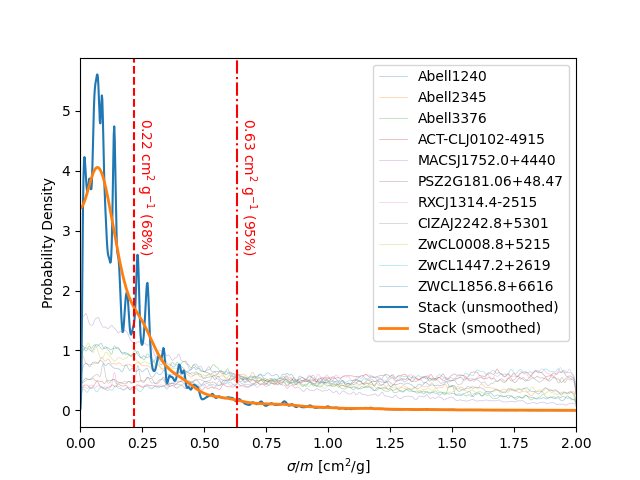}
\caption{\textbf{The double-radio-relic constraint on dark-matter self-interactions.} Posterior probability density of the self-interaction cross-section $\sigma/m$ from eleven merging clusters hosting symmetric double radio relics (thin curves), and their stacked constraint (thick curves; the smoothed stack in orange). The dashed and dash-dotted lines mark the resulting upper limits, $\sigma/m<0.22~\mathrm{cm^2\,g^{-1}}$ ($68\%$) and $<0.63~\mathrm{cm^2\,g^{-1}}$ ($95\%$), at the $\sim$1000--2000~km\,s$^{-1}$ collision velocities characteristic of clusters. KST's diffraction-limited lensing will extend this method from eleven systems to the $\gtrsim$1000 merging clusters of the Deep and Wide Surveys, projecting to $\sigma/m\lesssim0.02~\mathrm{cm^2\,g^{-1}}$. Adapted from \citet{Jee2026}.}
\label{fig:sics}
\end{figure}

A particularly robust new approach uses \textit{double radio relics} as a merger chronometer
\citep{Jee2026}, building on weak-lensing and X-ray analyses of individual double-relic systems
\citep{finner2021, ahn2025}. Symmetric double relics trace the outward-propagating merger shocks, whose
separation is nearly independent of the dark-matter physics, whereas the corresponding halo--halo
separation is suppressed by any self-interaction-induced drag. The ratio of the two constrains
$\sigma/m$ while marginalizing over the cluster mass, viewing angle, collision speed, merger phase,
impact parameter, and gas-profile slope. Applied to a sample of eleven merging clusters, this method
yields $\sigma/m < 0.22~\text{cm}^2\,\text{g}^{-1}$ (68\% confidence; $<0.63~\text{cm}^2\,\text{g}^{-1}$
at 95\%; Figure~\ref{fig:sics}) at the $\sim1000$--$2000~\text{km\,s}^{-1}$ collision velocities characteristic of clusters
\citep{Jee2026}, a high-velocity regime complementary to the $\lesssim200~\text{km\,s}^{-1}$ scales
probed in dwarf galaxies, and therefore essential for testing velocity-dependent self-interactions.
The limiting systematic of the method is the weak-lensing determination of the halo centroids and
masses, precisely the measurement KST is designed to deliver. The present bound derives from eleven
systems. Because the statistical precision improves as $N^{-1/2}$, the sample of $\gtrsim1000$ merging
clusters that KST's diffraction-limited lensing, paired with next-generation radio-relic surveys, will
characterize projects to $\sigma/m \lesssim 0.02~\text{cm}^2\,\text{g}^{-1}$, roughly an order of
magnitude below the current limit and well into the regime where dark-matter self-interactions would
leave a detectable imprint at cluster scales.

\subsubsection{Compact-Object Dark Matter from Microlensing}
\label{sec:dm_pbh}
\begin{figure}[t]
\centering
\includegraphics[width=\columnwidth]{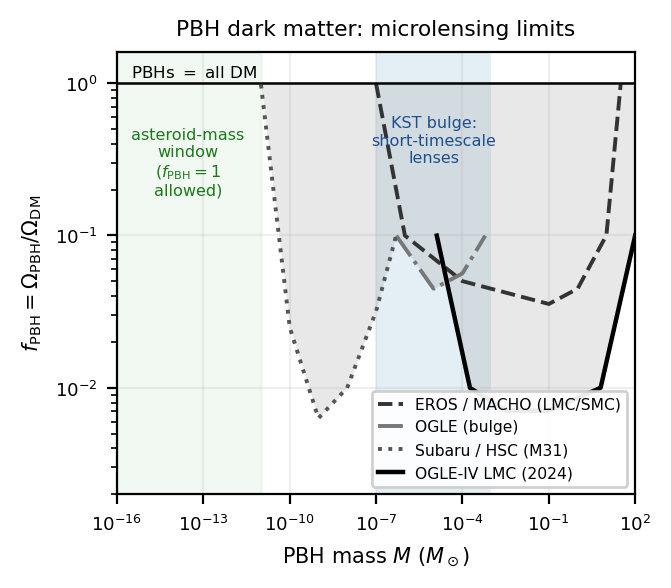}
\caption{\textbf{Primordial-black-hole dark matter.} Microlensing upper limits on the fraction of dark matter in compact objects, $f_\mathrm{PBH}=\Omega_\mathrm{PBH}/\Omega_\mathrm{DM}$, as a function of PBH mass. The region above each curve is excluded. Curves are schematic renderings of published limits from EROS/MACHO toward the Magellanic Clouds \citep{Tisserand2007}, OGLE toward the bulge, Subaru/HSC toward M31 \citep{Niikura2019}, and the 20-year OGLE survey of the Large Magellanic Cloud \citep[black;][]{Mroz2024Nature}, which together exclude PBHs as more than $\sim$1\% of the dark matter over much of the planetary-to-stellar range, while the asteroid-mass window (green) remains open. The blue band marks the planetary-mass range ($\sim10^{-7}$--$10^{-3}\,M_\odot$) of hour-long events that KST's 15-minute bulge cadence resolves. KST's quantitative limit in this range is to be established in the mission-definition phase.}
\label{fig:pbh}
\end{figure}

If dark matter is composed wholly or in part of primordial black holes (PBHs) or other compact
objects, it betrays itself not by lensing the light of smooth halos but by the transient,
achromatic brightening of background stars as a compact lens crosses the line of sight
\citep{CarrKuhnel2020}. The event timescale scales as $t_{\rm E}\propto M^{1/2}$. Monitoring a
large stellar population over a wide range of cadences maps the compact-object mass function.
Existing surveys toward the Magellanic Clouds, the Galactic bulge, and M31 (MACHO, EROS, OGLE, and
Subaru Hyper Suprime-Cam, HSC; \citealt{Tisserand2007,Niikura2019,Mroz2024Nature}) already exclude PBHs as the \textit{entirety} of the dark matter across most of the
planetary-to-stellar mass range, leaving the low-mass windows, most notably the asteroid-mass
regime, as the principal surviving territory (Figure~\ref{fig:pbh}).

KST's high-cadence bulge microlensing program (Section~\ref{sec:survey:microlensing}) extends these
searches as a by-product of the dedicated campaign that also conducts the exoplanet census. The
20-year OGLE survey of the Large Magellanic Cloud already limits compact objects of
$1.8\times10^{-4}$--$6.3\,M_\odot$ to less than 1\% of the dark matter \citep{Mroz2024Nature}. KST therefore does not aim to
improve the limit in that range. Its distinct contribution lies at short event
timescales. Space-based photometric stability and a 15-minute cadence over $>$$10^{8}$ monitored
bulge stars resolve the hour-long events of planetary-mass lenses ($\sim10^{-7}$--$10^{-3}\,M_\odot$;
Figure~\ref{fig:pbh}), where ground-based coverage is interrupted by weather and daylight, and its NIR
reach penetrates the dust-obscured bulge. Toward the bulge such events must be separated
statistically from the free-floating-planet population, and the result is therefore an upper limit on any
compact-object contribution in this mass range, complementary to the Subaru/HSC \citep{Niikura2019} and
OGLE limits. Its quantitative depth is a mission-definition-phase deliverable. The same observations
that conduct the cold-exoplanet census (Section~\ref{sec:exo}) thus double as a compact-object
dark-matter experiment.

\subsubsection{Synthesis: Dark Matter Across the Mass Spectrum}
No single observation isolates the dark-matter particle, but KST's lensing data address the problem
on three independent scales whose systematics are largely orthogonal (Table~\ref{tab:dm_probes}).
A free-streaming cutoff would appear consistently as suppressed small-scale shear power, a truncated
subhalo mass function, \textit{and} altered low-mass halo abundances. A non-zero $\sigma/m$ would
appear as cluster-core offsets and cored profiles while leaving the linear power spectrum intact.
The concordance (or tension) among these probes, all obtained from the same survey and the same
diffraction-limited PSF, is what allows KST to discriminate among cold, warm, fuzzy, and
self-interacting dark matter. Gravitational microlensing complements these gravitational-lensing
probes by independently constraining the compact-object (primordial-black-hole) end of the
dark-matter mass spectrum.

\begin{table}[ht]
\centering
\caption{KST probes of the nature of dark matter.}
\label{tab:dm_probes}
\begin{tabular}{@{}>{\raggedright\arraybackslash}p{2.3cm}>{\raggedright\arraybackslash}p{2.6cm}>{\raggedright\arraybackslash}p{2.6cm}@{}}
\hline
\textbf{Scale} & \textbf{KST Probe} & \textbf{Sensitive To} \\ \hline
Linear / mildly non-linear & Cosmic-shear power spectrum & WDM/FDM cutoff; baryon feedback \\ \addlinespace
Galaxy / subhalo & Flux-ratio anomalies; gravitational imaging & Subhalo mass function ($10^{7}$--$10^{8}\,M_\odot$); WDM truncation \\ \addlinespace
Local Group & Ultra-faint satellite counts; cold stellar-stream gaps & Subhalo mass function; WDM free-streaming cutoff \\ \addlinespace
Cluster core & Strong+weak lensing vs.\ galaxies/ICL & Self-interaction $\sigma/m$; core vs.\ cusp \\ \addlinespace
Compact objects & High-cadence bulge microlensing (short timescales) & Planetary-mass compact-object fraction \\ \hline
\end{tabular}
\end{table}

\subsection{Pillar III: Dark Sirens}
\label{sec:dark_sirens}

A compact-binary coalescence is a self-calibrating ``standard siren'': general relativity alone
fixes the relation between the observed inspiral waveform and the luminosity distance $d_L$. The
distance thus comes with the signal, with no appeal to any rung of the cosmic distance ladder
\citep{Schutz1986}. Since the first detection of a
binary black-hole merger \citep{Abbott2016}, GW astronomy has opened a route to the Hubble constant
whose systematics are completely orthogonal to both the \textit{Planck} CMB and the
Cepheid/SNe~Ia ladder, exactly the third, independent anchor needed to arbitrate the Hubble
tension introduced in Pillar~I (Section~\ref{sec:dark_energy}). A GW signal alone, however, does not
supply a redshift. The detector-frame masses and $d_L$ are degenerate with cosmological redshift, and
each event must be paired with an independent $z$. That redshift is obtained either from an
electromagnetic (EM) counterpart (a ``bright'' siren) or, statistically, from a galaxy redshift
catalog covering the GW localization volume (a ``dark'' siren). In both channels the decisive
ingredient is the same, a deep, wide, NIR galaxy survey with reliable photometric
redshifts. This is precisely what the KST Wide Survey provides, making KST a purpose-built engine
for siren cosmology.

\subsubsection{Bright Sirens: Kilonova Counterparts}
The binary neutron-star merger GW170817, with its kilonova AT2017gfo localized to NGC~4993, gave
the first bright-siren measurement, $H_0 = 70.0^{+12.0}_{-8.0}~\mathrm{km\,s^{-1}\,Mpc^{-1}}$
\citep{AbbottMM2017, AbbottSiren2017}. A single such event already reaches $\sim15\%$ precision, and a
few tens would be competitive with the distance-ladder and CMB anchors. The bottleneck is not the
GW data but the EM follow-up. Confident counterparts have proven rare, and no comparable event has
been secured since, because identifying a fading kilonova within a GW localization region of tens to
hundreds of square degrees demands rapid, deep, wide-field imaging. Kilonovae, particularly
their lanthanide-rich ``red'' component, also peak and persist longest in the NIR.

KST is well matched to restore this channel. Its $0.25~\text{deg}^2$ field of view and simultaneous
VIS imaging and NIR spectrophotometry allow tiling of well-localized GW error regions while capturing the rapid red-ward
spectral evolution that distinguishes a kilonova from contaminating transients. The $R\sim100$ LVF
adds spectrophotometric sampling of the candidate concurrently with the imaging, building up a low-resolution spectrum as the tiling scan proceeds (the simultaneity advantage of Section~\ref{sec:simultaneity}). Realizing this requires
the rapid-response engineering developed in Section~\ref{sec:design} (target-of-opportunity settling
in $<2$~min and prioritized data downlink within $<15$~min; Table~\ref{tab:req_matrix}). Each secure
counterpart pins down the host galaxy and hence a spectroscopic-quality redshift, yielding a direct
$d_L$--$z$ measurement, while the kilonova photometry itself constrains $r$-process nucleosynthesis
and neutron-star physics.

\subsubsection{Dark Sirens: Statistical $H_0$ from the KST Host Catalog}
For the majority of events (in particular binary black-hole mergers, which have no expected EM
counterpart), the redshift cannot be measured directly. The dark-siren technique instead assigns a
statistical redshift by weighting the candidate host galaxies within the three-dimensional GW
localization volume using a galaxy redshift catalog, each host contributing a redshift prior
weighted by the GW posterior support, and stacking many events sharpens $H_0$ \citep{DelPozzo2012,
Gray2020}. The current state of the art, combining the O1--O3 observing-run events (the third Gravitational-Wave Transient Catalog, GWTC-3) with galaxy catalogs
and the GW170817 counterpart, yields $H_0 \approx 68^{+13}_{-7}~\mathrm{km\,s^{-1}\,Mpc^{-1}}$
\citep{AbbottGWTC3cosmo2023}, a $\sim15$--$20\%$ measurement, not yet competitive with the
$\sim6~\mathrm{km\,s^{-1}\,Mpc^{-1}}$ ($\sim8\%$) Hubble tension, which requires $\sim1$--$2\%$ precision to arbitrate, but improving rapidly as the event count grows.

The dominant limitation is the galaxy catalog. When the catalog is shallow, most true hosts at the
distances of GW sources (hundreds of Mpc to $\sim$Gpc) are missing, and the method defaults to an
uninformative completeness correction. Inaccurate redshifts bias the inferred $H_0$. KST's deep
Wide Survey relieves this bottleneck within its footprint. Reaching $r\approx25.6$, it is complete to
roughly $0.1\,L^*$ out to $z\sim1$, well beyond the distances of O4/O5 events, with NIR-anchored
photometric redshifts reaching $\sigma_z/(1+z)\lesssim0.01$ for the NIR-detected subset. Because the
footprint covers only $\sim$5\% of the sky, KST contributes directly only for events localized within
it, and its deep, well-calibrated catalog also serves as a completeness and redshift-calibration
reference for shallower wide-area host catalogs. In the era of next-generation detectors, the eventual precision will be set by the galaxy
survey rather than the GW data \citep{ChenHolz2018, Maggiore2020}---an observational, not
instrumental, bottleneck that KST is built to relieve.

\begin{figure}[t]
\centering
\includegraphics[width=\columnwidth]{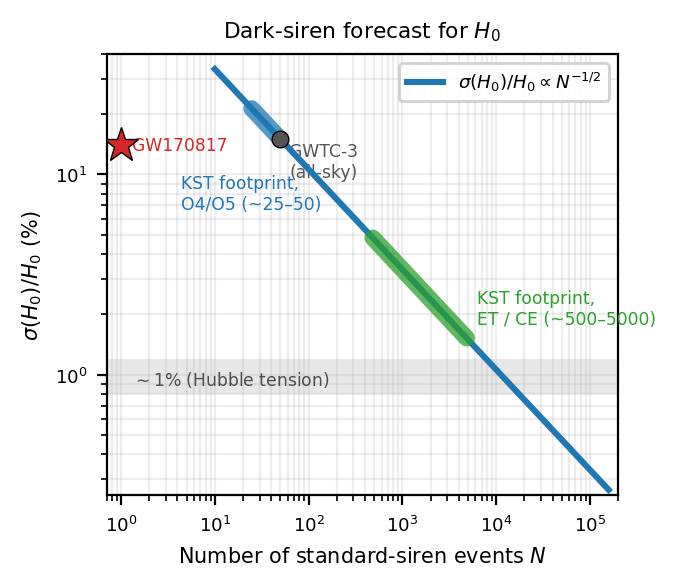}
\caption{\textbf{Dark-siren forecast for the Hubble constant.} Projected fractional precision on
$H_0$ versus the number of standard-siren events, under the idealized $\sigma(H_0)/H_0\propto
N^{-1/2}$ scaling anchored to the GWTC-3 combined result ($\sim$15\% from a few tens of events;
\citealt{AbbottGWTC3cosmo2023}). The single bright siren GW170817 is shown for reference. The
highlighted segments count only events localized within the $2{,}000$~deg$^2$ KST footprint
($\sim$5\% of the sky): $\sim$25--50 in the O4/O5 era ($\sim$15--20\%) and $\sim$500--5000 in
the Einstein Telescope/Cosmic Explorer era ($\sim$1.5--5\%). The scaling is a statistical
idealization. See Section~\ref{sec:forecast} for the forecasting methodology.}
\label{fig:siren_h0}
\end{figure}

Quantitatively, the standard-siren precision improves as $\sigma(H_0)/H_0\propto N^{-1/2}$ in the
number of well-localized events. Anchoring to the GWTC-3 combined measurement ($\sim$15\% from a few
tens of events; \citealt{AbbottGWTC3cosmo2023}) and counting only events localized within the KST
footprint ($\sim$5\% of the sky), the O4/O5 era ($\sim$hundreds of events all-sky, $\sim$25--50 in the
footprint) reaches $\sigma(H_0)/H_0\sim15$--$20\%$, comparable to the current all-sky value but
from a far more complete catalog, while the next-generation-detector era (the Einstein Telescope and
Cosmic Explorer; $10^{4}$--$10^{5}$ events all-sky, $\sim$500--5000 in the footprint) reaches
$\sim$1.5--5\% (Figure~\ref{fig:siren_h0}). This idealized scaling is optimistic (in practice a
floor is set by catalog incompleteness, peculiar velocities, and the assumed source population),
but it shows that KST's dark-siren contribution grows with the event rate of next-generation
detectors and is greatest in combination with wider-area host catalogs that KST helps calibrate.

\subsubsection{An Independent Verdict on the Cosmic Expansion}
The two siren channels are complementary. Bright sirens deliver few but exquisitely clean
$d_L$--$z$ pairs, while dark sirens harness the full, much larger population of mergers through the
KST host catalog. Together they provide a local measurement of $H_0$ whose systematics are largely
independent of both the CMB and the distance ladder. Combined with the lensed-supernova time-delay $H_0$ from
Pillar~I, a geometric anchor whose only ladder dependence enters through the SN
absolute-magnitude calibration (Section~\ref{sec:dark_energy}), KST contributes multiple, largely
independent constraints on the present expansion rate. It is precisely this convergence of
orthogonal probes, with KST supplying the redshift information for each, that is required to confirm or refute the
Hubble tension and, with it, the standard cosmological model. The same rapid-response, dual-channel
capability underpins the broader multi-messenger program of Section~\ref{sec:mma}, which develops
the kilonova spectroscopy, r-process nucleosynthesis, and short gamma-ray-burst science cases that
share this infrastructure.

\section{Legacy Science Portfolio}
\label{sec:legacy}
The Wide and Deep surveys that drive the dark-sector pillars simultaneously yield a rich multiwavelength, time-domain dataset for the 2030s. From the \textit{same} data, at no additional observing cost beyond the dedicated bulge microlensing campaign (which serves both the exoplanet census and the compact-object dark-matter test), KST delivers a broad Legacy Science portfolio. The programs below are not the mission's primary justification, but represent high-impact by-product science enabled by KST's combination of diffraction-limited VIS imaging and simultaneous NIR $R\sim100$ spectrophotometry. Their principal projected yields are collected in Table~\ref{tab:legacy_yields}. Multi-messenger astrophysics, which draws on the target-of-opportunity reserve and is closely tied to the dark-siren pillar (Section~\ref{sec:dark_sirens}), is presented separately in Section~\ref{sec:mma}.

\begin{table*}[t]
\centering
\caption{Indicative Legacy Science yields over the nominal six-year mission. Entries are order-of-magnitude estimates obtained by scaling the survey area, depth, and source/star density against existing surveys (the forecasting methodology is summarized in Section~\ref{sec:forecast}). All except the microlensing yields, which come from the dedicated bulge campaign, are by-products of the dark-sector survey at no additional observing cost.}
\label{tab:legacy_yields}
\begin{tabular}{@{}>{\raggedright\arraybackslash}p{3.2cm}>{\raggedright\arraybackslash}p{13.0cm}@{}}
\hline
\textbf{Program} & \textbf{Projected 6-year yield} \\ \hline
Galaxy evolution & a few $\times 10^8$ galaxies with resolved morphology and photometric redshifts ($\sim$$10^{7}$ at $z>2$); hundreds of protoclusters at $z>1.5$; $\sim$$10^{4}$ ultra-diffuse galaxies; deep intracluster-light maps of the ten Deep-field clusters and shallower maps for hundreds of Wide-Survey clusters \\ \addlinespace
Milky Way & $\sim$$10^{7}$ stars with NIR spectrophotometric metallicities; $\sim$$10^{5}$--$10^{6}$ RR Lyrae, Cepheid, and Mira variables; $\sim$$10^{3}$--$10^{4}$ brown dwarfs; tens of newly characterized bulge star clusters; cold stellar streams in the halo \\ \addlinespace
Active galactic nuclei & $\sim$$10^{6}$ variability- and SED-selected AGN; AGN luminosity function to $z\sim3$ \\ \addlinespace
Strong lenses & $\sim$$10^{4}$--$10^{5}$ galaxy-scale strong lenses \citep{Collett2015}; $10$--$20$ multiply imaged supernovae in the Deep fields; lensed quasars for time-delay cosmography \\ \addlinespace
Exoplanets (microlensing) & Several hundred bound planets, with sensitivity down to approximately an Earth mass, plus $\lesssim100$ free-floating planets (scaled from \citealt{Penny2019,Johnson2020}) \\ \hline
\end{tabular}
\end{table*}

\subsection{Galaxy Evolution across Cosmic Time}
\label{sec:galaxy_evolution}

Understanding how galaxies form and evolve across cosmic time is one of the central goals of
modern astrophysics. Galaxies are not static objects. They are the visible tracers of dark matter
halos, shaped by a complex interplay of hierarchical mass assembly, cold gas accretion, stellar and
AGN feedback, and environmental interactions. The cosmic star formation rate density peaked near
$z \sim 2$—the epoch known as ``Cosmic Noon''—and has declined by more than an order of magnitude
to the present day \citep{Madau2014}. Yet the physical mechanisms driving this decline, including morphological and environmental quenching, remain poorly
constrained, particularly at intermediate and high redshift.

A fundamental challenge has been the absence of a single facility capable of simultaneously
delivering: (1) the statistical depth and area to characterize rare populations and trace the full
diversity of environments; (2) the angular resolution to measure galaxy morphology and structural
parameters at cosmologically relevant redshifts; and (3) the NIR photometric and
spectroscopic coverage needed to penetrate dust and measure redshifts reliably across the
``Redshift Desert'' ($1.5 < z < 3$). Ground-based facilities are fundamentally limited by
atmospheric seeing and OH airglow emission in the NIR, while existing space missions such as
\textit{HST} and \textit{JWST} offer exquisite depth over pencil-beam fields that are too small
for statistical studies of large-scale structure and environment.

\begin{figure}[t]
\centering
    \includegraphics[width=\linewidth]{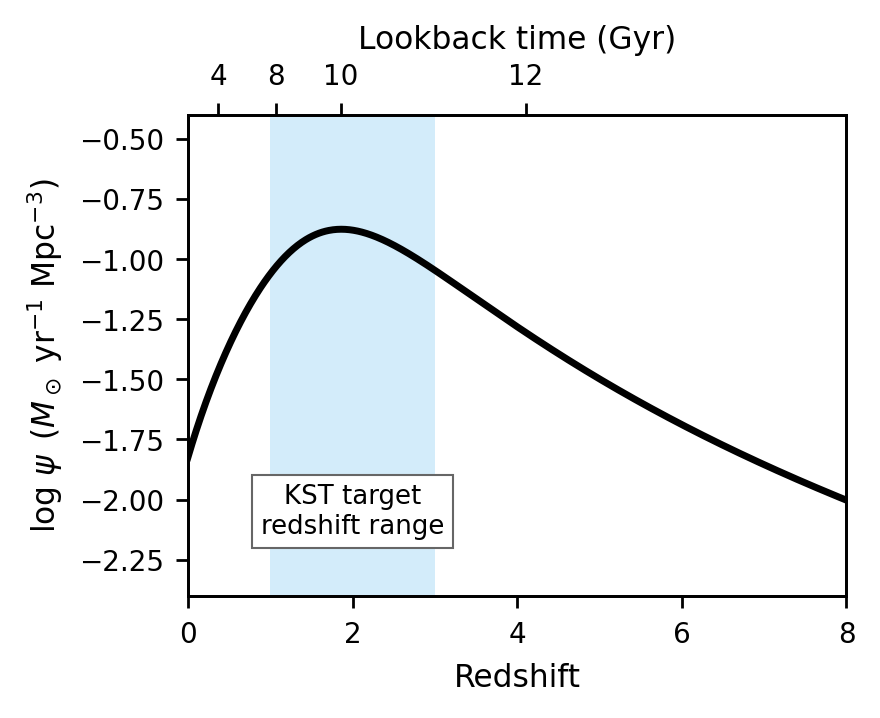}
\caption{The cosmic star formation history. The shaded region ($1\lesssim z\lesssim3$) marks the redshift range in which KST's $1.1$--$1.8~\mu$m channel samples the rest-frame optical continuum and the 4000~\AA\ break, anchoring photometric redshifts and stellar-population estimates around the star-formation peak. The curve shows the fitting function of \citet{Madau2014}.}
\label{fig:csfr}
\end{figure}

The Korean Space Telescope (KST) is optimized to bridge this gap. 
For galaxy-evolution science, KST combines the 2{,}000~deg$^{2}$  Wide Survey with the 10~deg$^{2}$ Cluster-Targeted Deep Survey. Together with its simultaneous dual-channel
(VIS+NIR) imaging and LVF  spectrophotometry ($R \sim 100$),
this creates a dataset unlike any available today. 
With a diffraction-limited PSF of $\sim 0.15''$, a $5\sigma$
Wide-Survey imaging depth of $r \approx 25.6$ (scaled from the 1-hr values of Table~\ref{tab:kst_specs}), and
simultaneous $R \sim 100$ NIR spectrophotometry reaching
$m_{AB} \approx 22.5$ per resolution element (1\,hr benchmark; Table~\ref{tab:kst_specs},
Figure~\ref{fig:sensitivity}; per-mode survey depths in Table~\ref{tab:lvf_modes}), KST will deliver resolved galaxy morphologies from
the VIS imaging together with NIR-anchored photometric
redshifts for a few $\times 10^8$ objects from $z \sim 0$ to $z \sim 3$, reaching $\sigma_z/(1+z) \lesssim 0.01$ for the NIR-detected subset (see Figure \ref{fig:csfr}).

This section describes how KST will address the following questions in galaxy
evolution:
\begin{itemize}
    \item \textbf{How does environment regulate star formation and morphological transformation from Cosmic Noon to the present?}
    \item \textbf{How do intracluster stars, ultra-diffuse galaxies, and tidal debris encode the assembly history of clusters?}
    \item \textbf{How has the global stellar mass density of the Universe been built up, and what quenches star formation?}
    \item \textbf{How do stellar halos and diffuse outskirts trace the hierarchical accretion history of galaxies?}
\end{itemize}

\subsubsection{Environmental Dependence of Galaxy Properties and Its Evolution}
\label{subsec:environment_evolution}

Galaxy evolution does not proceed in isolation. The physical properties of galaxies (including
stellar mass, star formation rate (SFR), morphology, internal structure, and nuclear
activity) are strongly influenced by their surrounding environment. Seminal observational
studies established that star formation rates are suppressed and early-type fractions are elevated
in dense environments in the local Universe \citep{Dressler1980, Kauffmann2004}, yet the
dominant quenching mechanisms and their dependence on halo mass, redshift, and large-scale
structure remain debated (e.g. \citealt{lf2026}). Environmental processes span a wide range of physical scales, from
the cosmic web (tens to hundreds of Mpc), through galaxy groups and clusters (Mpc scales), down
to direct galaxy--galaxy interactions and ram-pressure stripping (kpc scales). Disentangling
these contributions requires a facility that can simultaneously map the large-scale density field
and resolve individual galaxies within it across a wide redshift baseline.

The combination of the KST Wide Survey, which will reconstruct the three-dimensional cosmic
web over cosmological volumes, and the Cluster-Targeted Deep Survey, which will probe the
densest environments in unprecedented detail, makes KST well suited to establish a
coherent, multi-scale framework for environmental galaxy evolution.

\paragraph{Defining Galaxy Environments with KST}

The KST Wide Survey will provide accurate photometric redshifts for a few $\times 10^8$ galaxies over 2{,}000~deg$^{2}$, enabling three-dimensional reconstruction of large-scale
structure, with a precision of $\sigma_z/(1+z) \lesssim 0.01$ for the NIR-detected subset. Using various estimators of galaxy environments (e.g. \citealt{muldrew12,Darvish2015}), we will classify galaxies according to their
location within the cosmic web, including voids, filaments, sheets, galaxy groups, and massive
clusters (e.g.  \citealt{Laigle2018, kra2018, Im2024}). This is a critical advance over purely photometric surveys in optical bands, which
suffer from catastrophic redshift failures for red galaxies at $z > 1$ in the absence of deep NIR
data.

In parallel, galaxy cluster/group catalogs and nearest-neighbor analyses will characterize environments on
intermediate and small scales \citep{ph09, pc09}. This multi-scale approach allows KST to
separate the effects of large-scale density, halo mass, and local neighbor interactions,
providing a physically motivated definition of galaxy environment that is essential for
distinguishing between ``mass quenching'' and ``environmental quenching'' \citep{Peng2010}.
KST will therefore enable:
\begin{itemize}
    \item Tomographic reconstruction of the cosmic web to $z \sim 2$ over cosmological volumes,
    \item Consistent environmental classification across wide areas and a wide redshift baseline,
    \item Separation of halo-scale and neighbor-scale effects via multi-scale density estimation,
    \item Unified environment metrics applicable to all observed galaxy populations.
\end{itemize}

\paragraph{Evolution of Galaxy Properties across the Cosmic Web}

With its wide sky coverage and NIR depth, KST will measure how galaxy properties vary as
functions of environment and redshift across the full range $0 < z \lesssim 3$. By combining
photometric redshifts with physical parameter estimates (stellar masses from spectral energy
distribution (SED) fitting, SFRs from rest-frame UV and NIR flux ratios, and rest-frame colors
from the broad VIS+NIR photometric baseline), KST will trace the evolution of star formation,
stellar mass growth, and the bimodal color distribution across different cosmic web environments.

A central unresolved question is whether star formation suppression in galaxies occurs
preferentially upon accretion into massive halos (``satellite quenching'') or is driven by
internal processes tied to halo mass (``mass quenching'') \citep{Peng2010, Wetzel2013}. Current
observational constraints on the star formation--density relation at $z > 1$ are hampered by
the small areal coverage of deep NIR surveys. KST will address this with a dataset that is
simultaneously wide and deep.
By combining star-formation-rate indicators from the UV–optical continuum with LVF-detected 
H$\alpha$ emission at $0.7\lesssim z \lesssim 1.7$, KST will determine how the
star formation--density relation evolves, including its possible reversal \citep{elbaz07,hwa2019} and whether environmental quenching was already established
at Cosmic Noon \citep{Muzzin2012}.

Key science outcomes include:
\begin{itemize}
    \item Evolution of the star formation--density relation from $z \sim 2$ to $z \sim 0$,
    \item Environmental quenching efficiencies as a function of halo mass and redshift,
    \item Growth of the red sequence in voids, filaments, groups, and clusters,
    \item Dependence of stellar mass assembly on cosmic web location.
\end{itemize}

\paragraph{Morphology--Environment Relation and Structural Transformation}

Galaxy morphology and internal structure are among the most sensitive tracers of environmental
influence. In the local Universe, early-type galaxies preferentially occupy dense environments,
forming the morphology--density--luminosity relation \citep{Dressler1980, park08}. At intermediate
redshift this relation weakens, indicating that morphological segregation is established
gradually over cosmic time \citep{Postman2005}. Yet the physical mechanisms responsible---cumulative gravitational/hydrodynamic interactions, ram-pressure stripping, or the preferential accretion
of already-quenched field galaxies into clusters \citep{bg06,hp09,hwa18}---are not well constrained because morphological
measurements at $z > 0.5$ require space-based angular resolution.

KST's diffraction-limited PSF of $\sim 0.15''$ corresponds to a physical scale of
$\sim 1.3$~kpc at $z = 1$, sufficient to resolve bulges, disks, and structural disturbances in
massive galaxies across the full redshift range of interest. Non-parametric morphological
indicators such as the Gini coefficient, $M_{20}$, concentration, asymmetry, and smoothness
\citep{Lotz2004, con2014}, as well as parametric S\'{e}rsic decompositions, will be
applied uniformly over the survey (e.g. \citealt{park2022, leejh2024}). The unprecedented combination of depth, area, and resolution
will allow KST to address:
\begin{itemize}
    \item Evolution of morphology fractions (early-type, late-type, irregular) with density and redshift,
    \item Morphology--mass--environment relations across the cosmic web,
    \item Disk survival and bulge growth in dense regions,
    \item Identification of transitional populations (green-valley, post-starburst).
\end{itemize}

\paragraph{Merger History, Interaction-Driven Evolution, and Small-Scale Galaxy Ecology}
 
Galaxy mergers and tidal interactions are primary drivers of structural transformation, mass
growth, and star formation activity. Major mergers (mass ratio $> 1{:}4$) can trigger centrally
concentrated starbursts, build classical bulges, and drive powerful AGN feedback
\citep{Hopkins2008,kar2012}. Minor mergers and fly-bys contribute to disk heating, outer stellar halo
growth, and the redistribution of angular momentum. Environmental influence also operates on
sub-halo scales through direct interactions between neighboring galaxies. Star formation rates,
morphologies, and colors can depend strongly on whether a galaxy lies within the virial radius
of a massive companion and on the type of that companion, a phenomenon known as ``galactic
conformity'' \citep{Weinmann2006, ann2008,pc09}. These small-scale signals, largely independent of the
large-scale density field, provide direct evidence for gravitational and hydrodynamic interactions operating at the two-galaxy level, and they represent the microscopic counterpart
to the large-scale environmental quenching discussed above.
 
KST will investigate this full hierarchy of interaction-driven evolution through three
complementary observables: (1) close-pair statistics, counting galaxy pairs within projected
separations of $<50$~kpc and 
consistent LVF/photometric redshift probabilities, with line-of-sight velocity criteria added where ancillary spectroscopic redshifts are available, samples that simultaneously probe merging systems and nearest-neighbor conformity (e.g. \citealt{pc09}); (2) disturbed morphologies
identified via non-parametric asymmetry measures; and (3) detection of deep tidal features
such as streams, shells, and extended tails, accessible from space due to their extremely low
surface brightness ($\mu_V \gtrsim 28$~mag~arcsec$^{-2}$; \citealt{byun2026}). The accurate photometric redshifts
and large galaxy sample of KST will enable systematic studies of these effects out to
$z \sim 1.5$ and beyond. Together, these diagnostics will enable:
\begin{itemize}
    \item Measurement of major and minor merger rates as functions of mass, environment, and redshift,
    \item Identification of merger-induced starbursts, subsequent quenching episodes, and AGN triggering,
    \item Reconstruction of individual galaxy accretion histories from tidal debris and close-pair statistics,
    \item Quantification of morphological and star-formation conformity as functions of neighbor mass, type, and large-scale environment.
\end{itemize}

\paragraph{Joint Analysis of Galaxy Environments}

By integrating cosmic-web reconstruction, morphology measurements, merger diagnostics, and
cluster deep-field observations, KST will connect the small-scale processes that transform
individual galaxies to the large-scale structure that assembles the cosmic web. In summary,
KST will deliver:
\begin{itemize}
    \item A timeline of environmental quenching from Cosmic Noon to the present,
    \item The physical origin of morphological segregation as a function of epoch,
    \item The survival and transformation pathways of disk galaxies in dense regions,
    \item Statistical links between galaxy evolution and the growth of dark matter halos and clusters.
\end{itemize}

\subsubsection{Cluster Deep Fields: Intracluster Light and Low-Surface-Brightness Science}
\label{subsec:icl}

The Cluster-Targeted Deep Survey over approximately 10~deg$^{2}$ is designed to probe the
densest environments in the Universe with unprecedented sensitivity to faint and diffuse
structures. These observations will enable comprehensive investigations of intracluster light
(ICL), low-surface-brightness (LSB) galaxies, tidal debris, and environmentally driven galaxy
transformation processes. By combining exquisitely deep imaging with lensing-based mass
measurements, KST will establish a physically coherent picture of baryonic and dark matter
assembly in galaxy clusters that has been difficult with previous facilities.

The scientific case for deep cluster observations is strong. Clusters are the most massive
gravitationally bound structures in the Universe, and their assembly history encodes
information about the growth of dark matter halos, the efficiency of baryonic physics, and the
long-term fate of accreted galaxies. The diffuse stellar components, the ICL and extended brightest cluster galaxy (BCG)
envelopes, are particularly powerful diagnostics because they are collisionless and therefore
preserve kinematic and chemical records of past accretion events \citep{Monachesi2019, mihos2017, jj2023, joo2025}.
Yet detecting these components from the ground is severely limited by atmospheric PSF wings
and sky-background fluctuations, which contaminate faint extended emission at
$\mu_V \gtrsim 26$~mag~arcsec$^{-2}$. KST's space-based environment and stable PSF provide a
decisive advantage.

\paragraph{ICL as a Record of Cluster Assembly}

Intracluster light originates from stars that have been gravitationally stripped from galaxies
during mergers, tidal interactions, and satellite disruption. The spatial distribution,
morphology, and stellar populations of the ICL therefore preserve a fossil record of cluster
assembly and dynamical evolution \citep{Murante2007, Contini2014,kj2018}. The color and metallicity
gradient of the ICL trace the sequence in which galaxies were accreted. ICL contributed by
early, massive mergers tends to be metal-rich and concentrated, while that contributed by
recent tidal stripping of lower-mass satellites is more metal-poor and extends to larger radii.

KST will systematically map ICL surface brightness profiles, isophotal shapes, and color
gradients over wide radial baselines, extending from the brightest cluster galaxy (BCG) to the
cluster outskirts. These measurements will allow discrimination between different ICL formation
channels: BCG major mergers, gradual stripping of infalling satellites, and accretion of
pre-processed intragroup light \citep{mihos2017}. Key scientific outcomes include:
\begin{itemize}
    \item Measurement of ICL fractions as functions of cluster mass, redshift, and dynamical state,
    \item Identification of diffuse substructures (shells, streams, plumes) associated with recent interactions,
    \item Constraints on the timescales of cluster relaxation and the dominant ICL formation channel.
\end{itemize}

\paragraph{ICL and Dark Matter Coupling}

Because the ICL is dynamically collisionless, its distribution is expected to trace the cluster
gravitational potential more closely than the discrete galaxy population \citep{Monachesi2019}.
This makes ICL an independent and unbiased probe of the total mass distribution, complementary
to weak-lensing mass maps constructed from background galaxy shear \citep{yoo2022, hyeonghan2024a}. Comparing ICL surface
brightness profiles with weak-lensing convergence maps provides a powerful probe of the
baryon--dark matter coupling in clusters and a test of the self-similar nature of dark matter
halos predicted by CDM.

KST will quantify spatial alignments, centroid offsets, and substructure correlations between
ICL distributions and dark matter mass reconstructions from weak lensing. In dynamically
unrelaxed clusters, such as post-merger ``bullet cluster'' analogs, such comparisons will
reveal departures from equilibrium and constrain cluster dynamical histories (e.g., \citealt{cha2025, hyeonghan2024b, hyeonghan2025}). 
These measurements provide an independent baryon–mass comparison in dynamically disturbed clusters and complement the dark-matter tests developed in Section \ref{sec:dark_matter}, including searches for offsets between the lensing mass distribution and collisionless stellar tracers.
KST will enable:
\begin{itemize}
    \item High-precision spatial alignment tests between ICL and lensing mass maps,
    \item Detection of ICL--dark matter centroid offsets in unrelaxed and merging clusters,
    \item Identification of ICL streams associated with accreted dark matter subhalos.
\end{itemize}

\paragraph{BCG Growth and Extended Stellar Envelopes}

Brightest cluster galaxies (BCGs) represent the most massive stellar systems in the Universe
and grow primarily through ``dry'' (gas-poor) mergers and accretion of satellite galaxies
within the cluster potential. Their extended outer stellar envelopes form a transitional
component between the compact inner galaxy and the diffuse ICL. Cosmological simulations
predict that BCG stellar masses should roughly double from $z \sim 1$ to the present via
accretion \citep{DeLucia2007}, but observational constraints on BCG growth remain uncertain by
factors of several \citep{Lidman2012}.

Deep KST imaging will enable multi-component photometric decompositions of BCGs, separating
the inner S\'{e}rsic component, intermediate stellar envelope, and the diffuse ICL. Surface
brightness profiles will be traced to very faint levels ($\mu_V \sim 30$~mag~arcsec$^{-2}$) over a large sample of clusters spanning $0 < z \lesssim 1$. Measurements of envelope
growth, size evolution, and stellar population color gradients will directly constrain the
relative roles of in-situ star formation and accretion-driven (``ex-situ'') assembly. These
observations will:
\begin{itemize}
    \item Quantify BCG stellar mass and size growth since $z \sim 1$ for a statistically large sample,
    \item Trace the buildup of extended stellar envelopes through color gradients and S\'{e}rsic decompositions,
    \item Link BCG evolution directly to cluster assembly history and dynamical state.
\end{itemize}

\paragraph{Low-Surface-Brightness Galaxies and Dwarf Populations}

Galaxy clusters host abundant populations of low-surface-brightness (LSB) galaxies, including
ultra-diffuse galaxies (UDGs), objects with effective radii comparable to the Milky Way but
with stellar masses two to four orders of magnitude lower \citep{vanDokkum2015, Koda2015}.
UDGs are highly sensitive to environmental processing. Their unusually large sizes and diffuse
stellar distributions are thought to originate from either the ``failed galaxy'' scenario (high
spin halos with suppressed star formation) or from galaxies whose disks were puffed up and
subsequently quenched by cluster tidal and hydrodynamic forces \citep{Amorisco2016}. Their
survival or disruption thus provides a sensitive test of tidal and feedback models in dense
environments \citep{park2026}.

KST will conduct a census of UDGs and dwarf galaxies across the many clusters in the
Wide Survey footprint, reaching full co-added depth in the ten Deep-field clusters, an
ensemble spanning a wide range of masses and dynamical states, measuring their
spatial distributions, S\'{e}rsic parameters, and stellar population colors from the VIS+NIR
photometry. By linking UDG populations to the overall ICL budget, KST will clarify the
contribution of low-mass galaxy disruption to the total diffuse stellar mass in clusters. KST
will address:
\begin{itemize}
    \item Environmental dependence of UDG abundance and structural properties as a function of cluster mass,
    \item Tidal disruption rates of dwarf satellites and their contribution to the ICL,
    \item Connection between the properties of disrupted satellites and the diffuse stellar component.
\end{itemize}

\paragraph{Environmental Transformation within Clusters}

Dense cluster environments drive rapid galaxy transformation through a suite of physical
processes: ram-pressure stripping of cold gas reservoirs \citep{GunnGott1972}, cumulative gravitational/hydrodynamic interactions~\citep{hwa18},
galaxy harassment through high-speed encounters \citep{Moore1996}, removal of hot gas halos,
and repeated minor mergers. These processes leave distinct and observable signatures in galaxy
structure, star formation, and the distribution of stripped material. A key open question is
the relative importance of these mechanisms and how they depend on cluster mass, orbit, and
galaxy mass.

The KST cluster deep fields will enable measurements of radial gradients in star formation
rate, morphology, and stellar mass surface density from the cluster virial radius down to the
dense cluster core. KST will detect stripped stellar and gaseous debris, fading disk
components, and morphologically disturbed systems, providing direct evidence for
environmentally driven transformation processes. These data will enable:
\begin{itemize}
    \item Mapping of star formation quenching fronts as a function of cluster-centric radius,
    \item Identification of specific transformation pathways (ram-pressure stripping, tidal interactions, or removal of gas reservoirs) through morphological and color signatures,
    \item Quantification of the pre-processing of galaxies in infalling groups before they enter the cluster virial radius.
\end{itemize}

\paragraph{Tidal Debris, Streams, and Fossil Substructures}

Tidal streams, shells, and diffuse stellar plumes preserve fossil records of recent and ancient
accretion events. In the cluster environment, these features trace the ongoing redistribution
of stellar mass from disrupted satellites into the ICL reservoir. Their detection requires
both a stable, well-characterized PSF to suppress scattered light contamination and a deep
surface brightness limit, conditions met naturally by KST's space-based architecture.

KST will detect and characterize tidal substructures down to surface brightness levels of
$\mu_V \sim 30$~mag~arcsec$^{-2}$, enabling reconstruction of cluster assembly histories on
timescales of several Gyr. Combined with kinematic information from spectroscopic follow-up
and projected mass distributions from lensing, these data will provide time-resolved views of
cluster growth. KST will enable:
\begin{itemize}
    \item A census of tidal debris in the ten deeply imaged Deep-field clusters, extended to brighter features across the Wide-Survey clusters,
    \item Identification of infalling group remnants and their contribution to the ICL,
    \item Statistical constraints on satellite disruption timescales as a function of orbital parameters.
\end{itemize}

\paragraph{Low-Surface-Brightness Systematics and Analysis Strategy}

Robust LSB science requires careful control of both instrumental and astrophysical systematics.
The primary observational challenges are (1) the extended wings of bright star and galaxy PSFs
contaminating faint emission, (2) the subtraction of the sky background and any flat-field
gradients, and (3) contamination from unresolved background galaxies and galactic cirrus.
KST's space-based environment provides critical advantages for all three. The absence of
atmospheric scattering reduces PSF wings, the stable thermal environment eliminates
large-scale sky-gradient variations, and the excellent seeing significantly reduces the blending
of faint background sources.
The quoted surface-brightness performance is a mission-level target and will be validated through end-to-end simulations, in-flight calibration, and empirical characterization of scattered light, flat-field residuals, and sky-subtraction systematics.

Dedicated calibration strategies, advanced PSF reconstruction (including crowded-field
deconvolution using simultaneously observed PSF stars), and multi-scale decomposition
techniques will be applied to separate galaxy light from halos and ICL. End-to-end simulations
using synthetic cluster light profiles injected into real survey images, and machine-learning-assisted segmentation will be used to quantify surface-brightness-dependent detection biases and measurement uncertainties \citep{Duc2015,chun2026}. These efforts are designed to enable:
\begin{itemize}
    \item Characterized recovery of faint diffuse structures approaching $\mu_V \sim 30$~mag~arcsec$^{-2}$, subject to the final calibration performance,
    \item Accurate ICL fraction measurements across a wide range of cluster masses and redshifts,
    \item Well-characterized systematic uncertainties enabling meaningful comparisons with simulations.
\end{itemize}

\paragraph{Integrated View of Cluster Assembly}

By combining ICL mapping, galaxy population studies, UDG and dwarf censuses, and
weak-lensing mass reconstructions, KST will trace how baryons assemble within dark-matter
potentials and how environments shape galaxy populations over cosmic time. The survey will
provide matched catalogs of cluster galaxies, diffuse light, and lensing mass distributions.

\subsubsection{Cosmic Star Formation and Mass Assembly}
\label{subsec:sfh_mass}

The buildup of stellar mass and the evolution of star formation activity represent the
fundamental backbone of galaxy evolution. Observations have established that the cosmic star
formation rate density (SFRD) peaked near $z \sim 2$ and has declined by roughly an order of
magnitude to the present \citep{Madau2014,lin2026}. Over the same period, the comoving stellar mass
density has grown by a factor of $\sim$4--5 since $z\sim2$, and by more than an order of
magnitude since $z\sim4$ \citep{Ilbert2013}, reflecting the
integrated history of star formation, enrichment, and mass exchange through mergers and
outflows. Yet the physical mechanisms governing this transformation (including cold gas
accretion from cosmic filaments, quenching by AGN and stellar feedback, and the role of
environment in halting star formation) remain incompletely understood.

A key limitation of current surveys is the difficulty of constructing complete, mass-selected
galaxy samples at $z > 1.5$ with reliable physical parameter estimates. Deep optical surveys
from the ground (e.g., COSMOS) are restricted to small fields ($<10$~deg$^{2}$) and
suffer from significant photometric redshift uncertainties for red, dusty galaxies whose
4000~\AA\ break has redshifted beyond 1~$\mu$m. KST overcomes both limitations. Its $R \sim
100$ LVF spectrophotometry provides a direct spectral identifier for the 4000~\AA\ break and
major emission lines (H$\alpha$, \text{[O III]}) over the full 2{,}000~deg$^{2}$ Wide Survey
footprint, enabling photometric redshifts with precision $\sigma_z/(1+z) \lesssim 0.01$ out to
$z \sim 3$ for galaxies bright enough in the NIR (Table~\ref{tab:lvf_modes}).

Using these measurements, KST will construct one of the largest homogeneous galaxy samples spanning $0 < z \lesssim 3$ with jointly measured morphology, broad-band SEDs, and LVF-based redshift information, enabling systematic studies of mass growth across
cosmic time. KST will enable:
\begin{itemize}
    \item Precision measurements of the evolution of the cosmic star-formation-rate density from $z \sim 3$ to $z \sim 0$, separated by galaxy mass and environment,
    \item Determination of the redshift-dependent stellar mass function and its faint-end slope across the epoch of peak activity,
    \item Mapping of the star formation main sequence \citep{Noeske2007,elbaz2011} and its scatter as a function of mass, environment, and redshift,
    \item Identification of transitioning (green-valley) and post-starburst (E+A) populations that trace the pathways from active to passive.
\end{itemize}

By combining mass assembly histories with environmental information from the cosmic web
reconstruction and morphological measurements, 
KST will constrain the relative importance of gas accretion, stellar feedback, mergers, and environmental processes in regulating star formation and quenching across diverse galaxy populations. The complementary role of AGN feedback and black-hole growth is developed in Section \ref{subsec:agn}.

\subsubsection{Galaxy Clusters and Early Assembly}
\label{subsec:proto}

At $z \gtrsim 1.5$, the progenitors of today's massive clusters (often called
``protoclusters'') are rapidly assembling through hierarchical accretion. Protoclusters are
predicted by simulations to span tens of comoving Mpc and to contain the seeds of the most
massive galaxies observed today \citep{Overzier2016,Andrews2025}. Yet their observational characterization
remains incomplete because they are extended, diffuse, and require wide-area surveys with deep
NIR sensitivity to identify the overdense galaxy populations within them \citep{ram2024}.

KST will detect protocluster candidates through overdensities in photometric redshift space
and through the coherent large-scale modes recovered from its cosmic web reconstruction. 
The VIS broad-band data will identify Lyman-break galaxy overdensities,
while the LVF data will refine their redshift distributions and identify
emission-line overdensities, including [O~III]- and H$\alpha$-selected
structures where the relevant lines fall within the $1.1$--$1.8~\mu$m
bandpass.
By comparing the galaxy populations and morphologies within these structures to
field samples at the same redshift, KST will enable:
\begin{itemize}
    \item Measurements of the onset and accelerated pace of quenching in overdense environments at $z > 1.5$ (``early quenching''),
    \item Studies of rapid morphological transformation and the early emergence of the morphology--density relation,
    \item Detection of nascent diffuse stellar light in protoclusters, constraining the epoch of first ICL formation.
\end{itemize}

These observations will directly link the early assembly stages of protoclusters to the
structural and population properties of present-day clusters, providing a complete picture of
cluster formation across cosmic time.

\subsubsection{Stellar Halos and Diffuse Outskirts}
\label{subsec:halos}

The stellar halos of galaxies (faint, extended envelopes of stars that surround the main
body of a galaxy at surface brightnesses below $\mu_V \sim 27$~mag~arcsec$^{-2}$) are fossil
records of hierarchical mass assembly and satellite accretion \citep{bj05, Helmi2008}.
In the $\Lambda$CDM paradigm, stellar halos are built primarily through the tidal disruption
of accreted dwarf galaxies whose stars are dispersed into the outer regions of the host halo. The total mass, metallicity, color, and spatial extent of the
stellar halo therefore encode information about the number, mass ratio, and epoch of past
accretion events. The Milky Way's stellar halo, with its many identified streams (e.g., \citealt{myeong2019}), provides a benchmark that KST will generalize to a statistical ensemble
of external galaxies.

The stability of space-based observations (with a PSF that does not vary with atmospheric
conditions and a dark sky background free from airglow) enables KST to detect extended halos
and diffuse outskirts to extremely low surface brightness levels that are unreachable from the
ground. KST will measure halo extent, mass fraction, ellipticity, and color gradients for
statistically large samples of galaxies across a range of masses, morphologies, and
environments. These observations will allow:
\begin{itemize}
    \item Rigorous tests of $\Lambda$CDM predictions for stellar halo growth as a function of merger history,
    \item Constraints on the mass, number, and epoch of accreted satellites through halo mass and color gradients,
    \item Studies of halo structural diversity as a function of host mass, morphology, and environment.
\end{itemize}

By combining halo properties with lensing-based total mass measurements, KST will connect the
stellar halo to the underlying dark matter structure and provide critical observational
benchmarks for cosmological hydrodynamic simulations \citep{Pillepich2018,lee2021}.

\subsubsection{Synergy with International and Korean Facilities}
\label{subsec:synergy_ge}

The scientific return of KST's galaxy evolution program will be substantially amplified through
strategic synergies with complementary ground-based and space-based facilities. KST serves as
the high-resolution, wide-area NIR anchor in a coordinated network spanning photometric
calibration, spectroscopic follow-up, and multi-wavelength characterization.

\paragraph{7-Dimensional Telescope (7DT)}

The 7-Dimensional Telescope (7DT) is a Korean-led wide-field imaging system composed of
twenty 0.5~m telescopes operating in concert at the La Silla Observatory, covering
$\sim$1{,}000~deg$^{2}$ with 40 medium-band filters spanning $0.4$--$0.9~\mu$m at a
spectral resolution of $R \sim 50$. The 7DT's medium-band optical photometry is a natural
and powerful complement to KST's NIR LVF spectrophotometry. Together, the two surveys will
deliver continuous spectrophotometric coverage from $0.4$ to $1.8~\mu$m at $R \sim 50$--$100$
for every object in their common footprint, a capability that no single facility provides.
This joint SED will break the age--dust--metallicity degeneracy that limits photometric
stellar population estimates, enabling robust determinations of stellar mass, SFR, and stellar
age for hundreds of millions of galaxies. The 7DT's medium-band photometric redshifts,
precise to $\sigma_z/(1+z) \sim 0.005$ at $z < 0.5$ \citep{kim2024}, will also calibrate
and validate the KST LVF-based photometric redshifts at low redshift, where the 7DT has
superior optical sensitivity. As a Korean facility operating concurrently with KST, the 7DT
represents the most direct ground-based counterpart to KST's galaxy evolution program.

\paragraph{Vera C. Rubin Observatory / LSST and SPHEREx}

The Vera C. Rubin Observatory's LSST will survey $\sim$18{,}000~deg$^{2}$ to $r \approx
27.5$ in $ugrizy$, providing the deepest wide-area optical baseline available. KST
complements Rubin by supplying the diffraction-limited morphological measurements
($0.15''$ PSF) and NIR photometry ($1.1$--$1.8~\mu$m) that LSST cannot provide, while
Rubin's $u$-band depth constrains rest-frame UV fluxes for SFR estimates at $z < 0.5$
\citep{Salvato2019}. NASA's SPHEREx extends the combined wavelength baseline further, with
all-sky LVF spectrophotometry at $R \sim 40$--$130$ out to $5~\mu$m. Although SPHEREx's
$6.2''$ pixel scale cannot resolve galaxy substructure, its $3$--$5~\mu$m coverage constrains
stellar masses \citep{leejh2025},  reverberation mapping for AGN-host galaxies \citep{kimm2021} and polycyclic aromatic hydrocarbon (PAH)-based SFRs \citep{shim2023} in the regime beyond KST's $1.8~\mu$m cutoff. The
combination of LSST (deep optical), KST (resolved NIR morphology and spectrophotometry), and
SPHEREx (all-sky near- to mid-infrared) creates an unprecedented panchromatic survey capability for
galaxy evolution science.

\paragraph{Korea Microlensing Telescope Network (KMTNet) and Gemini Observatory}

KMTNet's three 1.6~m telescopes provide continuous wide-field optical time-series monitoring
of the Galactic bulge and Magellanic Clouds, complementing KST's bulge campaign for
variable-star and distance-ladder work (Section~\ref{sec:milky_way}). The twin 8.1~m Gemini telescopes (North
and South) provide the high-resolution spectroscopic follow-up that KST's $R \sim 100$ LVF
cannot deliver. Gemini instruments (GNIRS, FLAMINGOS-2, GIRMOS) will measure stellar velocity
dispersions, gas kinematics, and stellar population diagnostics (age, $[\mathrm{Fe/H}]$,
$[\alpha/\mathrm{Fe}]$) for high-priority KST-selected subsamples, including protoclusters,
BCGs, post-starburst galaxies, and strong lensing arcs at $z > 1$.

\paragraph{Nancy Grace Roman Space Telescope, JWST, and Multi-wavelength Facilities}

Roman's High Latitude Wide Area Survey ($\sim$2{,}000~deg$^{2}$, $0.11''$ pixel scale,
$0.5$--$2.3~\mu$m) shares significant scientific territory with KST. Roman's deeper NIR
imaging ($H \approx 26.5$) and extended wavelength coverage to $2.3~\mu$m will complement
KST's comparable-area survey, simultaneous VIS+NIR architecture, and LVF-based spectrophotometric coverage.
\textit{JWST} will characterize in detail the
rarest KST-selected targets (massive quiescent galaxies at $z > 3$, obscured AGN, and
multiply lensed arcs) in pencil-beam fields. \textit{eROSITA}'s all-sky X-ray catalog will
provide mass-selected cluster and AGN samples for direct cross-matching with the KST Wide
Survey, while ALMA will supply molecular gas masses and star formation efficiencies for
targeted high-$z$ subsamples, completing the baryonic picture of galaxy evolution that KST
alone cannot provide.

\subsubsection{Legacy Value}
\label{subsec:legacy}

KST will provide an enduring legacy dataset for galaxy evolution by combining wide-area
statistical power with deep cluster observations, in a single, homogeneous, and well-calibrated
survey architecture. The survey data products will include photometric redshift catalogs for
hundreds of millions of galaxies over 2{,}000~deg$^{2}$, reaching $\sigma_z/(1+z) \lesssim 0.01$ for the NIR-detected subset,
structural parameter catalogs (S\'{e}rsic fits, morphology classifications, merger flags) from
the $0.15''$ PSF imaging, and LVF spectrophotometric catalogs with $R \sim 100$ spectra in the
1.1--1.8~$\mu$m range. Further products are deep ICL and stellar halo surface brightness maps for the ten Deep-field clusters, with
shallower maps for hundreds of Wide-Survey clusters, and variability catalogs for AGN and transients in the Deep fields. These datasets will
support Korean-led and international science programs for decades and will serve as the primary
calibration and training set for the next generation of machine-learning-based photometric
redshift and morphological classification pipelines.

\subsection{Active Galactic Nuclei and Black Hole Growth}
\label{subsec:agn}
\label{sec:agn}

Supermassive black holes (SMBHs) are ubiquitous, at least at the centers of massive galaxies, and are known to play an important role in galaxy evolution, largely driven by the tight correlation between SMBH mass and host galaxy stellar mass \citep[e.g.,][]{kormendy2013}. While the majority of SMBHs are dormant without a significant amount of accreting material, a fraction of them become active due to rapid gas accretion toward the central SMBH—a phenomenon known as an active galactic nucleus (AGN). As AGNs are among the most energetic objects in the universe, their study is critical for several reasons:

\begin{itemize}
    \item \textbf{Black Hole Physics}: AGNs are the only laboratories where we can directly observe SMBH growth and accretion physics across cosmic time, probing extreme gravity, relativistic jets, and accretion disk physics.

    \item \textbf{Co-evolution with Galaxies}: There is a tight correlation between SMBH mass and host galaxy properties (e.g., stellar mass and stellar velocity dispersion), suggesting that black holes and galaxies co-evolve. This likely occurs through kinetic and radiative AGN feedback, which can quench or trigger star formation, making AGNs a key driver of galaxy evolution.

    \item \textbf{Cosmic History of Black Hole Growth}: AGN luminosity and black hole (BH) mass functions, along with their evolution, trace how and when SMBHs assembled their mass, directly connecting to the cosmic star formation history.    

    \item \textbf{Probes of Large-Scale Structure}: AGNs are highly luminous objects visible across the distant universe, making them powerful tracers of large-scale structures, including dark matter halos and cosmic web filaments. They are also frequently used to probe cosmology through baryon acoustic oscillations \citep[e.g.,][]{DESI2025}.
\end{itemize}

The combination of wide-field imaging, low spectral resolution, and NIR coverage makes KST particularly powerful for statistical studies of AGN demographics and their physical properties across cosmic time.

\subsubsection{AGN Host Galaxy Decomposition}
The properties of host galaxies provide crucial constraints on the physical mechanisms that drive and regulate AGN activity, as well as the pathways of BH-host coevolution. For example, the luminosity of the stellar bulge component is particularly informative, as it traces the depth of the gravitational potential well and encodes the mass of the central supermassive black hole through well-established scaling relations such as the $M_{\rm BH}-L_{\rm bulge}$ correlations \citep{magoriaan1998}. In addition, the stellar population age of the host, derived from spectral energy distribution (SED) fitting or absorption-line diagnostics, reflects the integrated star formation history and can reveal whether AGN feedback has quenched or enhanced star formation on galactic scales \citep{Kauffmann2003}. Furthermore, host galaxy morphology—whether disk-dominated, elliptical, or exhibiting signs of tidal interaction—constrains the dynamical pathways through which gas is funneled to sub-parsec scales to fuel the central engine. Disturbed or merging systems show elevated AGN fractions at high luminosity, while Seyfert-level activity is preferentially found in spiral hosts with prominent bulges \citep{kim2017}, suggesting that secular processes and major mergers occupy distinct regimes of the AGN luminosity function. Together, bulge luminosity, stellar age, and morphology provide complementary diagnostics that, when combined with multiwavelength AGN characterization, enable a more complete picture of the coevolution of black holes and their host galaxies across cosmic time.

\begin{figure}
\centering
    \includegraphics[width=\linewidth]{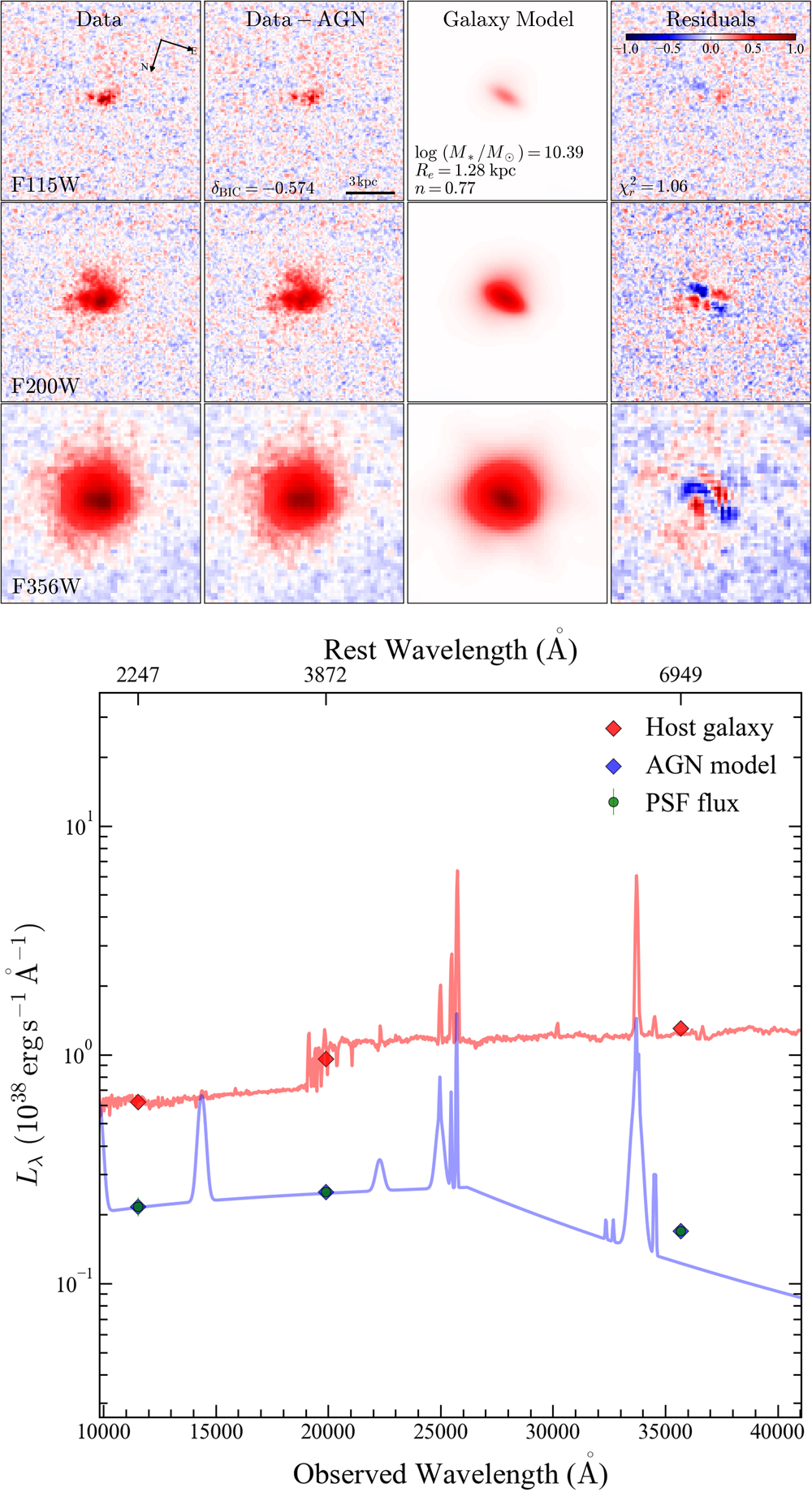}
    \caption{Demonstration of simultaneous two-dimensional imaging and SED decomposition of a high-$z$ AGN with \texttt{GalfitS}, shown here with \textit{JWST}/NIRCam data (F115W, F200W, and F356W imaging; the fitted host-galaxy and AGN SEDs extend to $4~\mu$m) from \citet{li2026}. KST will apply the same technique to its own data, replacing the three NIRCam broadbands with diffraction-limited optical $griz$ imaging and contiguous $\sim$48-channel LVF spectrophotometry over $1.1$--$1.8~\mu$m, to disentangle host galaxies from their central nuclei for the intermediate-redshift AGN samples described in the text.}
\label{fig:agn_galfits}
\end{figure}

In this context, KST will enable in-depth studies of AGN host-galaxy properties. The combination of high spatial resolution and spectrophotometric capabilities across a wide wavelength range, spanning from the optical to the NIR, will enable us to decompose host galaxies from their bright central nuclei effectively (Figure~\ref{fig:agn_galfits}; \citealp{li2026}). This decomposition yields robust estimates of bulge luminosity, stellar age, and merger features of AGN hosts, uncovering their physical connection to the properties of the central engine. In particular, the LVF will provide the spectral data essential for robustly estimating bulge mass and age, offering an unprecedented opportunity to explore the BH–host connection across a large sample at intermediate redshifts. 

\subsubsection{AGN Variability and Reverberation Mapping}
Repeated LVF observations can measure variability in the NIR continuum and broad emission lines of sufficiently bright AGNs. At these wavelengths, the dominant continuum contributors shift from the optical accretion disk to the innermost regions of the dusty torus, where thermal emission from hot dust grains near the sublimation temperature ($T_\mathrm{sub} \sim 1500$--$1800~{\rm K}$) peaks around the H-band ($\sim1.6~\mu\text{m}$; \citealp{Barvainis1987}). Reverberation mapping (RM) in the NIR exploits the time delay between UV/optical continuum fluctuations and the lagged response of this hot dust emission to constrain the dust sublimation radius, providing an independent geometric distance estimator and a potential extragalactic standard candle \citep{koshida2014}. Dust RM lags observed across this bandpass follow the luminosity--lag relation $\tau \propto L^{0.5}$ predicted by simple torus models \citep{Suganuma2006}, though growing observational evidence for chromatic lag structure, in which the H-band response lags the Y/J bands, implies a stratified, wavelength-dependent dust emission geometry that challenges single-sublimation-temperature torus models and motivates higher-cadence, multi-epoch NIR monitoring programs \citep{lyu2019}.

Within the $1.1$--$1.8~\mu\text{m}$ bandpass, several broad emission lines are accessible depending on the source redshift, including H$\alpha$, Pa$\beta$ ($\lambda1.282~\mu\text{m}$), and Pa$\gamma$ ($\lambda1.094~\mu\text{m}$) from the broad-line region (BLR). These offer a determination of BLR size that is unaffected by the heavy dust obscuration complicating optical H$\beta$-based RM \citep{lafranca2015}. The $R \sim 100$ spectral resolution is sufficient to isolate these lines from the underlying continuum and neighboring features, while simultaneously enabling efficient cadenced monitoring of large AGN samples. This configuration is exceptionally well-suited to wide-field photometric variability surveys with KST. Continuum and emission-line lags would constrain the sizes of the hot-dust region and the BLR.

\subsubsection{Physical Properties in Intermediate-$z$ AGNs}

\begin{figure}
\centering
    \includegraphics[width=\linewidth]{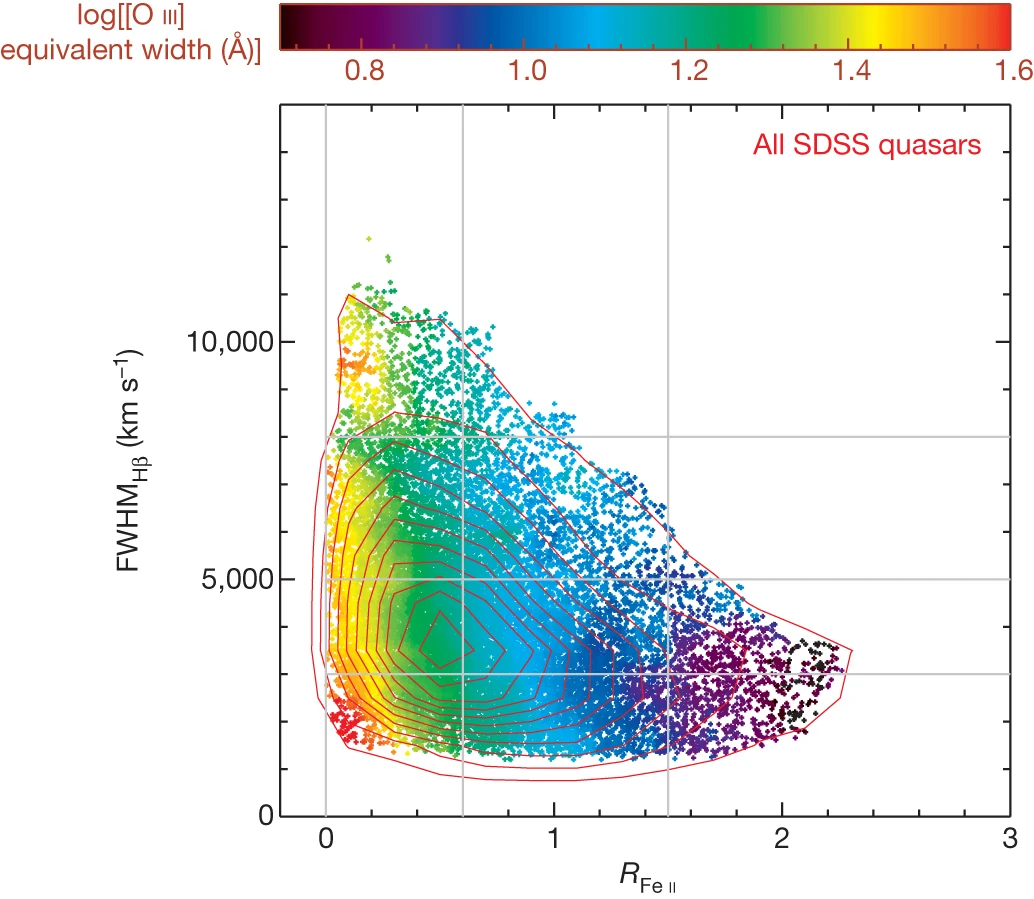}
    \caption{Comparison between the relative strength of \text{Fe\,\textsc{ii}} ($R_{\rm Fe\,II}$) and the FWHM of broad H$\beta$ for low-$z$ QSOs, adapted from \citet{shen2014}. The color coding denotes the equivalent width of [\text{O\,\textsc{iii}}], indicating that it traces EV1 along with $R_{\rm Fe\,II}$.}
\label{fig:agn_shen}
\end{figure}

\citet{boroson1992} applied principal component analysis to a sample of low-$z$ Palomar-Green (PG) quasars, isolating a dominant variance vector known as Eigenvector 1 (EV1) that organizes the diverse spectral properties of Type~1 AGNs into a coherent ``quasar main sequence.'' EV1 defines a prominent anti-correlation between the optical iron emission strength, $R_{\mathrm{Fe\,\textsc{ii}}} \equiv \text{Fe\,\textsc{ii}}\,\lambda4570/\text{H}\beta$, and the full width at half-maximum of the broad $\text{H}\beta$ line, $\text{FWHM}(\text{H}\beta)$ (i.e., strong \text{Fe\,\textsc{ii}} is associated with narrower $\text{H}\beta$ lines). Theoretically, EV1 is primarily driven by the Eddington ratio; AGNs with high Eddington ratios can drive strong outflows, altering the ionization structure to favor \text{Fe\,\textsc{ii}} production; such sources typically host lower-mass black holes with high accretion rates and show narrower broad lines. Conversely, the scatter around EV1 may be governed by secondary parameters, such as the viewing angle and black hole mass \citep[Figure~\ref{fig:agn_shen};][]{shen2014}. Consequently, this spectral framework has become a standard tool for quasar taxonomy. However, it has been validated comprehensively only for low-$z$ AGNs with optical spectra. 

KST provides an ideal framework to extend EV1 studies to intermediate-$z$ AGNs. The strong optical \text{Fe\,\textsc{ii}} complex ($\lambda\lambda4500$--$5400\,\text{\AA}$) at $z \sim 1.4$--$2.3$ shifts directly into the LVF coverage window. At $R \sim 100$, \text{Fe\,\textsc{ii}} blends can be decomposed from adjacent features, enabling EV1 analysis at cosmic noon. With this statistical dataset, we can systematically explore the physical origin of EV1 and characterize its intrinsic scatter across an extended parameter space.

\subsubsection{SMBH Demographics}

The black hole mass function is key to understanding the formation mechanisms and evolution of SMBHs across cosmic time. Directly measuring BH masses for large samples of quiescent galaxies remains exceptionally challenging, as resolving the SMBH sphere of influence is currently feasible for only a few dozen nearby, massive systems. However, for unobscured AGNs, the presence of broad-line regions under the gravitational influence of the central engine enables us to reliably estimate BH masses using single-epoch spectral measurements of broad emission lines across large datasets. 

Specifically, H$\alpha$ at $z \sim 0.7$--$1.7$ and H$\beta$ at $z \sim 1.3$--$2.7$ fall within the coverage window of the LVF, enabling single-epoch virial black hole mass estimates across cosmic noon. In addition, Mg\,\textsc{ii}\,($\lambda2798\,\text{\AA}$) at $z \sim 2.9$--$5.4$ provides access to black hole masses extending back toward the high-redshift universe. At $R \sim 100$, the FWHM of these broad lines can be reliably measured for sources with line widths $\gtrsim 3000$--$4000~{\rm km\,s}^{-1}$ (typical of Type~1 AGNs), allowing us to construct robust black hole mass functions over a wide redshift baseline.

\subsubsection{AGN Luminosity Functions \& Number Counts}
 
The robust estimation of AGN luminosity functions across cosmic history is essential to understanding the evolution of BH growth and their roles in galaxy evolution, by comparing them with the star formation history. A combination of deep optical broadband photometry and $R \sim 100$ LVF spectroscopic data substantially improves SED fitting compared to standard broadband photometry alone. 

This leads to:
\begin{itemize}
    \item Cleaner separation of the AGN from the inactive galaxy, which is important to the pre-selection of AGN candidates.
    
    \item Extending AGN selection beyond X-ray or mid-IR methods, recovering
          low-luminosity and obscured broad-line AGN, and improvement of the completeness of Type~1 AGN selection.
    
    \item Better photometric redshift precision for AGN by detecting prominent emission lines, reducing scatter in luminosity function measurements at $1 < z < 3$.
    
\end{itemize}

\subsubsection{Obscured AGNs} \label{sssec:obscured_agns}
According to the evolutionary scenario of AGNs, dust-obscured AGNs may reflect the early phase of merger-driven AGNs, in which AGNs play an important role in shaping the host galaxies through outflows \citep[e.g.,][]{sanders1988}. In this light, the demographics of obscured AGNs compared to unobscured AGNs are crucial to achieving a complete view of AGN evolution and their physical connection with host galaxies. Because NIR light penetrates dust more effectively than optical wavelengths, the NIR regime enables the effective identification of (partially) obscured AGNs via detections of Pa$\beta$ ($\lambda1.282\,\mu\mathrm{m}$) and Pa$\gamma$ ($\lambda1.094\,\mu\mathrm{m}$), which are accessible at $z \lesssim 0.4$. This will enable the identification of AGNs missed by optical spectroscopic surveys and allow for a robust estimation of line ratios between optical and NIR hydrogen recombinations. This is essential to investigate the dust extinction toward the central AGNs and the physical properties of the BLR.

\subsubsection{AGN Outflows of Intermediate-$z$ AGNs} \label{sssec:agn_outflows}

The narrow-line region (NLR) of AGNs provides critical insights into the energetic scaling relations and hydrodynamic feedback processes connecting AGNs to their host galaxies. Probing this ionized gas phase often utilizes the [O III] $\lambda5007$ transition in low-redshift samples. With KST, the NIR [S III] $\lambda\lambda9069, 9532$ doublet at moderate redshifts ($z \sim 0.2$--$0.9$) can be detected. Even at the low spectral resolution of $R \sim 100$ intrinsic to the LVF architecture, the high equivalent widths of these forbidden lines in luminous AGNs ensure robust detectability. This configuration facilitates statistical mapping of extended ionization-cone geometries and a systematic census of galactic-scale, AGN-driven outflows, which are crucial for assessing the kinetic feedback responsible for quenching star formation \citep{belli2024}. Furthermore, the [S III] doublet serves as a powerful physical anchor within the LVF bandpass; because its theoretical flux ratio is fundamentally invariant under atomic physics, it establishes an indispensable tool for empirical flux calibration, attenuation corrections, and secure line identification across large-sky spectroscopic surveys.

\subsubsection{High-Redshift Quasars ($z > 5$)} \label{sssec:highz_quasars}

The discovery and characterization of luminous QSOs at the dawn of cosmic time ($z > 5$) are vital to mapping the growth of the earliest SMBHs and tracking the reionization history of the intergalactic medium (IGM). While recent wide-field optical and NIR imaging surveys have significantly expanded the high-redshift census, traditional ground-based color selections suffer from severe completeness drops due to contamination from brown dwarfs and compact, dusty galaxies. Spectroscopic follow-up of these rare targets is heavily resource-intensive. 

While the SPHEREx dataset has been effectively utilized to identify high-redshift QSOs above $z \sim 5$, detecting these distant targets remains limited during the SPHEREx mission due to the shallowness of its all-sky survey strategy and severe source confusion within its deep fields \citep[e.g.,][]{davies2026}. Thanks to the significantly greater survey depth and superior spatial resolution of KST, this facility will enable efficient detection and confirmation of high-$z$ QSOs. In particular, prominent ultraviolet (UV) broad emission lines shift into the LVF window at redshifts set by the $1.1$--$1.8~\mu$m bandpass: \text{Mg II} ($\lambda2800$) at $z\lesssim5.4$, \text{C IV} ($\lambda1550$) at $z\gtrsim6.1$, and Ly$\alpha$ ($\lambda1216$) only at $z\gtrsim8$ (at lower redshifts Ly$\alpha$ falls in the VIS $i$ and $z$ bands). These lines can be resolved at the intrinsic spectral resolution of $R \sim 100$. This specific configuration allows us not only to determine secure redshifts but also to estimate underlying black hole masses and Eddington ratios, providing critical inputs for evaluating the physical seeding and growth mechanisms of SMBHs in the early universe. Furthermore, tightly constraining the bright end of the UV luminosity function will clarify the exact cosmic contribution of high-$z$ QSOs to the global hydrogen reionization budget.

\subsection{Exoplanet Discovery via Microlensing} \label{sec:exo}
\begin{figure}
\centering
    \includegraphics[width=\linewidth]{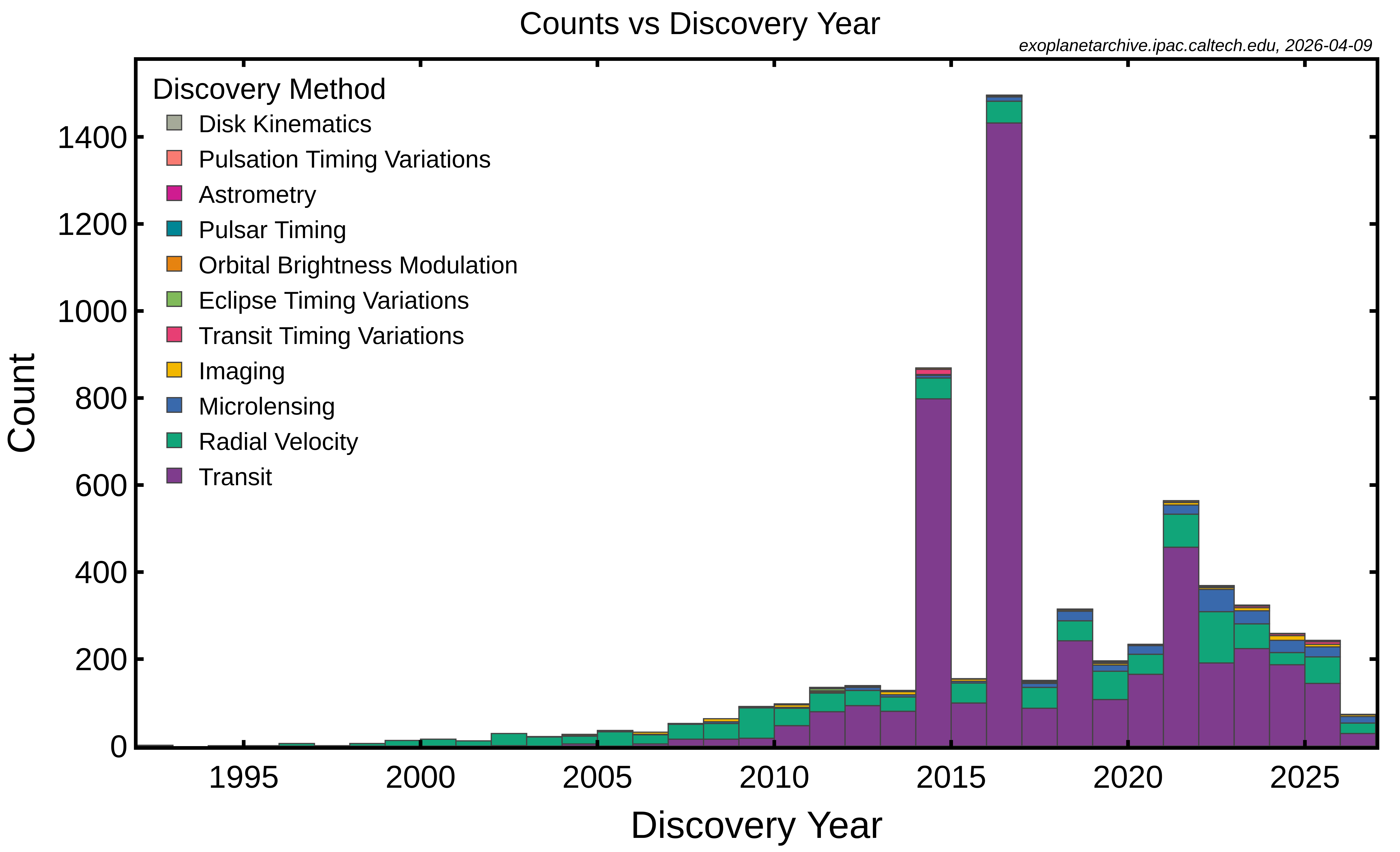}
    \caption{Histogram of confirmed exoplanets by discovery year and detection method. The annual number of confirmed exoplanets is shown, color-coded by detection method. Transit surveys dominate the overall discovery yield, with radial-velocity surveys providing the second-largest contribution, while microlensing, imaging, astrometry, and other methods probe complementary regions of parameter space. The rapid increase in confirmed planets in the mid-2010s illustrates the transformative role of space-based missions such as Kepler and K2. Adapted from the NASA Exoplanet Archive.}
\label{fig:exo_number}
\end{figure}

The discovery and characterization of exoplanets represents one of the major frontiers in modern astronomy. Exoplanet discovery has expanded rapidly over the past two decades, with the NASA Exoplanet Archive listing more than 6,300 confirmed planets as of September 2026. Most exoplanet discoveries have been made through transit photometry, with radial-velocity monitoring providing the second-largest contribution, while gravitational microlensing, astrometry, direct imaging, and other methods probe complementary regions of parameter space. Figure~\ref{fig:exo_number} shows the number of confirmed exoplanets discovered by different detection methods as a function of discovery year. 

Among these methods, transit photometry has been the most productive, accounting for more than 70 percent of all known exoplanets. This success has been driven by NASA’s space missions, particularly the Kepler space telescope \citep{Borucki2010} and its extended K2 mission \citep{Howell2014}. The Kepler space telescope, launched in 2009 and decommissioned in 2018, identified more than 2,600 confirmed planets through continuous monitoring of a single 105 square degree field. Transit surveys are particularly powerful for detecting short-period planets, including hot Jupiters and compact multi-planet systems. However, because the transit method requires a nearly edge-on orbital alignment and its sensitivity decreases rapidly with increasing orbital separation, it is much less effective at detecting planets on wide orbits.

Astrometry and microlensing probe discovery spaces that are complementary to those of transit surveys. Astrometry is particularly well suited to detecting planetary systems with orbital orientations that are favorable for motion in the plane of the sky \citep{Perryman2014}. Although only a modest number of planets have been confirmed through astrometry to date, forthcoming Gaia data releases, based on long-baseline high-precision astrometric time series, are expected to significantly expand this domain. In contrast, microlensing is uniquely sensitive to cold planets located at or beyond the snow line and remains the only method sensitive to old, low-mass free-floating planets (down to Earth mass), often referred to as rogue planets, that are not gravitationally bound to any host star \citep{Gaudi2012, Mroz2023}. Although fewer planets have been discovered through microlensing than through transit surveys, microlensing probes a fundamentally different region of parameter space, namely cold planets at orbital separations of approximately 1–10 au and beyond. This regime remains largely inaccessible to transit and radial-velocity techniques. 

In particular, the sharp increase in confirmed exoplanet discoveries in 2014 and 2016 in Figure~\ref{fig:exo_number}, driven primarily by the Kepler and K2 missions, clearly demonstrates the transformative power of continuous, stable, space-based photometric monitoring. In this context, the continuous high-cadence space-based monitoring capability of the proposed KST mission would open a powerful new window for exoplanet discovery through microlensing, thereby bridging the gap between close-in and wide-orbit planetary populations.

\subsubsection{Microlensing Detection} \label{sec:exo:microlensing}

\begin{figure}
\centering
    \includegraphics[width=\linewidth]{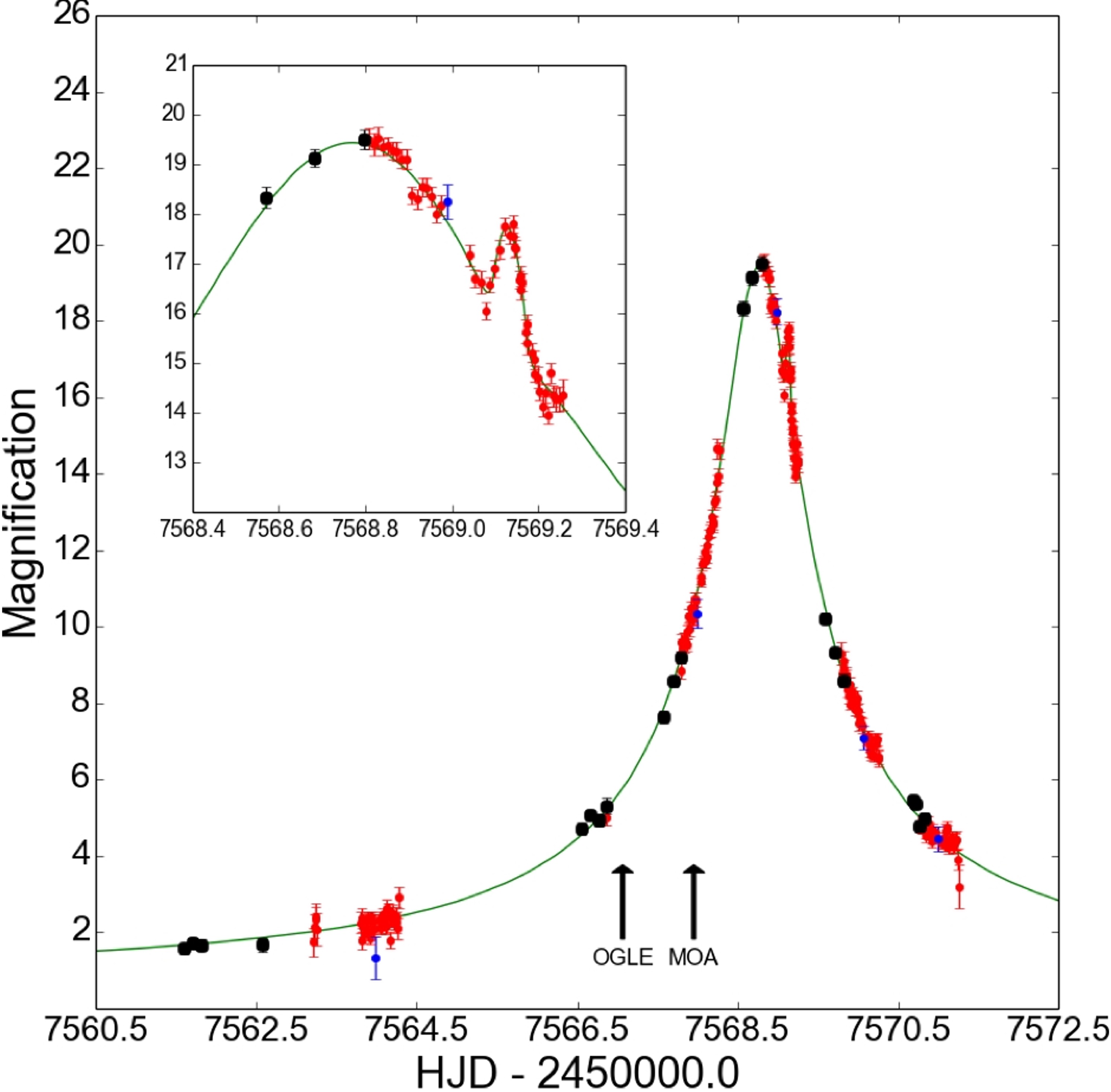}
    \caption{Light curve of the planetary microlensing event OGLE-2016-BLG-1195. The brief anomaly near peak magnification indicates the presence of a planetary companion. Adapted from \citet{Bond2017}.}
\label{fig:exo_lightcurve}
\end{figure}

Gravitational microlensing occurs when a foreground object passes sufficiently close to the line of sight to a more distant background star. The foreground object acts as the lens and the background star as the source. Because the individual lensed images are typically unresolved, the observable signature is not image splitting but a time-dependent magnification of the source flux. If the lens hosts a planet, the light curve may exhibit a short-duration deviation from the standard single-lens form, which is the key signature used to detect exoplanets through microlensing (see Figure~\ref{fig:exo_lightcurve}).

The characteristic angular scale of a microlensing event is set by the angular Einstein radius ($\theta_{\rm E}$), and the event duration is described by the Einstein timescale,
\begin{equation}
t_{\rm E}=\frac{\theta_{\rm E}}{\mu_{\rm rel}},
\end{equation}
where $\mu_{\rm rel}$ is the relative lens-source proper motion.
The Einstein radius itself is given by
\begin{equation}
\theta_{\rm E}=\sqrt{\kappa M_{\rm L}\pi_{\rm rel}},
\end{equation}
where $M_{\rm L}$ is the lens mass, $\pi_{\rm rel}$ is the relative lens-source parallax, and $\kappa$ is a constant that converts the lens mass and relative parallax into the angular Einstein radius. These relations show that the light-curve timescale depends not only on the lens mass but also on the lens-source geometry and relative motion.

For a single lens and a single source (1L1S), the light curve is smooth and symmetric to first order and is commonly described by three basic parameters: the time of closest approach ($t_{0}$), the impact parameter ($u_{0}$), and the Einstein timescale ($t_{\rm E}$).
Planetary microlensing events require a binary-lens single-source (2L1S) model, in which the host star and its planet act as a binary lens system. In addition to the basic 1L1S parameters, the planetary system is characterized primarily by the mass ratio ($q = M_{\rm planet }/M_{\rm host~star}$), the projected separation ($s$) in units of $\theta_{\rm E}$, and the source trajectory angle relative to the binary axis ($\alpha$). In cases where finite-source effects are important, the normalized angular source radius ($\rho$) also becomes an important parameter. The mass ratio determines the strength of the perturbation, while $s$ and $\alpha$ define the caustic structure and the morphology of the anomaly. These parameters are fitted simultaneously with the 1L1S parameters to reproduce both the underlying magnification curve and the short-duration planetary deviation.

A major strength of microlensing is that the directly measured model parameters can be translated into physically meaningful quantities when $\theta_{\rm E}$ and the microlensing parallax ($\pi_{\rm E}$) are constrained. The microlensing parallax is defined as
\begin{equation}
\pi_{\rm E}=\frac{\pi_{\rm rel}}{\theta_{\rm E}}.
\end{equation}
The host and planet masses can then be written as
\begin{equation}
M_{\rm host}=\frac{\theta_{\rm E}}{\kappa\pi_{\rm E}}, 
\end{equation}
\begin{equation}
M_{\rm planet}=q M_{\rm host}.
\end{equation}
Similarly, the projected physical star-planet separation at the time of the event is
\begin{equation}
r_\perp = s D_{\rm L}\theta_{\rm E} = s r_{\rm E},
\end{equation} 
where $D_{\rm L}$ is the distance to the lens and $r_{\rm E}$ is the physical Einstein radius. 
Thus, the observables $q$ and $s$ can be transformed into the physical planet mass and projected separation once $\theta_{\rm E}$, $\pi_{\rm E}$, and $D_{\rm L}$ are determined or constrained.
In practice, $\theta_{\rm E}$ is often obtained from finite-source effects, while $\pi_{\rm E}$ can be measured either in long-duration events, where the Earth's orbital motion induces subtle deviations in the microlensing light curve, or more powerfully through simultaneous observations from two well-separated observatories, such as ground- and space-based facilities. 
More detailed discussions of the microlensing formalism, parameter interpretation, and recent observational developments can be found in \citet{Gaudi2012}, \citet{Mroz2023}, and \citet{Han2025}.

\begin{figure}
\centering
    \includegraphics[width=\linewidth]{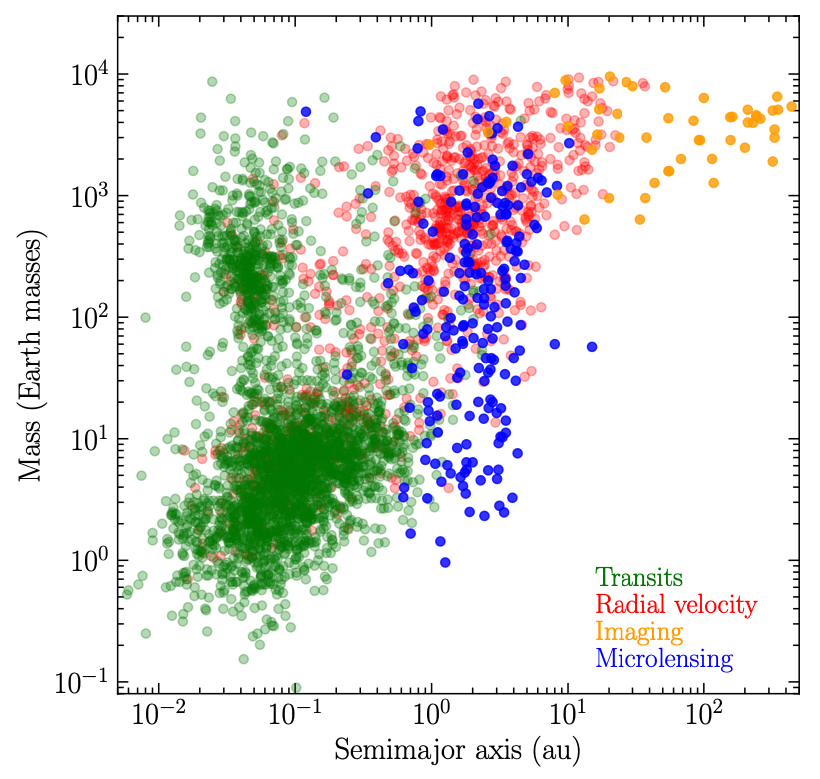}
    \caption{Distribution of known exoplanets in the planet-mass versus semimajor-axis plane, color-coded by detection method. Microlensing occupies a complementary discovery space relative to transit and radial-velocity surveys by probing planets at wider separations, particularly cold planets located near and beyond the snow line. This sensitivity makes microlensing a key method for studying planetary populations in the regime most relevant to giant-planet core formation. Adapted from \citet{Mroz2023}.}
\label{fig:snowline}
\end{figure}

Microlensing is especially powerful because it is most sensitive to planets whose projected separations are comparable to the Einstein radius. For typical Galactic bulge events, this physical scale is of order a few au, corresponding to the region around and beyond the snow line. 
This is a key regime for planet formation theory, because the snow line marks the region beyond which core accretion is expected to proceed more efficiently, promoting the growth of icy planetary cores and the seeds of giant planets \citep{Kennedy2008, Lambrechts2012}. Unlike transit and radial-velocity surveys, which are most sensitive to short-period planets, microlensing probes cold planets on orbits of order 1-10 au and beyond, including low-mass planets around faint or distant host stars (see Figure~\ref{fig:snowline}). This discovery space is highly complementary to those of transit and radial-velocity surveys and is particularly well matched to a space-based high-cadence survey such as KST.

\subsubsection{Current Ground-Based Microlensing Surveys} \label{sec:exo:ground}
Three major ground-based microlensing surveys are currently in operation:
\begin{itemize}
    \item \textbf{Optical Gravitational Lensing Experiment (OGLE):} a 1.3 m telescope in Chile with a $\sim$1.4 ${\rm deg}^{2}$ FoV \citep{Udalski2015}.
	\item \textbf{Microlensing Observations in Astrophysics (MOA):} a 1.8 m telescope in New Zealand with a $\sim$2.2 ${\rm deg}^{2}$ FoV \citep{Sumi2003, Sako2008}.
	\item \textbf{Korea Microlensing Telescope Network (KMTNet):} three 1.6 m telescopes in Chile, South Africa, and Australia, each with a $\sim$4 ${\rm deg}^{2}$ FoV \citep{Kim2016, Kim2018}.
\end{itemize}

These surveys have established microlensing as a major exoplanet discovery technique through high-cadence monitoring of dense stellar fields toward the Galactic bulge. OGLE and MOA played central roles in the early development of planetary microlensing discoveries, while KMTNet was specifically designed to provide near-continuous longitudinal coverage of bulge events without relying exclusively on follow-up observations. In particular, KMTNet monitors approximately 100 deg$^{2}$ of the bulge with a tiered cadence strategy ranging from 0.2 to 8 observations per hour, with the highest cadence applied to prime fields of greatest microlensing sensitivity \citep{Kim2018}. This capability has enabled KMTNet to achieve major successes in the detection and characterization of exoplanets \citep[e.g.,][]{Shvartzvald2017, Kim2021, Dong2026}.

Despite their success, all ground-based microlensing surveys are fundamentally limited by the atmosphere. Seeing broadens the PSF and increases source blending in the extremely crowded bulge fields, while sky background reduces sensitivity to faint source stars. In addition, interruptions of coverage by weather and the day--night cycle can compromise the detection or characterization of short-duration planetary anomalies, especially those produced by low-mass planets. These limitations are particularly important for planetary microlensing because the anomaly often lasts only a few hours or less. Incomplete temporal coverage can therefore significantly reduce detection efficiency and parameter recovery. These limitations define the key regime in which a space-based microlensing mission can deliver substantial gains.

\subsubsection{Advantages of the KST Mission} \label{sec:exo:KST}
The current generation of ground-based microlensing surveys has demonstrated the power of high-cadence monitoring in discovering planets toward the Galactic bulge. At the same time, these surveys also reveal the key performance gaps that a space-based mission can address. The proposed KST mission, with its balanced aperture-etendue design (see Section~\ref{sec:etendue}) and space-based observing environment, provides several decisive advantages for planetary microlensing.

\begin{itemize}
    \item \textbf{Stable, uninterrupted high-cadence monitoring:}
    KST can provide stable and continuous monitoring without interruptions from weather, seeing variations, or the day-night cycle. This will improve the detection and characterization of short-duration anomalies, which are often produced by low-mass planets and may last only a few hours or less. 

    \item \textbf{Deeper limiting magnitude and improved access to dwarf source stars:}
    The absence of atmospheric background and the much smaller PSF in space enable significantly deeper photometry in crowded bulge fields than is possible from the ground. Even a gain of a few magnitudes in limiting depth would substantially increase the number of accessible main-sequence source stars, whose smaller angular radii reduce finite-source suppression of planetary signals. This is expected to improve sensitivity to low-mass planets and increase the number of detectable microlensing events.

    \item \textbf{Reduced blending in extremely crowded bulge fields:}
    Space-based imaging greatly reduces source confusion relative to ground-based observations, leading to cleaner light curves and more robust anomaly detection. In favorable cases, this also enables more precise measurements of finite-source parameters, thereby improving constraints on the angular Einstein radius and planetary properties.

    \item \textbf{Robust microlensing parallax constraints:}
    Combined ground- and space-based observations can provide robust constraints on the microlensing parallax (see Section~\ref{sec:exo:microlensing}), which is critical for estimating the physical properties of planetary systems, including host mass, planet mass, and projected separation.

    \item \textbf{Unique optical and NIR survey synergy:}
    A distinctive advantage of KST is its capability for simultaneous optical and NIR observations. This would enable an optical microlensing survey of the bulge to be carried out in parallel with a NIR spectrophotometric survey using the LVF. Such simultaneous observations would enhance the overall scientific return of the mission by combining time-domain exoplanet discovery with NIR characterization of stellar populations in the inner Galaxy, even during periods when no strong microlensing anomaly is detected (see Section~\ref{sec:MW:bulge}).
\end{itemize}

Taken together, these advantages make KST particularly well suited to expanding the microlensing discovery space toward fainter source stars, lower-mass planets, and shorter-timescale anomalies than are currently accessible from ground-based surveys alone.
The observational strategy required to realize these advantages, including field selection, cadence, and seasonal mission allocation, is described in Section~\ref{sec:survey:microlensing}.

\subsubsection{Synergy with Other Surveys} \label{sec:exo:synergy}
KST will operate in strong synergy with ongoing and future exoplanet surveys. Its scientific return will be substantially enhanced through coordinated observations and complementary data sets from ground-based microlensing programs, Roman, and Gaia.

\begin{itemize}
    \item \textbf{Ground-based microlensing surveys:}
    Joint observations with ground-based microlensing programs such as KMTNet will provide a major advantage for the physical characterization of planetary events. Simultaneous ground- and space-based observations naturally enable measurements of the microlensing parallax through the long projected baseline between observatories. When combined with constraints on the angular Einstein radius, this allows direct estimates of lens masses and distances for a substantial fraction of detected systems, transforming microlensing events from statistical detections into physically characterized planetary systems. This synergy is especially important for short-duration and low-mass planetary events, for which concurrent ground-based data improve parallax sensitivity and increase confidence in event interpretation.

    \item \textbf{Roman Space Telescope:}
    KST is also complementary to NASA's Nancy Grace Roman Space Telescope. Roman's Galactic Bulge Time-Domain Survey is designed to measure exoplanet demographics through microlensing and is expected to discover more than a thousand bound microlensing planets over its planned bulge survey fields \citep{Penny2019, Johnson2020}. In particular, while Roman will monitor six fields over a total area of about 1.7 deg$^{2}$, the KST microlensing survey is designed to cover about 1 deg$^{2}$. The KST fields can be selected to complement the Roman footprint, providing additional coverage of the Galactic bulge and enabling coordinated observations of overlapping or adjacent fields. In this broader context, KST can strengthen the overall space-based microlensing effort through complementary wavelength coverage, observing strategy, survey flexibility, and additional space-ground parallax baselines.

    \item \textbf{Gaia astrometry and targeted follow-up:}
    Synergy with Gaia astrometry will further improve the characterization of selected lens and source stars through proper-motion and parallax information. Future Gaia releases are expected to expand and refine the available astrometric information, improving the identification and characterization of selected systems. In favorable cases, follow-up spectroscopy or radial-velocity observations of bright lens stars may provide additional constraints on host-star properties. Together, these complementary data sets will strengthen the physical characterization of planetary systems, including planet masses, lens distances, and projected separations, thereby maximizing the scientific return of the KST microlensing program.
\end{itemize}

\subsubsection{Scientific Impact} \label{sec:exo:science}
The KST microlensing program will probe planetary populations that remain poorly constrained by other detection techniques. In particular, it will:

\begin{itemize}
    \item \textbf{Constrain the occurrence rate of cold planets near and beyond the snow line.}
    Microlensing is uniquely sensitive to projected separations of order a few au, the snow-line regime introduced above. Measuring the planet frequency there provides a direct test of core-accretion predictions for the growth of icy cores and giant-planet seeds \citep[e.g.,][]{Gould2010, Winn2015} and anchors the cold end of the occurrence-rate distribution.

    \item \textbf{Extend sensitivity to Earth-mass planets in wide orbits.}
    The combination of space-based cadence, deeper limiting magnitude, and reduced blending will improve sensitivity to low-mass planets, whose short-duration anomalies are often difficult to detect from the ground. This capability is essential for probing Earth-mass and super-Earth planets in the cold outer regions of planetary systems \citep[e.g.,][]{Beaulieu2006, Han2021}. Such planets provide crucial information on the low-mass end of the planet population, where current observational constraints remain limited but where the initial assembly of planetary cores is expected to be especially important.

    \item \textbf{Probe Solar-system analog architectures.}
    Microlensing is one of the few techniques capable of detecting multiple planets in wide-orbit configurations around distant stars. It therefore provides a unique opportunity to study planetary architectures analogous to the outer Solar System and to assess how common such systems are in the Galaxy \citep[e.g.,][]{Gaudi2008, Howe2025}.

    \item \textbf{Detect and characterize free-floating planets.}
    Microlensing remains the only established technique for detecting old, low-mass planetary-mass objects that are not detectably bound to host stars (direct imaging finds young, self-luminous ones) \citep[e.g.,][]{Mroz2020, Dong2026}. The abundance and mass distribution of free-floating planets provide a unique probe of planetary dynamics and planet formation, including planet-planet scattering, ejection, and the possible formation of isolated planetary-mass objects.
\end{itemize}

Together, these capabilities will allow KST to bridge the parameter space between close-in planets discovered by transit surveys and wide-separation planets accessible to direct imaging. The KST microlensing survey will therefore provide critical insight into planet formation efficiency, migration, and the dynamical evolution of planetary systems throughout the Galaxy.

\begin{figure}[t]
\centering
\includegraphics[width=\linewidth]{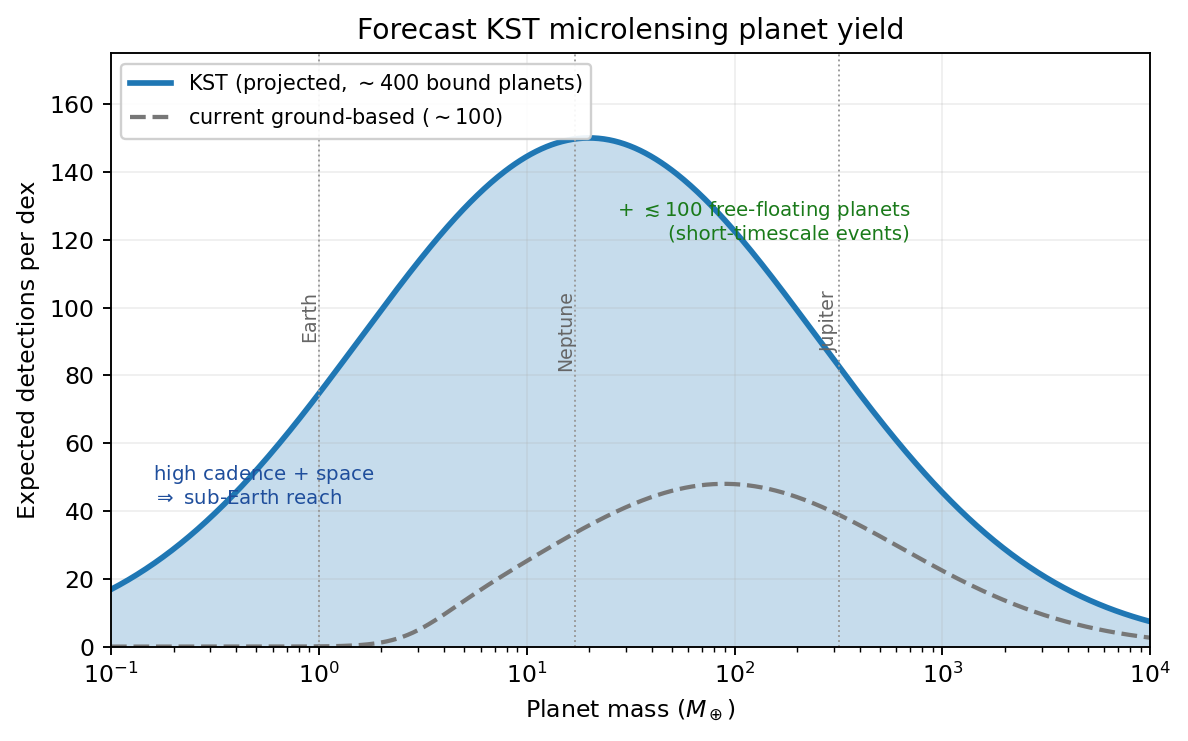}
\caption{Projected KST microlensing planet yield as a function of planet mass (illustrative; per logarithmic mass interval), scaled from the demonstrated yields of ground-based surveys and space-based bulge-survey predictions. KST's high cadence and stable space-based photometry extend sensitivity down to approximately an Earth mass, while deeper observations and reduced blending in crowded bulge fields raise the overall detection efficiency above current ground-based surveys; an additional $\lesssim100$ free-floating planets are expected from short-timescale events.}
\label{fig:exo_yield}
\end{figure}

\paragraph{Projected yield}
Scaling the demonstrated planet-detection rates of the ground-based surveys (OGLE, MOA, and KMTNet) and the predicted yields of space-based bulge surveys by KST's monitored star-hours, 15-minute cadence, and deep, stable space-based photometry, the two-season microlensing campaign is projected to detect of order several hundred bound planets, with the sensitivity extending down to approximately an Earth mass owing to the high cadence and stable space-based photometry, together with $\lesssim100$ free-floating planets identified through their short-timescale events (Figure~\ref{fig:exo_yield}). For comparison, the \textit{Roman} Galactic Bulge Time-Domain Survey, covering $\sim$1.7~deg$^2$ (six fields) at $\sim$12-minute cadence, is forecast to yield $\sim$1000--1500 bound planets \citep{Penny2019} and $\sim$250 free-floating planets \citep{Johnson2020}; KST's smaller $\sim$1~deg$^2$ microlensing footprint therefore targets a proportionally scaled, complementary yield. This would roughly double the current cold-planet inventory and, together with Roman, populate the Earth-to-super-Earth regime at wide orbital separations that is inaccessible to transit and radial-velocity methods. The yield distribution shown is illustrative and scaled to existing surveys; the basis for all forecasts is summarized in Section~\ref{sec:forecast}.

\subsubsection{Transit Survey Capability} \label{sec:exo:transit}
Although the primary exoplanet science driver of KST is gravitational microlensing, its stable high-cadence time-series photometry can also enable a complementary transit survey. Continuous monitoring of Galactic bulge fields will naturally accumulate dense light curves for a large number of foreground disk stars, providing sensitivity to short-period transiting planets.

With a 1.4 m aperture and a space-based observing environment free from atmospheric systematics, KST is expected to deliver stable photometric monitoring well suited to the detection of short-period planetary transits around moderately bright stars. The adopted cadence of 15 minutes is well matched to the durations of many short-period transits, allowing robust identification and phase-folding of repeated signals over the course of the observing campaign. Although the relatively modest field of view will not permit the statistical yield of dedicated transit missions such as Kepler, KST can probe a complementary stellar population toward the inner Galaxy. In particular, it can help constrain the occurrence of short-period planets around foreground disk stars along the bulge line of sight, thereby extending transit studies into Galactic environments that remain comparatively less explored.

The coexistence of microlensing and transit sensitivity within the same observing program provides an additional opportunity to investigate planetary populations across a broad range of orbital separations within a unified survey framework. This complementary capability enhances the overall exoplanet science return of KST without compromising its primary microlensing objectives.
\subsection{Milky Way Astronomy: Galactic Archaeology in the Near-Infrared}
\label{sec:milky_way}
KST combines broadband optical imaging with NIR LVF spectrophotometry. 
Rather than an all-sky survey, KST builds its Galactic dataset from two components: the Galactic Bulge, where extreme dust extinction and source crowding defeat optical surveys, observed through the high-cadence fields of the microlensing program (Section~\ref{sec:survey:microlensing}); and the high-latitude $2{,}000~\text{deg}^2$ Cosmology Wide Survey, which simultaneously contributes the stellar halo, cold streams, and substellar populations.
The combination of simultaneous visible/NIR observations, the superb spatial resolution ($\sim0.15''$ in the optical and $\sim0.2''$ in the NIR) afforded by the 1.4m aperture, and the low-resolution (R$\sim$100) NIR spectra provided by the LVF system will yield a distinctive dataset for Galactic archaeology. The expected scientific impacts across subfields are detailed below.

\subsubsection{Mapping the Galactic Bulge} \label{sec:MW:bulge}
The central Bulge, Bar, and inner regions of the Thin Disk are obscured by thick interstellar dust, creating a ``Zone of Avoidance'' for optical telescopes. KST will pierce this dust veil through the 1.1–1.8$\mu$m NIR window, and its sharp NIR PSF ($\sim0.2''$) will overcome the severe source confusion that plagued even the 0.4~m Wide-field Infrared Survey Explorer (WISE) observations in the Galactic center. By resolving $\sim$$10^{5}$ Red Clump (RC) giants in the $\sim$1~deg$^2$ bulge fields and utilizing them as standard candles \citep{Girardi2016}, KST will map the three-dimensional structure of the Bulge, including its X-shaped morphology \citep{Wegg2013}, at high spatial resolution. This dataset requires no dedicated survey time. Because the Galactic bulge is the primary field for microlensing planet searches (Section~\ref{sec:exo}), the repeated and deep monitoring of that program will naturally produce not only microlensing light curves but also homogeneous stellar photometry and low-resolution NIR spectral information for a vast number of stars in the inner Galaxy, a rich legacy dataset for Milky Way astronomy.

\subsubsubsection{The Bulge as the Fossil Record of the Milky Way} \label{sec:MW:bulge:background}
The Galactic bulge contains a large fraction of the stellar mass in the inner Galaxy and preserves key information about the earliest phases of Galactic formation. Because stars in the bulge are generally old and densely concentrated, they provide a fossil record of the chemical enrichment, dynamical evolution, and structural assembly of the Milky Way. Understanding the bulge is therefore essential for understanding the formation and evolution of the Milky Way. The Milky Way bulge also serves as a nearby laboratory for studying bulge-dominated stellar systems, such as early-type galaxies. Because it can be studied star by star, the Galactic bulge allows us to connect individual stellar properties to the global properties of a bulge-dominated system. It therefore provides a direct bridge between resolved stellar population studies in the Milky Way and the broader question of bulge formation in galaxies.

Despite its importance, the Galactic bulge remains significantly less well understood than the Galactic disk and halo, owing to the extinction and crowding described above. Previous NIR surveys have made major progress in overcoming these limitations. Large-scale photometric surveys such as the Two Micron All Sky Survey \citep[2MASS;][]{Skrutskie2006} and the VISTA Variables in the Via Lactea survey \citep[VVV;][]{Minniti2010} have provided deep NIR photometry of the bulge and have revealed its complex structure, stellar populations, and many previously unknown stellar cluster candidates. Spectroscopic surveys such as the Apache Point Observatory Galactic Evolution Experiment \citep[APOGEE;][]{Majewski2017} have further provided detailed chemical and kinematic information for selected bulge stars based on high-resolution NIR spectra. However, an important gap remains between these two approaches. Photometric surveys provide wide spatial coverage but limited spectral information, while high-resolution spectroscopic surveys provide detailed abundances and velocities but only for relatively bright and sparse target samples. A wide-field, space-based, NIR spectrophotometric survey of the bulge would fill this gap, and KST is well positioned to provide such a homogeneous dataset for characterizing individual stellar properties and mapping stellar populations across the inner Galaxy.

\subsubsubsection{KST as an Optical-to-NIR Spectrophotometric Bulge Survey} \label{sec:MW:bulge:survey}
KST will provide a new type of Galactic bulge survey by combining wide-field optical imaging with NIR LVF spectrophotometry in a single space-based mission. The two channels together give the broad wavelength coverage needed for stellar spectral energy distributions, the LVF adds low-resolution spectra for individual stars, and the sharp, stable space-based PSF permits reliable photometry and spectrophotometry in the densest fields, including the low-latitude bulge, the inner bar, and the immediate vicinity of the Galactic center. The resulting homogeneous dataset will bridge the gap between wide-field NIR photometric surveys and sparse high-resolution spectroscopic surveys, establishing a new observational framework for the science topics described below.

\subsubsubsection{Science Cases} \label{sec:MW:bulge:science}

\begin{itemize}
    \item \textbf{Stellar populations and bulge structure:}
    Large-scale optical-to-NIR spectrophotometric mapping of bulge stars will enable systematic stellar population studies across the inner Galaxy. In particular, population-sensitive diagnostics derived from the NIR LVF data will allow the construction of spatially resolved stellar population maps of the Galactic bulge. Such maps are critical for understanding the multiple stellar populations that compose the bulge. Previous studies suggest that the bulge contains multiple components, including a metal-poor spheroidal population and a metal-rich population associated with the boxy or X-shaped structure \citep[e.g.,][]{Zoccali2017, Lim2021, Hong2024}. Figure~\ref{fig:MW_bulge} illustrates how optical-to-NIR photometry can separate these bulge populations and reveal their different structures. The metal-poor and metal-rich populations may reflect different formation channels, with the former often associated with a classical bulge component and the latter linked to a pseudo-bulge component formed through subsequent secular evolution. With low-resolution spectrophotometry for a vast number of bulge stars, KST will enable improved separation of major bulge populations and provide a homogeneous dataset for investigating their spatial distributions, stellar properties, and formation history.

    \begin{figure}
    \centering
    \includegraphics[width=\linewidth]{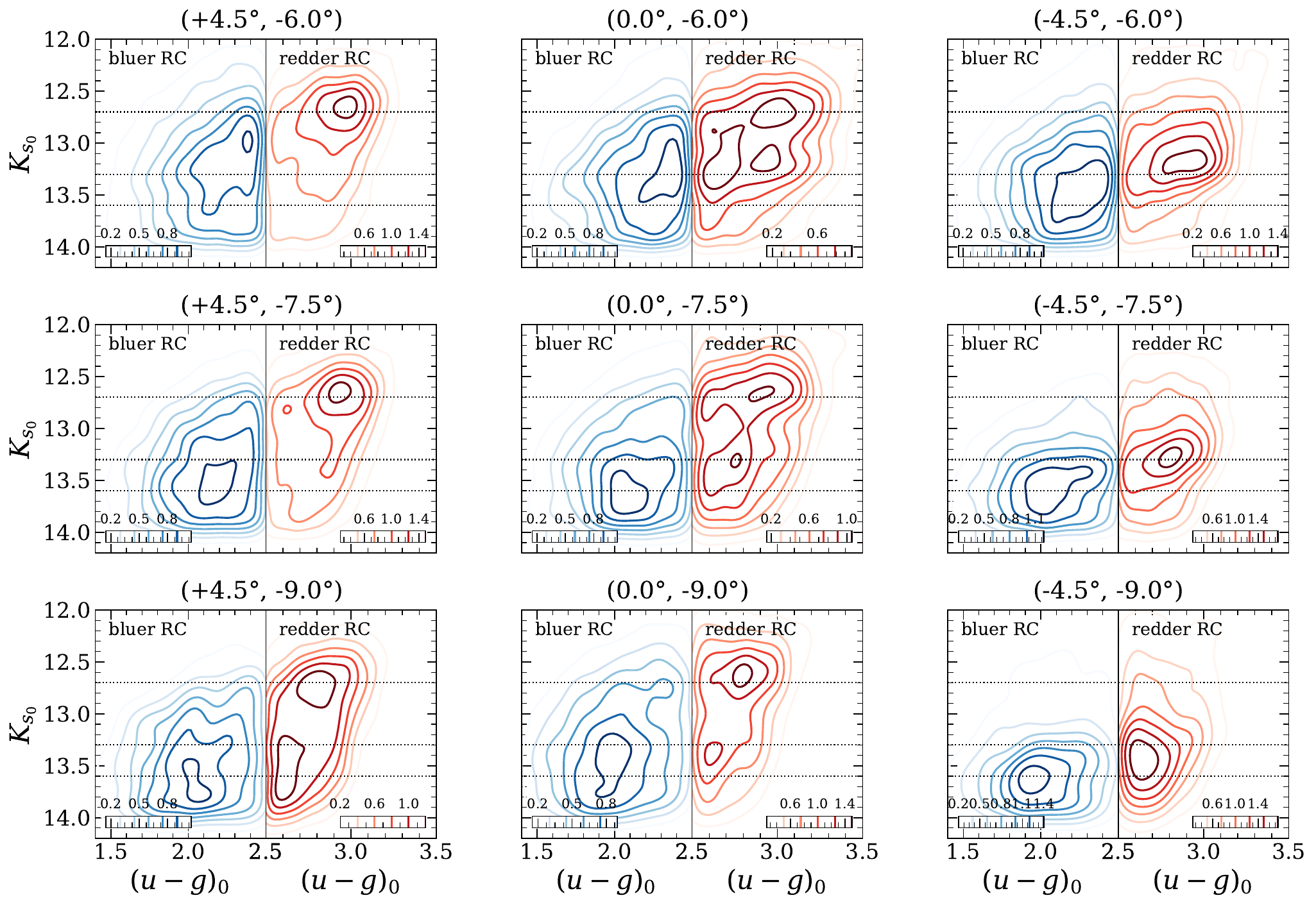}
    \caption{Density contours of RC stars in the $(K_{\rm s}, u-g)_{0}$ color--magnitude diagram (CMD), illustrating the population-dependent structure of the Galactic bulge with ground-based photometry in bands ($u$, $K_{\rm s}$) that KST does not carry. The bluer RC population, associated with a more metal-poor component, shows a single clump consistent with a relatively spheroidal distribution, while the redder RC population, associated with a more metal-rich component, shows a double RC feature that traces the X-shaped bulge. The separation rests on a wide optical-to-NIR wavelength baseline. KST will form analogous population-sensitive diagnostics from its own data, combining $griz$ colors with the $\sim$48-channel LVF spectrophotometry over $1.1$--$1.8~\mu$m, and extend such bulge population and structure studies to more crowded and highly reddened regions of the inner Milky Way. Adapted from \citet{Lim2021}.}
    \label{fig:MW_bulge}
    \end{figure}
    
    \item \textbf{Discovery and characterization of new stellar clusters:}
    Recent infrared surveys such as 2MASS, WISE, and VVV have revealed a large number of previously unknown stellar cluster candidates in the Galactic bulge region. Although about 150 globular clusters are currently known in the Milky Way, more than one hundred additional cluster candidates have been identified in recent infrared surveys, particularly toward highly reddened regions near the Galactic center \citep[e.g.,][]{Borissova2018, Bica2024}. Many of these candidates remain poorly characterized, and it is still unclear whether they represent genuine globular clusters, open clusters, associations, or chance stellar overdensities. Determining the true nature of these systems requires information on their stellar populations, metallicities and kinematics. KST will provide valuable spectrophotometric information for stars in and around these cluster candidates. Its high angular resolution will be particularly important for resolving crowded cluster cores and reducing source confusion in the inner Galaxy. The NIR LVF data will provide spectroscopic information and stellar population diagnostics for cluster candidates, helping to distinguish real stellar systems from field-star overdensities. These data will also provide a homogeneous basis for selecting promising targets for follow-up high-resolution spectroscopy. Beyond individual cluster studies, the confirmation and characterization of new star clusters in the bulge will have broad implications for understanding the formation and assembly history of the Milky Way \citep[e.g.,][]{Romero-Colmenares2021, Lim2024}.

    \item \textbf{Nuclear star cluster and nuclear stellar disk:}
    The Galactic center hosts a distinctive nuclear environment composed of the nuclear star cluster (NSC) and the nuclear stellar disk (NSD). The NSC is one of the most massive and densest star clusters in the Milky Way and surrounds the central supermassive black hole, Sagittarius A*. Observations have revealed multiple stellar populations and recurrent star formation episodes in the NSC, indicating a complex assembly history \citep[e.g.,][]{Nogueras-Lara2020}. The NSD, extending over several hundred parsecs, represents a flattened stellar component that is closely linked to the central molecular zone and may have formed through recurrent gas inflow and star formation in the Galactic center \citep[e.g.,][]{Schultheis2021}. The formation and evolution of these nuclear components are closely connected to star formation, gas inflow, and dynamical processes in the central regions of galaxies. KST will provide a new optical-to-NIR dataset across the nuclear region, enabling systematic studies of stellar population variations in and around the NSC and NSD. These measurements will help clarify how the nuclear components formed and how they are linked to the larger-scale Galactic bulge.

    \item \textbf{Stellar variability:}
    The repeated monitoring of the Galactic bulge fields for the microlensing survey will naturally provide a valuable time-domain dataset for Milky Way astronomy. These observations will enable the identification and characterization of a wide variety of variable stars, including RR Lyrae stars, Cepheids, Mira variables, and eclipsing binaries (see also Section~\ref{sec:MW:variables}). Such variable stars are powerful tracers of Galactic structure, distance, and stellar evolution in the inner Milky Way. In particular, RR Lyrae stars trace the old, metal-poor component of the bulge, while long-period variables such as Mira stars provide complementary probes of evolved, relatively metal-rich stellar populations \citep[e.g.,][]{Kunder2016, Catchpole2016}.

    \item \textbf{Additional science opportunities:}
    The same dataset may also reveal rare or unusual objects in the Galactic center region, including candidate high-velocity stars, peculiar variables, and transient sources. When combined with Gaia astrometry and follow-up spectroscopy, these objects can provide additional constraints on the dynamical environment of the Galactic center. The KST bulge survey will also provide a homogeneous basis for identifying unusual stellar populations and candidate dynamical substructures in the inner Galaxy.

\end{itemize}

\subsubsection{Galactic Rotation and Dark Matter}
The Galactic rotation curve is a direct tracer of the dark matter distribution \citep{BlandHawthorn2016, Eilers2019}. By utilizing the LVF system to classify the broad spectral types of millions of stars, KST will derive accurate absolute magnitudes and distances. This enables the construction of 3D spatial maps using specific tracer stars (e.g., K-type giants) deep within the Bulge or the outer disk. When combined with the high-precision proper motion data from the Gaia satellite (see Section~\ref{sec:MW:Gaia}), this dataset will dramatically enhance the resolution of gravitational potential models used to map the density distribution of the dark matter halo, even in regions where radial velocity data is sparse.

\subsubsection{Baryonic Content of the Galaxy: M Dwarfs and Brown Dwarfs}
A significant fraction of the Galaxy's baryonic mass is hidden in cold, low-mass objects such as M dwarfs and substellar brown dwarfs (L, T, Y types). While invisible in the optical regime, their spectral energy distributions peak in the 1.1–1.8 $\mu$m band targeted by KST. Furthermore, this wavelength range harbors strong molecular absorption bands (such as H$_2$O and CH$_4$) that are critical diagnostic features for classifying brown dwarfs \citep{Kirkpatrick2005}. With full $R\sim100$ spectra to $\approx$21~AB per channel and synthetic medium-band photometry $\approx$1~mag fainter in a Wide passage (Table~\ref{tab:lvf_modes}), the LVF data can identify and classify of order $10^{3}$--$10^{4}$ brown dwarfs over the $2{,}000$~deg$^2$ footprint, including nearby isolated planetary-mass objects, directly from their spectral energy distributions and without ground-based spectroscopic follow-up. This spectroscopic census is complementary to the dynamically-selected free-floating-planet sample from the microlensing survey (Section~\ref{sec:exo}), and together they constrain the lower end of the Galactic Initial Mass Function (IMF) \citep{Chabrier2003}.

\subsubsection{Chemo-dynamical Map of the Galaxy}
The evolutionary history of the Galaxy is reconstructed by linking the ages, chemical abundances, and kinematics of its stellar populations. Although KST’s LVF provides low resolution (R$\sim$100), it is sensitive to the broad molecular absorption features within its $1.1$--$1.8~\mu$m window (CN near $1.1~\mu$m, the second-overtone CO bands near $1.6~\mu$m, and H$_2$O) prevalent in cool giants. These features serve as robust proxies for stellar metallicity (i.e. [Fe/H]). Procuring metallicity estimates for of order $10^{7}$ stars brighter than the per-channel limits across the surveyed Bulge and high-latitude fields will be instrumental in mapping the ``Chemo-dynamical'' history of the Galaxy \citep{Freeman2002}, allowing us to trace phenomena such as radial migration \citep{Schonrich2009} and dynamical heating.

\subsubsection{Star Clusters}
Numerous open and globular clusters lie hidden along the Galactic equatorial plane, highly obscured by dust. Even known clusters are notoriously difficult to study from the ground due to extreme stellar crowding in their cores. Leveraging its $\sim 0.2''$ NIR spatial resolution, KST will resolve the dense cores of these clusters into individual stars, opening their multiple stellar populations to detailed study \citep{BastianLardo2018}. Deep photometry will extend their CMDs well past the Main Sequence Turn-Off (MSTO) and, for nearby clusters, toward the low-mass end of the main sequence. This will not only yield highly accurate cluster ages but also allow for the direct observation of dynamical mass segregation, providing vital clues about the disruption processes of clusters in the Galactic disk. Candidate clusters toward the bulge, and their spectrophotometric confirmation, are discussed in Section~\ref{sec:MW:bulge}.

\subsubsection{Stellar Variabilities}
\label{sec:MW:variables}
Time-domain coverage comes from the microlensing bulge fields (two 4-month seasons at 15-minute cadence) and the high-cadence Deep fields; the single-pass Wide Survey provides no light curves. Because NIR wavelengths suffer only a fraction of the dust extinction seen in the visible, the bulge time series are well suited to Cepheids and Mira variables embedded in the inner Galaxy \citep{Whitelock2008}. Contemporaneous optical and NIR time series will allow us to measure the wavelength dependence of pulsation amplitudes, test stellar atmosphere models, and calibrate period--luminosity relations within the Milky Way.

\subsubsection{Asteroseismology}
As demonstrated by the Kepler mission \citep{Borucki2010}, extreme photometric precision enables the detection of minute luminosity variations caused by internal oscillations in red giants \citep{ChaplinMiglio2013}. If KST achieves sufficient observational cadence in specific fields, its visible CMOS detectors can be utilized for asteroseismology of luminous red giants, constraining stellar masses, radii, and ages, subject to a feasibility assessment of the achievable photometric precision. The LVF will concurrently provide effective temperatures ($T_{\rm eff}$), metallicities, and a coarse surface-gravity classification for the same stars, which can help reduce the uncertainties in asteroseismic scaling relations and the ages derived from them.

\subsubsection{Stellar Streams and Magellanic Clouds}
The Large and Small Magellanic Clouds (LMC/SMC) lie outside the baseline survey footprint but are natural targets for ToO-reserve or extended-mission pointings. As these satellites interact with the Milky Way's massive gravitational potential, they shed long trails of gas and stars (e.g., the Magellanic Stream and Leading Arm) \citep{DOnghiaFox2016}. While the gaseous components have been mapped by radio telescopes, the accompanying, extremely faint stellar streams remain largely undetected. Such pointings would allow KST to search for faint main-sequence stars associated with these streams in the vicinity of the Magellanic Clouds, which would provide direct evidence regarding the accretion and destruction histories of satellite galaxies within the Local Group. Gaps and perturbations in these cold streams also serve as a probe of dark-matter substructure \citep{Bonaca2019}, contributing to the Local-Group subhalo-mass-function measurement of Pillar~II (Section~\ref{sec:dm_dwarfs}).

\subsubsection{Synergies with the Gaia Astrometric Legacy} \label{sec:MW:Gaia}
For stars observed by both missions, KST photometry and spectrophotometry can be combined with Gaia astrometry \citep{GaiaMission2016}. Over its decade-long operation, Gaia has revolutionized astronomy by providing ultra-precise positions, parallaxes, and proper motions for nearly 2 billion stars. However, Gaia is fundamentally limited by its reliance on optical wavelengths, preventing it from probing the dust-obscured Galactic plane, and its high-resolution spectroscopy is restricted to relatively bright stars.

The 1.4m KST will provide the ``missing piece'' in the Gaia mission to complement and dramatically extend Gaia's capabilities:
\begin{itemize}
   \item \textbf{Completing the 6D Map Beyond the Dust Veil:} KST will distinctly resolve the Bulge and Thin Disk regions hidden from Gaia using the 1.1–1.8 $\mu$m NIR band. For overlapping targets, KST stellar-population estimates and Gaia astrometry will support studies of Galactic structure and dynamics.

   \item \textbf{Unprecedented Proper Motion Precision via Baseline Extension:} If KST observes the same optical sky ($g, r, i, z$) approximately 10 to 20 years after Gaia's initial epoch, the resulting ``long time baseline'' will be a valuable asset for Milky Way astronomy. By utilizing the epoch difference between the two missions, the proper motions of stars can be derived with extreme precision, even if KST's single-epoch astrometric precision is inferior to Gaia's. This is absolutely essential for tracking the orbits of ultra-faint halo stars and globular clusters to trace the formation history of the Milky Way.
\end{itemize}

\subsection{Multi-Messenger Astrophysics}
\label{sec:mma}
The start of gravitational-wave (GW) astronomy in 2015 made it possible to probe the Universe in an unprecedented way. Yet, most of the detected GW events to date have been binary black hole (BBH) mergers that are not expected to be accompanied by electromagnetic (EM) signals. The potential of GW astronomy can broaden significantly with the detection of events with EM signals that  originate from binary neutron star (BNS) mergers in the form of kilonovae (KNe; singular KN) and gamma-ray bursts (GRBs), the domain of multi-messenger astronomy (MMA).  The first BNS merger GW event in 2017, GW170817, demonstrated the transformative value of a secure EM counterpart. Its optical/NIR counterpart, AT2017gfo, enabled the localization of the merger in NGC 4993, established the connection between BNS mergers, kilonovae, and short gamma-ray bursts, and provided direct evidence that neutron-star mergers synthesize r-process material \citep[e.g.,][]{Abbott2017GW170817, Pian2017, Kasen2017, Watson2019}. Furthermore, the GW170817 event showed that MMA can be a powerful tool for cosmological studies.

The number of GW detections will grow rapidly in the 2030s. The planned O5 run of the International Gravitational-Wave Observatory Network (IGWN; formerly LIGO--Virgo--KAGRA) is expected to yield a sample of BNS mergers substantially larger than that of previous runs. The subsequent A\# network and the next-generation Einstein Telescope (ET) and Cosmic Explorer (CE) observatories will extend BNS merger detections from tens--hundreds per year to $10^{4}$--$10^{5}$ per year.

However, kilonovae are intrinsically faint and evolve on timescales of hours to days. For a representative event at 300~Mpc, the optical counterpart is expected to reach approximately $r\simeq22$--23~mag and can fade beyond $r\simeq25$~mag within several days. At $z\sim0.5$, only the brightest and/or most favorably viewed events are expected to be detectable in the optical near peak. At $z\gtrsim0.8$, typical KNe are fainter than $r\simeq26$ AB mag and thus beyond the 1-hour planning depths adopted below. 
In NIR, the situation is somewhat better, with the ejecta component dominating the NIR emission for an extended period ($\sim$ week) but they are still faint. Therefore, in such a revolutionary era of MMA, sensitive facilities to detect and follow up GW optical counterparts will be essential.

KST, with its wide-field, high resolution, highly sensitive, simultaneous optical and NIR observational capability, can become a key instrument to identify and characterize electromagnetic (EM) counterparts of gravitational wave (GW) events for MMA in the 2030s.
In the optical channel, we adopt conservative 1-hour, 5$\sigma$ planning depths of $g=26.1$, $r=25.7$, $i=25.3$, and $z=24.9$ AB mag, which budget for host-galaxy subtraction and non-optimal extraction. The corresponding point-source limits are $\sim$1.1--1.5 mag deeper (Table~\ref{tab:kst_specs}). These depths give robust detection of GW170817-like KNe at $\sim300$~Mpc around peak and permit multi-epoch measurements of brighter events and favorable viewing angles at substantially larger distances. Diffraction-limited $griz$ imaging is also important in crowded host-galaxy regions, where accurate image subtraction and separation from host-galaxy light are essential for identifying a rapidly fading counterpart.
 The wide-field imaging data will also be useful as reference frames for uncovering new transients in the GW localization region. The dual-channel VIS and NIR architecture is crucial for capturing the rapid spectral evolution of kilonovae. NIR spectroscopic capability provides insights into heavy element nucleosynthesis and neutron star physics.

Table~\ref{tab:mma_kn_yield} gives planning-level yields for KST. These are counterpart yields, not raw GW-trigger rates. They assume that KST can tile the selected localization region on a timescale short compared with the optical evolution, that approximately one half of triggers are inaccessible because of Sun-angle constraints, and that the event is observed near peak. 

\begin{table*}
\centering
\caption{Illustrative annual KST yield of optically identified BNS kilonovae. CR90 denotes the area of the 90\% credible localization region.}
\label{tab:mma_kn_yield}
\begin{tabular}{p{2cm}p{3.5cm}p{2cm}p{7cm}}
\hline
\textbf{GW network / era} & \textbf{Relevant GW sample} & \textbf{KST optical KN yield} & \textbf{Basis} \\
\hline
O5 & Follow-up-accessible BNS alerts & $2$--$10$ yr$^{-1}$ &
Current O5 optical-follow-up simulations predict $2$--$19$ detectable BNS KNe yr$^{-1}$ for a deep, targeted program; the KST range adopts a conservative tiling and visibility fraction \citep{Kunnumkai2025}. \\ \addlinespace
A\# & $\mathrm{CR90}<10~\mathrm{deg}^2$, $D\lesssim300$ Mpc & $\sim5$--$8$ yr$^{-1}$ &
A forecast of $\sim20$ BNS events in 1.5 yr with $\mathrm{CR90}<10~\mathrm{deg}^2$, together with distance and Sun-angle cuts, gives $\sim7$ accessible nearby triggers yr$^{-1}$; most should be within KST optical reach near peak \citep{Andreoni2024Roman}. \\ \addlinespace
ET alone & $\mathrm{CR90}<10~\mathrm{deg}^2$ & $\sim1$--$5$ yr$^{-1}$ &
Only a few to $\sim15$ BNS events yr$^{-1}$ are expected to be localized this well with ET alone; the optical sample is dominated by the nearby subset \citep{Colombo2025}. \\ \addlinespace
ET+2CE & $\mathrm{CR90}<10~\mathrm{deg}^2$ & $\sim15$--$50$ yr$^{-1}$ &
Starting from $\sim3000$ well-localized BNS events yr$^{-1}$, this estimate assumes that 3--10\% are at $z\lesssim0.5$, and applies 50\% visibility, 50\% schedulability/tiling, and 60\% optical-recovery factors \citep{Colombo2025}. \\
\hline
\end{tabular}
\end{table*}

Here, we outline several key science cases for KST in the area of MMA.

\subsubsection{Identification of faint, distant KNe}

As outlined above, the O5 and A\# runs (late 2020s and early 2030s) and the next-generation ET and CE observatories will increase BNS-merger detections from tens--hundreds to $10^{4}$--$10^{5}$ events per year, with still larger numbers of binary black holes. However, most next-generation BNS events will be at cosmological distances, where their EM counterparts will be too faint for routine optical identification. A sensitive, rapid-response facility is therefore needed to exploit the well-localized subset for which kilonovae can be detected and characterized.

Although distant GW events will often have large localization regions, a scientifically valuable subset of BNS and neutron-star--black-hole (NSBH) mergers will have 90\% credible localization areas of $\mathrm{CR90}<10~\mathrm{deg}^{2}$. These events are the principal targets for KST. The KST optical channel provides diffraction-limited $griz$ imaging at the 1-hour planning depths stated above (Table~\ref{tab:kst_specs}). The high angular resolution is particularly valuable for image subtraction and for recovering a faint, rapidly evolving counterpart against structured host-galaxy light.

The Vera C. Rubin Observatory will be a powerful discovery facility for wide GW localization regions, reaching approximately $r\simeq24$ mag in short exposures \citep{Andreoni2024}. Nevertheless, Rubin follow-up is limited by ground-based sky visibility, weather, cadence, and competition with its primary survey program. KST provides complementary, stable space-based observations with rapid target-of-opportunity (ToO) response. It can observe fields inaccessible from Chile or during unfavorable ground-based conditions, and can obtain the contemporaneous optical and NIR information required to distinguish KNe from supernovae and other transients. Other wide-field space facilities can also contribute, but their ToO scheduling and the number of rapidly executable follow-up programs may be limited. KST's short-response ToO capability is thus especially valuable for KNe, whose optical emission fades on timescales of days.

At $z\sim0.5$, representative KN models span approximately $g,r\simeq24$--26 AB mag near peak, depending on ejecta mass, composition, and viewing angle. The optical emission is fast. A source near $r=26$ may remain detectable for only about one day, whereas the redder $z$-band emission can persist for $\sim2$ days. KST can detect an $r=24$ source in approximately 4 minutes, while its 1-hour $r=25.7$ limit reaches the bright end of the $z\sim0.5$ population. Reaching $r=26$ at the same significance requires approximately 1.7 hours per pointing. Thus, rapid optical mapping is most effective for relatively bright, favorably viewed KNe, while deeper integrations are best used once a candidate has been well localized.

For a planned 3~deg$^{2}$ optical mapping unit, KST can obtain $griz$ imaging to approximately $r\simeq25.7$ in a 1-hour integration per pointing. This depth is sufficient to detect ordinary KNe near peak for well-localized events at $z\lesssim0.5$, subject to the diversity of KN luminosities. For a brighter counterpart with $r\simeq24$, the same observing time can be distributed over approximately 15 such fields, corresponding to $\sim45$--75~deg$^{2}$ depending on overheads and the adopted multi-band cadence. Consequently, KST can use a depth-versus-area strategy matched to the GW posterior, with deep mapping for $\mathrm{CR90}\lesssim3~\mathrm{deg}^{2}$ events and shallower, wide mapping for bright nearby events with larger localization regions.

The A\# era is particularly favorable because the majority of relevant BNS triggers are expected within $\sim1$~Gpc ($z\lesssim0.2$). At these distances, typical KNe are expected to peak at approximately 22--24 mag in the optical, where KST can tile $\sim10~\mathrm{deg}^{2}$ rapidly to a depth adequate for discovery and color measurement. The simultaneous NIR spectral-mapping mode can then obtain early NIR spectra for counterparts bright enough for spectroscopy, linking optical identification directly to r-process abundance studies. For more distant events, the optical channel remains essential for finding the blue/less-obscured component and measuring rapid color evolution, while the NIR channel is more favorable for red, lanthanide-rich, or edge-on KNe.

A realistic planning yield is several KST optical counterparts per year during O5 and A\#, increasing to tens per year in the ET+CE era. Current O5 simulations for a deep targeted program predict approximately 2--19 detectable BNS KNe per year \citep{Kunnumkai2025}. KST can contribute to the well-localized and observable portion of this sample. For A\#, the forecast of approximately 20 BNS events in 1.5 years with $\mathrm{CR90}<10~\mathrm{deg}^{2}$ implies roughly 5--8 KST-accessible nearby optical KNe per year after visibility and distance cuts \citep{Andreoni2024Roman}. With ET+2CE, forecasts of $\sim3000$ BNS events per year localized to $\mathrm{CR90}<10~\mathrm{deg}^{2}$ imply that even a small low-redshift, observable fraction can provide a KST sample of approximately 15--50 optically identified KNe per year \citep{Colombo2025}. 

Therefore, we expect about 5--8 accessible events per year with $\mathrm{CR90}<10~\mathrm{deg}^{2}$ during A\# and 15--50 per year during the ET+2CE era.  Scaling the Wide Survey imaging rate ($0.074$~deg$^{2}$~hr$^{-1}$ in full $griz$ to $r\approx25.6$) to single-filter search observations gives $\approx3$~deg$^{2}$ per day at $r\approx26$, consistent with the short-GRB estimate below. A $\mathrm{CR90}=10~\mathrm{deg}^{2}$ localization thus requires about three days per epoch. With at least two epochs per event, the detection rates correspond to KST ToO requirements of 30--48 or 90--300 days per year for the A\# and ET+2CE runs, respectively. The nominal allocation adopted below is 30 days per year, which matches the lower end of the A\# expectation only if the shared ToO/calibration reserve is charged entirely to ToO. The resulting shortfall is addressed at the end of this section.  

KST will therefore fill a distinct observational niche. It will identify fast optical counterparts in well-localized GW fields that are outside the Rubin coverage, measure their multi-band evolution with diffraction-limited imaging, and immediately connect these discoveries to NIR spectral mapping with a reasonable amount of observation time. This capability will build a homogeneous sample of KNe from the A\# era through the next-generation GW era, enabling population studies of merger ejecta, viewing-angle effects, host environments, and r-process element production.

\subsubsection{Cosmic origin of r-process heavy elements: uncovering gold mines in the Universe}

Obtaining spectra of KNe in the early phase ($< 1$ week) is important. During the early phase, the ejecta are still in the photospheric phase, and r-process elements can imprint relatively interpretable broad absorption features \citep{Watson2019,Domoto2023}. KN spectra at later phases ($> 1$ week) are still useful, but using them for inferring abundances is much less straightforward with the disappearance of the photosphere, Non-LTE effects in place, and uncertainties in heating sources. Hence, with a rapid ToO observation, KST can play a crucial role in uncovering abundances of r-process heavy elements produced by kilonovae. 

The diagnostic value of early NIR spectroscopy arises from the wavelength-dependent bound--bound opacity of individual r-process ions. In the photospheric phase, we can analyze the spectra with radiative-transfer models that specify the ejecta mass, velocity/density structure, temperature and ionization state, and elemental abundance pattern. The model line list is then used to predict the blended, Doppler-broadened absorption features expected from rapidly expanding ejecta ($v\sim0.1$--$0.3c$). Rather than resolving narrow lines, the comparison uses the positions, widths, relative depths, and temporal evolution of broad spectral complexes. By fitting multi-epoch spectra jointly with the optical/NIR light curve, and marginalizing over ejecta mass, velocity, viewing angle, and heating/thermalization uncertainties, we can infer posterior constraints on the lanthanide mass fraction, $X_{\rm lan}$, and on selected elemental abundances. The total lanthanide yield is then estimated as $M_{\rm lan}=X_{\rm lan}M_{\rm ej}$.

The Sr~II NIR triplet (1.0036, 1.0327, and 1.0914~$\mu$m rest frame), central to the strontium identification in AT2017gfo \citep{Watson2019}, is observed as a broad, blueshifted P~Cygni feature near $0.8$--$1.0~\mu$m and therefore lies blueward of KST's 1.1--1.8~$\mu$m LVF window. KST constrains it only through $z$-band photometry. Within the LVF window, and more importantly for the red component, radiative-transfer calculations predict strong NIR absorption from La~III near 1.41~$\mu$m and Ce~III near 1.585~$\mu$m. These transitions can account for broad features in the early spectra of AT2017gfo and provide direct sensitivity to rare-earth/lanthanide material \citep{Domoto2022}. Recent work further shows that Gd~III may contribute to the $\sim1.2~\mu$m feature, emphasizing that the abundance measurement should fit several features and their time evolution simultaneously rather than assign a single feature to one element \citep{Rahmouni2025}. Consequently, KST spectra will constrain both an aggregate lanthanide fraction and, where signal-to-noise permits, the relative abundances of specific ions such as Sr, La, and Ce.

A spectral resolution of $R\sim100$ is well matched to this objective. Kilonova features are intrinsically broadened by $\Delta\lambda/\lambda\sim v/c\sim0.1$. The principal information is therefore contained in broad NIR morphology rather than unresolved narrow atomic lines. Multi-epoch KST spectra at approximately 1--3 and 5--7 days after merger would distinguish the evolution of a lanthanide-poor, blue component from a cooler lanthanide-rich component, substantially reducing the mass--opacity--composition degeneracy inherent in broad-band photometry alone.

KST can obtain a complete low-resolution ($R\sim100$) spectral map as wide as 3 deg$^2$ in the NIR in one day. Because the LVF encodes wavelength as focal-plane position, mapping speed and per-channel depth trade directly against one another: with the $0.25$~deg$^2$ NIR field and $\sim$48 channels, a complete-spectrum mapping rate of 3 deg$^2$ per day corresponds to $\sim$120~s per channel and a $5\sigma$ limit of $\approx20.1$~AB, 1 deg$^2$ per day to $\approx21.2$~AB, and the $22.5$~AB benchmark of Table~\ref{tab:lvf_modes} to $0.10$~deg$^2$ per day. Pointed ToO stares on an already-localized counterpart reach the benchmark depth without the mapping penalty. At the $\approx20.1$--$21.2$~AB limits of the fast mapping modes, the reach for the reddest KNe with little wind ejecta and viewed nearly edge-on \citep[][Fig.~7]{Andreoni2024Roman} during the first 7 to 10 days is correspondingly $\approx100$--$165$~Mpc, extending to $< 300$~Mpc for the deeper, slower mapping or for pointed follow-up at the benchmark depth. KST can therefore obtain early ($< 1$ week) NIR spectra for a substantial fraction of KN types in events with localization areas of $< 10$ deg$^2$ at two epochs or more, with the accessible distance set by the mapping mode adopted. 
Since the LVF observation mode allows us to obtain spectra of all objects in the field of view, KST NIR spectral map can be used to identify KNe and investigate their spectral evolution even without knowing where the KNe are located in advance. This capability is important since the rapid dimming in optical can make the identification of KNe challenging even with Rubin whose observational capability depends on good weather and the right location in the sky. ToO observations with other space missions such as JWST and Roman can be highly limited in terms of the number of targets as well as the ToO triggering time scale of $\gtrsim 1$ week \citep{Andreoni2024Roman}, or no NIR imaging or spectroscopic capability for the case of CSST. We also expect that most of the wide-field follow-up activities from the ground will be limited to the optical wavelength, making KST a key facility for uncovering the red KNe and completing the KN census. 

For the NIR study of early KN spectra,  we propose the following observation strategy. 
We will execute ToO within 1 day of GW detection for events with CR90 of $< 10$ deg$^2$ and the distance at $< 300$ Mpc. The typical duration of the ToO campaign will range from 7 days ($<5$ deg$^2$ events) up to 14 days for the largest ($\approx$10 deg$^2$) localization areas. For the 5--10 deg$^2$ class we adopt a bin-averaged campaign duration of 10 days in the budget below. This includes observations to obtain reference spectral mapping images at a much later epoch when the KNe faded away. A recent forecast for A\# (1.5 year duration is assumed) predicts about 10 BNS events with a median distance of 230 Mpc for events with localization areas $< 5$ deg$^2$ and roughly twice this number for the events with localization areas $< 10$ deg$^{2}$ (a median distance of 274 Mpc; \citealt{Andreoni2024Roman}). These numbers could increase with improved sensitivities of the next generation GW observatories. Here, we neglect the number of NSBH events since the number of EM-bright NSBH events is likely to be low \citep{Andreoni2024Roman,Biscoveanu2023,Broekgaarden2021,Drozda2022,Fragione2021,RomanGarza2021}.

For events with CR90 $< 5$ deg$^2$, we expect them to be at $< 300$ Mpc mostly. For events with $5 < \rm{CR90} < 10$ deg$^2$, about 60 \% of the events would be at $< 300$ Mpc. Assuming that 50 \% of the events will be out of visibility (the direction being toward the Sun), the required observation time is 5 BNS events $\times$ 7 days = 35 days for $< 5$ deg$^2$ events, and 5 $\times$ 10 days $\times$ 0.6 = 30 days for the additional BNS events with localization areas between 5 and 10 deg$^2$. Therefore, in total, the BNS ToO observation requires about 65 days of observing time per 1.5 years (or 43 days per year during the GW run period), yielding early spectra of approximately 8 KNe per 1.5-yr run (the $10+6$ accessible events reduced by the 50\% Sun-angle cut), i.e., $\sim$5 per year. This time estimate is an upper bound, since ground-based facilities will often identify the counterpart before KST observes, allowing pointed follow-up of the localized source in place of a full spectral-mapping campaign. Hence, we request ToO time of 30 days per year during the GW run period. Modeling of KNe NIR spectra would produce the production rates of lanthanide elements.

Joint modeling of KST multi-epoch NIR spectra and broad-band light curves will constrain the lanthanide mass fraction and total lanthanide yield in kilonova ejecta, while testing for spectral signatures of individual r-process species across the $\sim$8 KNe per 1.5-yr run expected from this program. This will enable us to place constraints on the r-process elements production in the local universe, which can also serve as a stepping stone for understanding the cosmic production history of r-process elements such as gold.

\subsubsection{Unveiling red kilonovae associated with binary neutron star or neutron star-black hole mergers}

 BNS mergers are promising events to produce EM counterparts, but neutron star-black hole (NSBH) merger events can also produce EM counterparts. No optical counterpart of an NSBH merger has been reported, presenting an excellent opportunity for deepening our current understanding of the astrophysical phenomena surrounding these enigmatic compact mergers. Current theoretical studies suggest that only a minority (typically $\sim$10–20\% in representative population models) of NSBH mergers may produce observable EM counterparts. NSBH kilonovae are expected to be generally redder than BNS kilonovae, owing to their neutron-rich, lanthanide-rich ejecta and relatively weak blue, high-\(Y_e\) component. Their brightness will depend on the ejecta mass, $M_{ej}$, although it is quite uncertain how much $M_{ej}$ will be produced in NSBH mergers. 

With its red-sensitive instruments, KST can place useful limits on KNe properties of NSBH mergers and test current theoretical models. 
The lower limit of the volume rate of NSBH mergers is estimated to be similar to BNS merger rate while the upper limit is predicted to be about 10 times less than the upper limit of the BNS  merger rate. On the other hand, NSBH merger GW signals are intrinsically stronger than the BNS merger signals due to larger chirp mass, meaning that the search volume will be larger for NSBH merger events than BNS merger events. Combining these considerations, we expect a similar number of, or possibly a few times more, NSBH merger detections than BNS mergers in future GW runs, as has been the case for previous GW observing runs. 

In practice, KST ToO of NSBH GW events can follow the strategy similar to the above studies of optical identification of KNe. 
For early NIR spectroscopy of events at $< 300$ Mpc, we assume a volume rate about 3 times lower than the BNS rate, in which case about 15 days of ToO observation is needed to cover the nearby NSBH merger events. 
For identification of the optical counterparts for faint events, we can provide similar constraints as for the BNS merger follow-up observations, which brings the required ToO observing time to 10 days per year. Overall, we expect to follow up 15 NSBH events for the detection of KNe in those events per year. Among these, about three could reveal their optical and NIR counterparts. Therefore, several years of KST ToO observations will net about 10 optical counterparts, which will allow us to examine a rough statistical census of NSBH KNe such as ejecta mass, viewing-angle effects, and host environments. For a few brightest NIR counterparts, r-process element abundance can be inferred.

\subsubsection{Determination of cosmic expansion rate of the Universe}

An independent, distance-ladder-free measurement of the Hubble constant is a principal science return of this ToO program. The Hubble tension that motivates it, and the two standard-siren channels, are developed in Section~\ref{sec:dark_sirens}. Here we summarize what the kilonova counterparts contribute. Each secure counterpart converts a GW event into a bright siren, combining the luminosity distance encoded in the waveform with the host-galaxy redshift from EM observations. The per-event uncertainty in H$_{0}$ is dominated by the GW distance error, which is expected to improve from $\sim$10\% in O5 to 5--10\% in A\# and 1--5\% in the ET era \citep{ChenHolz2018,Maggiore2020}. A homogeneous sample of tens of bright sirens, accumulated from the well-localized and observable subset of events that KST follows up within the ToO allocation quantified above ($\approx$5--8 accessible events per year in the A\# era), would therefore approach percent-level precision on H$_{0}$. This bright-siren route is complementary to the catalog-based dark-siren scaling of Section~\ref{sec:dark_sirens} and Table~\ref{tab:forecast}. The two channels share the GW data but carry different electromagnetic selection effects, and together they strengthen the independent test of the cosmic expansion rate that the dark-siren pillar provides.

\subsubsection{Other Transients}
\label{subsubsec: othertransients}

Gamma-ray bursts (GRBs) are among the prime transient phenomena for intensive multimessenger astronomy (MMA) studies. Binary neutron star (BNS) mergers can produce short GRBs (sGRBs; GRBs with burst durations shorter than 2 s), and hence can generate signals detectable in both gravitational waves (GWs) and electromagnetic (EM) radiation. GRBs are observed when a relativistic jet is oriented sufficiently close to our line of sight, resulting in strong relativistic beaming of the observed emission. Short GRBs are regularly detected at a rate of approximately 40 events per year by current gamma-ray satellites such as \textit{Fermi} \citep{vonKienlin2020}, and the population of sGRBs with measured spectroscopic redshifts has a median redshift of approximately 0.5 \citep{Berger2014,Nugent2022}. Although \textit{Fermi}/GBM localizations are typically much larger than a few deg$^2$, a subset of sGRBs can be localized to much smaller areas through observations with instruments such as \textit{Swift}, \textit{Fermi}/LAT, or other high-energy facilities. Such accurate localizations are particularly valuable for identifying EM counterparts of GW events. Future third-generation GW observatories such as the Einstein Telescope (ET) will substantially increase the number and distance range of detectable BNS mergers, providing important opportunities for joint GW--gamma-ray observations \citep{ETScience2025}.

Furthermore, GRBs often exhibit bright optical afterglows, allowing their positions to be accurately determined through optical and near-infrared (NIR) observations. Ground-based telescopes can therefore play an important role in GRB MMA observations by detecting bright afterglows with rapid response times. In the past, optical and NIR follow-up observations have detected afterglows for approximately one-third of short GRBs. For example, \citet{Fong2015} reported 30 optical/NIR detections in a sample of 103 sGRBs, corresponding to an observation-constraint-corrected detection fraction of approximately 40\%. The lack of optical detections for the remaining events may be attributable to low circumburst densities, dust extinction, unfavorable viewing geometries, or high redshifts \citep{Berger2010,Berger2014,Fong2015,Laskar2022}.

KST can be a powerful tool for detecting kilonovae (KNe) and identifying their host galaxies in association with short GRBs that lack bright optical afterglows. Short GRBs with faint or undetected optical afterglows may still produce detectable KNe unless their emission is strongly attenuated by dust. Rapid and deep follow-up observations with KST could therefore reveal these KNe. For events that are also associated with GW signals, we expect KST to be capable of detecting KNe associated with short GRBs at $z<0.5$, as well as some events at $z \gtrsim 0.5$, particularly when accurate localizations are available from high-energy observations.

We expect approximately 10 short GRBs without detected optical counterparts per year to have good localizations ($<3$ deg$^2$). Approximately one day of KST observations can cover $\sim3$ deg$^2$ to a depth of $r\sim26$ mag, potentially allowing the detection of short-GRB-associated KNe at $z<0.5$. Therefore, by allocating approximately 1 day $\times$ 3 epochs $\times$ 10 events = 30 days per year, including observations required to construct reference images, we could identify the locations and potential kilonova counterparts of these short GRBs. Over the 6-year prime mission, such a program could provide a sample of up to $\sim$60 short GRBs with kilonova detections, depending on the actual kilonova detection efficiency and on the ToO time actually available (see below). The KN spectral energy distributions (SEDs) could be examined to constrain extinction and characterize their emission properties. If some of these events occur during GW observing runs and are independently detected in GWs, they would provide particularly valuable joint sGRB+GW+KN events.

Taken separately, the programs above request approximately 30, 10, and 30 days per year for BNS, NSBH, and short-GRB follow-up respectively, a total that exceeds the $\approx$30 days per year represented by the combined ToO/calibration reserve of Table~\ref{tab:time_budget}. That reserve must also cover routine calibration and maintenance, and it absorbs the small Wide Survey margin noted in Section~\ref{sec:survey}, so the time actually available for rapid response is strictly less than 30 days per year. At the search rate adopted above, the BNS program alone requires 30--48 days per year during A\# and 90--300 days per year in the ET$+$2CE era. The reserve is thus already fully committed by BNS follow-up in the A\# era and is substantially undersized for the next-generation detector era. BNS follow-up during gravitational-wave observing runs is accordingly the primary call on that reserve. The NSBH and short-GRB programs are executed opportunistically within the same allocation, and their full scope would require either an expanded reserve or an extended mission. Setting the relative priorities, and the resulting event yields, is a mission-definition-phase task.

Furthermore, even when GW observatories are not operating, follow-up observations of short GRBs can improve our understanding of the connection between short GRBs and KNe and prepare for MMA observations with future GW observatories.

KST provides an additional advantage for studying kilonova emission associated with short GRBs. At early times, the optical and NIR light curves of short GRBs are generally dominated by afterglow emission. However, from several days to approximately 10 days after the burst, kilonova emission can become prominent relative to the fading afterglow \citep{Tanvir2013,Berger2013,Troja2019}. For relatively nearby events ($z<0.2$), the $H$-band magnitude can be approximately 22.5 AB mag. For such events, KST's high-spatial-resolution spectral-mapping capability could enable spatially resolved NIR spectroscopic studies of kilonovae, helping to disentangle kilonova emission from contributions by the host galaxy and afterglow. Such observations could provide information on the production and abundance of heavy elements, as described in the previous section.

In addition to GRBs, several other potential GW sources could benefit from MMA studies. These include supernovae (SNe), tidal disruption events (TDEs), and systems containing binary supermassive black holes (SMBHs). Core-collapse SNe can produce GW signals if the collapse of the progenitor star is sufficiently asymmetric. Next-generation GW observatories could detect such signals from sufficiently nearby events, particularly those occurring within the Local Group. TDEs may also produce bursts or other forms of GW emission. In particular, stellar tidal disruptions by intermediate-mass black holes (IMBHs) can potentially produce both EM and GW signals detectable by future space-based or decihertz GW observatories \citep{Pfister2022,Toscani2022}.

Binary SMBHs are important targets for pulsar-timing arrays (PTAs) and future space-based GW observatories. EM counterparts may arise when binary SMBHs or IMBHs interact with gas-rich environments, particularly in active galactic nuclei (AGNs) \citep[e.g.,][]{Milosavljevic2005,Mangiagli2022,Bogdanovic2022}. Although the likelihood of detecting such events during the operational lifetime of KST may be relatively low, they should nevertheless be considered high-priority targets for target-of-opportunity (ToO) observations.

\section{Survey Strategy}
\label{sec:survey}
\begin{figure*}[t]
\centering
\includegraphics[width=0.92\textwidth]{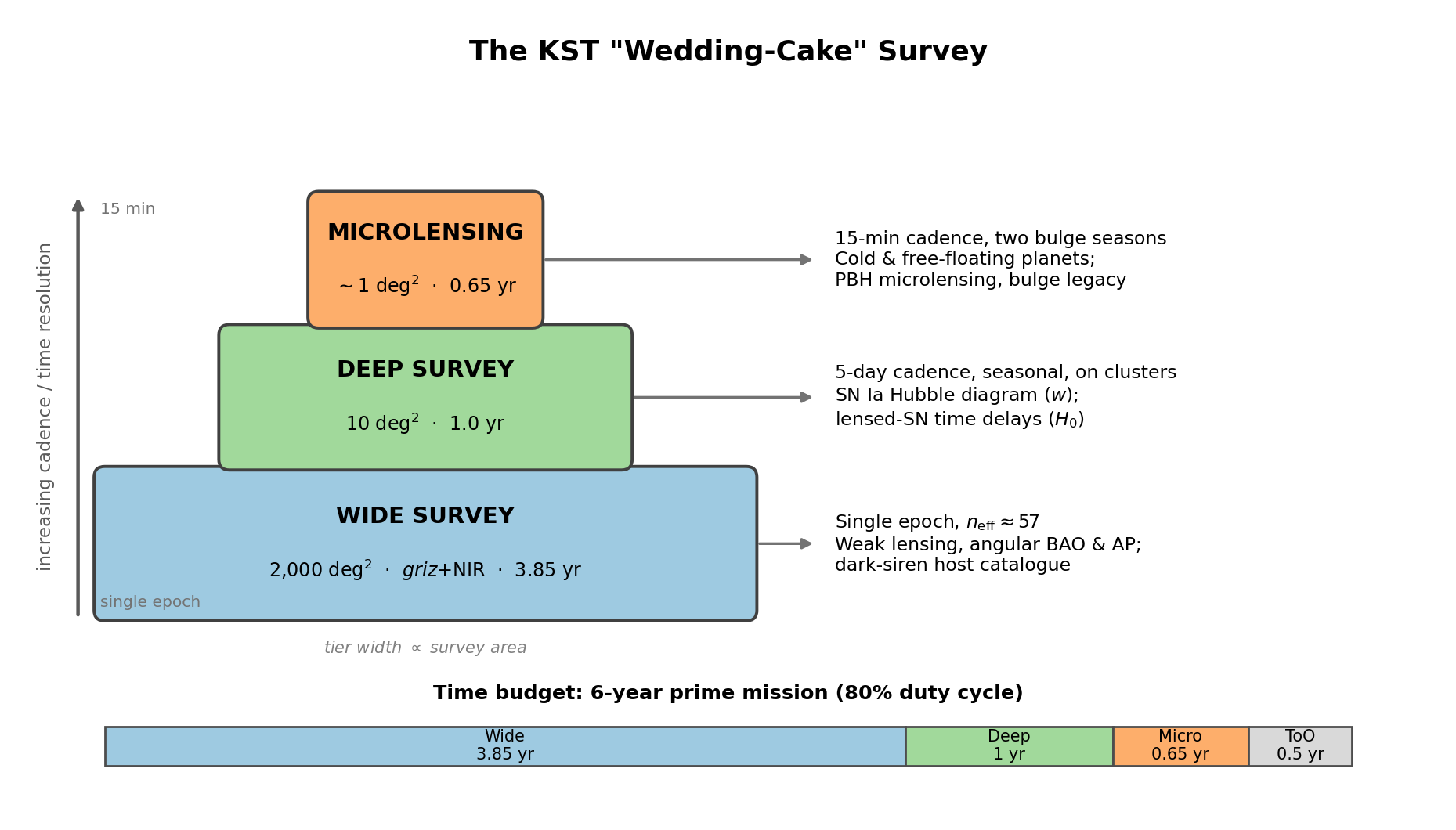}
\caption{The KST three-tier ``wedding-cake'' survey. Tiers are ordered by sky area (widest at the base) and by observing cadence (increasing upward): a $2{,}000~\text{deg}^2$ single-epoch Wide Survey for weak lensing and large-scale structure, a $10~\text{deg}^2$ cluster-targeted Deep Survey at 5-day cadence (in $\sim$8-month seasons) for supernova cosmology, and a $\sim1~\text{deg}^2$ Galactic-Bulge Microlensing campaign at 15-minute cadence. Call-outs list the primary science from each tier. The bottom bar shows the allocation of the 6-year prime mission across the four programs (Wide, Deep, Microlensing, and Target-of-Opportunity/calibration), which run in parallel and seasonally rather than strictly in series.\label{fig:weddingcake}}
\end{figure*}

KST executes a three-tier ``wedding-cake'' survey, trading sky area for cadence and depth as one ascends the stack (Figure~\ref{fig:weddingcake}). A wide, single-epoch \textbf{Cosmology Wide Survey} ($2{,}000~\text{deg}^2$ imaged in full $griz$ with simultaneous NIR spectrophotometry) provides the weak-lensing, BAO, and AP measurements that drive the dark-energy program and doubles as the deep NIR host catalog for dark sirens. Nested within it, a high-cadence \textbf{Deep Survey} of $10~\text{deg}^2$ centered on massive galaxy clusters builds the Type~Ia supernova Hubble diagram and captures strongly-lensed supernovae for time-delay $H_0$. A seasonal \textbf{Microlensing} campaign toward the Galactic Bulge ($\sim1~\text{deg}^2$ at 15-minute cadence) delivers the exoplanet and compact-object census, while a modest \textbf{ToO/calibration} reserve supports gravitational-wave follow-up and instrument maintenance. The multi-messenger programs of Section~\ref{sec:mma} request more than this reserve provides when costed separately. Their relative priority is therefore set in the mission-definition phase. The four components are accommodated within a 6-year prime mission at $80\%$ duty cycle (Table~\ref{tab:time_budget}). They run in parallel and seasonally rather than strictly in series, and the subsections below detail each tier in turn.

\subsection{Cosmology Wide Survey Strategy: Weak Lensing and LSS}
To achieve the statistical precision required for Stage IV dark energy constraints, the KST Cosmology Wide Survey is designed to cover a baseline area of $2,000~\text{deg}^2$ in all four optical bands ($g,r,i,z$) together with the simultaneous NIR LVF spectrophotometry. This footprint balances the source density for Weak Lensing (WL) and the comoving volume for Large-Scale Structure (LSS) probes, namely Baryon Acoustic Oscillations (BAO) and the Alcock--Paczyński (AP) test, against the time cost of full multi-band coverage.
\subsubsection{Survey Efficiency and Timeline}
The survey strategy leverages the simultaneous dual-channel capability of KST. Because the four optical bands occupy separate $0.25^\circ \times 0.25^\circ$ single-band sub-fields, the filters operate in parallel rather than in sequence. At every instant each sub-field images a different $0.0625~\text{deg}^2$ of sky, and the survey scan steps each sky position through all four sub-fields in turn. The time budget is therefore set by the per-band dwell time alone, not by the sum of the band exposures. The baseline adopts a uniform dwell of $0.85$~hr per band. One full field passage then takes $4 \times 0.85 \approx 3.4$~hours and one $0.0625~\text{deg}^2$ sub-field column is completed---in all four bands, with the NIR spectrum accumulating in parallel---per $0.85$~hr of open-shutter time, an effective multi-band survey rate of $0.0625/0.85 \approx 0.074~\text{deg}^2/\text{hr}$. At this dwell the shape bands reach $r \approx 25.6$ ($5\sigma$, $0.3''$ galaxy), scaled from the 1-hr depths of Table~\ref{tab:kst_specs}. The simultaneous NIR LVF spectrophotometry is recorded at no additional cost and supplies independent redshift leverage.
\begin{itemize}
    \item \textbf{Total Integration Time:} Completing the $2,000~\text{deg}^2$ survey in full $griz$ requires $(2,000~\text{deg}^2 / 0.0625~\text{deg}^2) \times 0.85~\text{hr} \approx 27,200$ hours of open-shutter time. Because the bands are acquired in parallel on separate sub-fields, this total is set by the per-band dwell alone, and no additional time is incurred for the remaining bands.
    \item \textbf{Mission Duration:} Assuming a standard observational efficiency of $80\%$ (accounting for slewing, dithering, and readout), the open-shutter total corresponds to $27,200/0.80 \approx 34,000$ hours of wall-clock time, i.e., $\approx$3.9 years of dedicated mission time, within $1\%$ of the $3.85$-yr allocation of Table~\ref{tab:time_budget} (the difference is absorbed by the ToO/calibration reserve), accommodated within the 6-year prime mission alongside the Deep, microlensing, and ToO programs.
\end{itemize}

\subsubsection{Data Yield: From Shear to 3D Clustering}
The scientific yield (summarized in Table~\ref{tab:survey_yield}) is driven by the combination of high spatial resolution (for WL) and deep NIR photometry (for Photo-z).
\begin{itemize}
    \item \textbf{Weak Lensing Sample:} Due to the diffraction-limited PSF ($\sim0.15''$), KST achieves a high source density of $n_{\rm eff} \approx 57~\text{arcmin}^{-2}$ at the survey depth. This yields a sample of $N_{\rm gal} \approx 4.1 \times 10^8$ galaxies for cosmic shear tomography.
    \item \textbf{LSS and High-z BAO/AP:} Unlike optical-only surveys, KST's NIR sensitivity allows it to map the galaxy density field into the ``redshift desert'' ($1.5 < z < 3.0$). The multi-band photometry ($g,r,i,z + \text{NIR}$) is designed to deliver photometric redshifts approaching the design goal of $\sigma_z/(1+z) \lesssim 0.01$ for galaxies detected in the NIR channels (Table~\ref{tab:lvf_modes}), with coarser, optical-dominated precision for the fainter majority. The number density of $1.5<z<3$ tracers meeting the goal is a mission-definition-phase deliverable. This precision allows for the tomographic slicing of the universe into thin shells, enabling angular BAO measurements of the comoving angular distance $D_M(z)$ at an epoch unprobed by current ground-based facilities, with the broad-bin AP anisotropy of the projected clustering providing complementary sensitivity to $H(z)$.
    \item \textbf{Data Volume:} The focal plane generates $\sim423$ Megapixels per exposure. With a 4-point dither pattern across four optical bands, the total raw data volume is $\sim450$ TB, managed via onboard Rice compression.
\end{itemize}

\begin{table}[ht]
\centering
\caption{Projected Yield of the KST Cosmology Wide Survey}
\label{tab:survey_yield}
\begin{tabular}{@{}p{3.5cm}>{\raggedright\arraybackslash}p{3.9cm}@{}}
\hline
\textbf{Metric} & \textbf{Value} \\ \hline
Survey Area & $2,000~\text{deg}^2$ ($griz$) \\ \addlinespace
Depth ($5\sigma$, $d{=}0.3''$ galaxy) & $r \approx 25.6$ AB (shape band, 0.85-hr dwell) \\ \addlinespace
Total Exposure Time & $\sim27,200$ hr open-shutter ($\approx 3.9$ yr at $80\%$ duty cycle; $3.85$-yr allocation) \\ \addlinespace
Effective Source Density ($n_{\rm eff}$) & $\sim 57~\text{arcmin}^{-2}$ \\ \addlinespace
Total Sample Size ($N_{\rm gal}$) & $\sim 4.1 \times 10^8$ \\ \addlinespace
Redshift Range (LSS) & $0 < z < 3.0$ (Optimized for $z > 1.5$) \\ \addlinespace
Key Observables & Cosmic Shear, Angular BAO, AP Distortion \\ \hline
\end{tabular}
\end{table}

\subsection{Cosmology Deep Survey Strategy: Type Ia Supernovae and Strong Lensing}
Complementing the Wide Survey, the KST Deep Survey adopts a deep-drilling strategy to map the expansion history of the universe. By targeting smaller fields with high cadence, KST leverages its simultaneous VIS+NIR architecture to construct a homogeneous, NIR-standardized Hubble diagram using Type Ia Supernovae (SNe Ia) and to measure the Hubble Constant ($H_0$) via time-delay cosmography.

\subsubsection{Survey Efficiency and Timeline}
The Deep Survey targets a total area of $10~\text{deg}^2$, nested within the Wide Survey footprint and organized as ten $1^\circ \times 1^\circ$ tiles, each centered on a known massive galaxy cluster. Fields are revisited with a 5-day cadence to critically sample the rise time and peak brightness of SN light curves.
\begin{itemize}
    \item \textbf{Cadence and Seasons:} Each tile is monitored at a 5-day cadence during its annual visibility window ($\sim$8 months per year), giving $\approx$48 epochs per year and $\approx$290 epochs over the 6-year mission. The Wide Survey fills the remainder of the schedule. The contemporaneous optical ($g,r,i,z$) and NIR ($1.1$--$1.8~\mu$m) coverage within each epoch provides a wide baseline for $K$-corrections and dust extinction measurements.
    \item \textbf{Time Budget:} One epoch covers all ten tiles in all four bands with a per-dwell integration of $0.086$~hr ($\approx$5 minutes): $10$ tiles $\times$ $28$ scan dwells per tile (four $0.25^\circ$-wide strips, including the band-assembly run-in and run-out at the field edges) $\times$ $0.086$~hr $\approx 24$~hr of open-shutter time per epoch. Over $\approx$290 epochs this totals $\approx$7,000~hr, which at the $80\%$ observing efficiency corresponds to $1.0$~yr of mission time, closing on the Deep Survey allocation of Table~\ref{tab:time_budget}.
    \item \textbf{Per-Epoch and Co-added Depth:} Each epoch integrates $0.086$~hr per band per sky position, reaching $r \approx 25.7$ ($5\sigma$, point source; $g \approx 26.3$, $i \approx 25.2$, $z \approx 24.7$), sufficient for densely sampled light curves of SNe~Ia to $z \approx 0.7$ and near-peak detections to $z \approx 1$ in $i,z$. The mission-co-added stack reaches $r \approx 28.7$ (point source) and $r \approx 27.4$ ($0.3''$ galaxy), and the accumulated LVF exposure ($\sim$2~hr per spectral channel) yields a co-added $R \sim 100$ spectrum to $\approx$22~AB per channel at every position in the Deep fields.
\end{itemize}

\subsubsection{Cluster-Targeted Fields: The Path to $H_0$}
A distinctive feature of the KST survey design is the placement of the Deep Survey fields. Rather than selecting random blank fields, the $10~\text{deg}^2$ footprint comprises ten $1^\circ \times 1^\circ$ tiles, each centered on a known massive galaxy cluster (e.g., HST/JWST lensing targets).
\begin{itemize}
    \item \textbf{Lensed Supernovae:} By monitoring these strong lensing clusters, KST is expected to detect of order $10$--$20$ multiply imaged supernovae over the mission (cluster-scale and galaxy-scale systems combined; a planning estimate to be refined by simulation). The 5-day cadence suits the weeks-to-months time delays ($\Delta t$) typical of cluster-scale images.
    \item \textbf{Breaking Degeneracies:} Unlike quasar time delays, lensed Type Ia SNe are standard candles. Their known luminosity breaks the ``Mass-Sheet Degeneracy,'' with a target precision of $\sim$1--2\% per system that microlensing and lens-model systematics may degrade to a few percent (Section~\ref{sec:dark_energy}).
    \item \textbf{Cluster Member SNe:} Simultaneously, this strategy yields a rich sample of SNe Ia within cluster environments, allowing for tests of environmental dependence on the standardization relation.
\end{itemize}

\subsubsection{Sample Statistics and Spectroscopic Yield}
KST utilizes its LVF to provide self-confirmation for high-redshift candidates (Table~\ref{tab:sn_yield}).
\begin{itemize}
    \item \textbf{Total Sample:} KST is projected to discover $\sim 5,000$ SNe Ia to $z \approx 1$: $\sim 4,000$ at $z \lesssim 0.7$, of which $\sim 3,000$ have densely sampled multi-band light curves, and $\sim 1,000$ at $z > 0.7$.
    \item \textbf{Spectral Confirmation:} Near-peak stacks of the accumulating LVF spectrum provide spectrophotometric typing of the brightest, low-redshift events through their rest-frame NIR spectral features. The Si~II $\lambda6355$ feature enters the LVF bandpass only at $z\gtrsim0.8$, where SNe~Ia are far fainter than the per-channel limits. Higher-redshift events are instead classified from their optical light curves and colors. The mission-co-added LVF spectrum ($\approx$22 AB per channel) supplies a spectrophotometric redshift and stellar-population diagnostics for hosts above this limit (with adjacent-channel binning for fainter hosts; Table~\ref{tab:lvf_modes}), providing host redshifts for the $z > 0.7$ sample.
    \item \textbf{Systematics Control:} For the $z\lesssim0.6$ SNe observed in the rest-frame NIR, the intrinsic scatter of SN~Ia luminosities drops to $\sim 0.10$ mag and dust extinction is reduced by a factor of $\sim$3--5 relative to the rest-frame $B$ band.
\end{itemize}

\begin{table}[ht]
\centering
\caption{Projected Yield of the KST Deep Survey}
\label{tab:sn_yield}
\small
\setlength{\tabcolsep}{3pt}
\begin{tabular}{@{}l>{\raggedright\arraybackslash}p{4.1cm}@{}}
\hline
\textbf{Metric} & \textbf{Value} \\ \hline
Survey Area & $10~\text{deg}^2$ (ten $1^\circ\times1^\circ$ cluster-centered tiles) \\
Cadence & 5 days ($\sim$8-month seasons; $\approx$290 epochs) \\
Per-Epoch Depth & $r \approx 25.7$ ($5\sigma$, point source) \\
Co-added Depth & $r \approx 28.7$ (point); $27.4$ ($0.3''$ galaxy) \\
Wavelength Coverage & $0.4$--$1.8~\mu$m (simultaneous VIS+NIR) \\
Target Objectives & Hubble Diagram ($w$) \& Time Delays ($H_0$) \\
Total SNe Ia ($0 < z \lesssim 1$) & $\sim 5,000$ \\
High-z Sample ($z > 0.7$) & $\sim 1,000$ \\
Lensed SNe (multiply imaged) & $10$--$20$ \\
Systematic Control & Intrinsic scatter $\sigma_{\rm int} \approx 0.10$ mag (NIR) \\ \hline
\end{tabular}
\end{table}

\subsection{Exoplanet Microlensing Survey Strategy}
\label{sec:survey:microlensing}
To fulfill the exoplanet census objectives (Section~\ref{sec:exo}), KST dedicates a specific observational window to the Galactic Bulge. Unlike the Wide Survey which maximizes area, this program requires high-cadence, continuous monitoring of a fixed stellar population to detect the transient brightening caused by planetary microlensing events.

\subsubsection{Survey Design and Cadence}
The microlensing survey utilizes the wide-field optical and NIR capability to pierce the dust of the Galactic plane, monitoring millions of source stars that are difficult to access from ground-based observations. 
\begin{itemize}
    \item \textbf{Target Fields:} The survey targets 4 dedicated fields ($\sim 1~\text{deg}^2$ total) located in the low-extinction windows of the Galactic Bulge ($|b| \approx 1^{\circ}-2^{\circ}$), optimized for maximum stellar density.
    \item \textbf{High-Frequency Monitoring:} During observation windows, KST executes a continuous tile pattern with a \textbf{15-minute cadence}. This rapid sampling is critical to resolve the short-duration caustic crossing features characteristic of low-mass and free-floating planets.
    \item \textbf{Seasonal Windows:} Observations are conducted in \textbf{two dedicated bulge seasons} (nominally mission years 2 and 4), each covering the full $\sim$4-month Galactic Bulge visibility window with uninterrupted 15-minute-cadence monitoring. Continuous in-season coverage is essential for detecting and characterizing planetary anomalies. During these two seasons KST prioritizes the bulge fields, returning to the Wide Survey outside them, while ground-based networks (e.g., KMTNet) cover the remaining bulge seasons.
\end{itemize}

\subsubsection{Mission Allocation and Synergy}
The microlensing program is highly synergistic with the cosmological surveys. Since the Galactic Bulge is only visible for a fraction of the year, the microlensing campaign does not compete for time year-round.
\begin{itemize}
    \item \textbf{Total Time Budget:} Two full 4-month seasons total 8 months $=0.67$~yr of dedicated mission time, closing on the \textbf{0.65-year} Microlensing allocation of Table~\ref{tab:time_budget}.
    \item \textbf{Data Yield:} This strategy is projected to monitor $>100$ million stars, detecting thousands of microlensing events and enabling a statistical census of cold exoplanets down to Earth mass, a regime inaccessible to the Transit or Radial Velocity methods.
    \item \textbf{Bulge Legacy Science:} The same repeated optical and NIR observations will also produce a homogeneous legacy dataset for Milky Way astronomy. The NIR LVF data will provide low-resolution spectrophotometric information for stars in the bulge fields, enabling studies of bulge structure, stellar populations, variable stars, and candidate stellar clusters as a natural by-product of the microlensing survey (see Section~\ref{sec:MW:bulge}).
\end{itemize}

\begin{table}[ht]
\centering
\caption{Indicative 6-Year Mission Time Budget}
\label{tab:time_budget}
\small
\setlength{\tabcolsep}{3pt}
\begin{tabular}{@{}>{\raggedright\arraybackslash}p{2.4cm} c c >{\raggedright\arraybackslash}p{2.3cm}@{}}
\hline
\textbf{Survey Component} & \textbf{Area} & \textbf{Time (Yrs)} & \textbf{Cadence Strategy} \\ \hline
Cosmology Wide Survey & $2,000~\text{deg}^2$ & 3.85 & Single epoch, $griz$ (Shape/Photo-z) \\
Deep Survey (SNe Ia) & $10~\text{deg}^2$ & 1.00 & 5-day revisit (seasonal) \\
Microlensing (Bulge) & $\sim1~\text{deg}^2$ & 0.65 & 15-min cadence (two 4-month seasons) \\
ToO / Calibration & -- & 0.50 & Rapid Response / Maintenance \\ \hline
\textbf{Total Prime Mission} & & \textbf{6.00} & \\ \hline
\end{tabular}
\end{table}

\subsection{Projected 6-Year Performance}
\label{sec:forecast}
Table~\ref{tab:forecast} consolidates the projected deliverables of the nominal six-year mission across the three dark-sector pillars and the Legacy portfolio.

The forecasts quoted throughout this paper are deliberately transparent, order-of-capability estimates rather than full statistical analyses. The end-to-end Fisher and simulation-based forecasts are a mission-definition-phase deliverable. They are obtained by two methods. \textit{(i) Survey-parameter scaling:} precision metrics and yields are scaled from published results and forecasts---the Stage~IV imaging-cosmology forecasts of \textit{Euclid} \citep{EuclidVII2020} and Rubin/LSST \citep{Ivezic2019}; the demonstrated microlensing yields of OGLE, MOA, and KMTNet \citep{Udalski2015,Sumi2003,Kim2016} together with the space-based bulge-survey predictions for Roman \citep{Penny2019,Johnson2020}; the strong-lens and lensed-supernova population forecasts of \citet{OguriMarshall2010}, \citet{Collett2015}, and \citet{Goldstein2019}; and the LIGO--Virgo--KAGRA standard-siren projections---using KST's collecting area, source density ($n_{\rm eff}\approx57~\text{arcmin}^{-2}$), depth, redshift range, and time allocation. \textit{(ii) Statistical $N^{-1/2}$ scaling:} where a current measurement exists, its uncertainty is extrapolated as the inverse square root of the sample size to KST's projected sample. This underlies the self-interaction ($\sigma/m$), primordial-black-hole ($f_{\rm PBH}$), and dark-siren ($H_0$) forecasts, in each case anchored to the cited measurement. The cosmological constraints additionally assume a \textit{Planck} CMB prior, multiplicative shear-bias control to $m<10^{-3}$, conservative small-scale cuts to limit baryonic and non-linear-modeling systematics, and the nominal six-year survey allocation of Table~\ref{tab:time_budget}. Individual entries are accurate to a factor of $\sim$1.5--2, and the basis of each is stated where it is introduced.

The headline expectation is that a single 1.4-m mission delivers, within six years, a dark-energy measurement approaching Stage~IV precision---a joint geometry-plus-growth constraint with $\sigma(w_p)\approx0.03$ at the pivot redshift ($z_p\approx0.35$) and $\sigma(w_a)\approx0.25$--$0.36$, corresponding to a Dark Energy Figure of Merit $\mathrm{FoM}\equiv[\sigma(w_p)\,\sigma(w_a)]^{-1}\approx80$--$180$ (the lower end marginalizing KST's full systematics budget, the upper end the statistics-limited case; Figure~\ref{fig:w0wa}), and an $\approx$1\% growth amplitude $\sigma(S_8)\approx0.008$---a population-level test of the dark-matter particle across three decades of halo mass, and an independent, distance-ladder- and CMB-free measurement of $H_0$ from gravitational-wave sirens: three orthogonal probes of the dark sector from one dataset.

\begin{table*}[t]
\centering
\caption{Projected performance of the nominal six-year KST mission. Data products and the supernova/dark-energy figures follow the science and survey sections. The dark-matter and dark-siren constraints, the Dark Energy Figure of Merit, and $\sigma(S_8)$ are forecast-level estimates scaled from published Stage~IV and standard-siren forecasts. All entries assume the nominal allocation of Table~\ref{tab:time_budget}.}
\label{tab:forecast}
\begin{tabular}{p{2.7cm} p{6.2cm} p{6.4cm}}
\hline
\textbf{Science Pillar} & \textbf{6-Year Data Products} & \textbf{Projected Constraints} \\ \hline
\textbf{Dynamical Dark Energy} (Sec.~\ref{sec:dark_energy}) & $4.1\times10^{8}$ weak-lensing galaxies ($n_{\rm eff}\approx57$~arcmin$^{-2}$); $\sim$5{,}000 SNe~Ia ($\sim$1{,}000 at $z>0.7$ with LVF host redshifts); angular BAO and AP across the $1.5<z<3$ redshift desert; $\sim$$10^{4}$ weak-lensing-selected clusters to $z\sim1.5$; a lensed-SN sample of $10$--$20$ systems for percent-level $H_0$ & $\sigma(w_p)\approx0.03$, $\sigma(w_a)\approx0.25$--$0.36$ (Dark Energy FoM $\approx80$--$180$); $\sigma(S_8)\approx0.008$ ($\approx$1\%); independent $\sigma_8,\Omega_m$ from cluster abundance; $H_0$ to $\approx$0.5--1\% from lensed-SN time delays (target vs.\ degraded per-system precision, 20 systems assumed) \\ \hline
\textbf{Nature of Dark Matter} (Sec.~\ref{sec:dark_matter}) & $\gtrsim$$10^{3}$ galaxy-scale strong lenses analyzed for substructure; joint strong$+$weak lensing of the ten Deep-field clusters; diffraction-limited cosmic-shear maps to small scales; ultra-faint satellite census and cold stellar streams in the Local Group; $>$$10^{8}$ bulge stars monitored at 15-minute cadence for short-timescale compact-object lensing & Subhalo mass function to $10^{7}$--$10^{8}\,M_\odot$ (lensing $+$ satellites $+$ streams); warm-dark-matter free-streaming bound tightened toward $m_{\rm th}\sim10$~keV; self-interaction $\sigma/m\lesssim0.02$~cm$^2$g$^{-1}$ from $\gtrsim$1000 merging clusters ($N^{-1/2}$ scaling from \citealt{Jee2026}); limits on planetary-mass compact objects from short-timescale bulge microlensing, complementary to OGLE and Subaru/HSC \\ \hline
\textbf{Dark Sirens} (Sec.~\ref{sec:dark_sirens}) & Deep NIR photometric-redshift host catalog over $2{,}000$~deg$^2$ ($\sim$5\% of the sky); $\sim$25--50 well-localized GW events in the footprint (O4/O5) and $\sim$500--5000 (ET/CE); rapid kilonova follow-up & Independent $H_0$ from in-footprint events to $\sim$15--20\% (O4/O5) and $\sim$1.5--5\% (ET/CE era; $\sigma(H_0)/H_0\propto N^{-1/2}$), with systematics largely independent of the distance ladder and the CMB; calibration reference for wider-area host catalogs \\ \hline
\textbf{Legacy Science} (Sec.~\ref{sec:legacy}) & Resolved morphologies for a few $\times10^{8}$ galaxies, with full $R\sim100$ SEDs for the $\sim$$10^{6}$--$10^{7}$ brighter than $\approx$21~AB in the NIR; deep intracluster-light maps of the ten Deep-field clusters and shallower maps for hundreds of Wide-Survey clusters; $\sim$$10^{7}$ stars with NIR spectrophotometric metallicities; several hundred bound planets and $\lesssim100$ free-floating planets; variability-selected AGN & By-product datasets for galaxy-evolution, Milky Way, and AGN science at no additional observing cost; exoplanet yields from the dedicated bulge campaign \\ \hline
\end{tabular}
\end{table*}

\section{Synergy with International Missions}
\label{sec:synergy}
KST will contribute imaging and galaxy-redshift data to joint analyses with other surveys and gravitational-wave observatories. Its complementarity is twofold: KST adds four-band optical shapes and contiguous NIR spectrophotometry to the combined dataset, and its overlap with the major surveys enables the mutual cross-calibration of systematics on which Stage~IV cosmology depends.

\subsection{Wide-Field Imaging Surveys}
\begin{itemize}
    \item \textbf{Vera C. Rubin Observatory (LSST):} Rubin will provide the deepest wide-area \textit{optical} imaging from the ground, but is seeing-limited ($\gtrsim0.7''$) and carries no NIR channel. KST supplies the space-based diffraction-limited resolution ($\sim0.15''$) needed for weak-lensing shapes and strong-lens substructure, and the $1.1$--$1.8~\mu$m photometry needed for photometric redshifts beyond $z\sim1$. The two surveys cross-calibrate shear and photo-$z$ systematics in their overlap.
    \item \textbf{Euclid:} Euclid measures shapes in a single broad optical band ($I_E$) and adds three NIR photometric bands ($Y,J,H$) in a separate instrument. KST instead covers $g,r,i,z$ and the NIR with a single instrument, recording $R\sim100$ NIR spectrophotometry simultaneously with every optical exposure. Euclid's NISP grism is likewise exposed in parallel with VIS, and both missions obtain full-field spectral information in parallel with imaging. The difference lies in what that information delivers. NISP's slitless spectra yield secure redshifts principally for emission-line galaxies above its line-flux limit, whereas the LVF provides contiguous $R\sim100$ continuum spectrophotometry at a fidelity set by source brightness (Table~\ref{tab:lvf_modes}) rather than by the presence of a detected line. Four-band shapes provide an independent shear estimate and a different systematics floor for cross-validation.
    \item \textbf{Nancy Grace Roman Space Telescope:} Roman reaches deeper in the NIR ($0.5$--$2.3~\mu$m) and adds slitless NIR spectroscopy in its high-latitude survey. KST's four-band optical shapes, simultaneous optical$+$NIR coverage, and independent systematics are complementary, and the overlapping footprints allow joint control of shear and photometric-redshift systematics.
    \item \textbf{CSST:} CSST, like KST, obtains imaging and slitless spectra simultaneously on a shared focal plane, but its coverage ends near $1~\mu$m. KST extends to $1.8~\mu$m and replaces grism dispersion with contiguous LVF continuum spectrophotometry, making the two missions complementary in wavelength and in the source populations each characterizes best.
\end{itemize}

\subsection{Spectroscopic Cosmology: DESI and Successors}
The dynamical-dark-energy hint that motivates Pillar~I originates in spectroscopic BAO from DESI combined with SNe~Ia (Section~\ref{sec:dark_energy}). KST tests the same physics with an \textit{orthogonal} dataset: imaging weak lensing plus angular BAO and a NIR-standardized SN~Ia Hubble diagram, with independent selection functions and systematics. A joint analysis of KST's geometry-plus-growth measurements with DESI's three-dimensional clustering tightens the $w_0$--$w_a$ constraint beyond what either provides alone.

\subsection{The Gravitational-Wave Detector Network}
Pillar~III depends on the facilities that produce the sirens: the LIGO--Virgo--KAGRA network in its O4/O5 runs, and ultimately the Einstein Telescope and Cosmic Explorer. These detectors supply the luminosity distances. KST supplies the redshift information---the deep NIR host catalog for dark sirens within its footprint and the rapid wide-field follow-up for kilonova bright sirens. KST thus acts as the optical-to-NIR counterpart-and-catalog node of the multi-messenger network, a role for which no current wide survey is optimized.

\section{Summary}
\label{sec:summary}
The Korean Space Telescope (KST) is conceived as a dedicated probe of the dark sector of the Universe. Optimized as a 1.4-meter wide-field surveyor, it leverages a strategic aperture--etendue balance and a simultaneous dual-channel (VIS+NIR) architecture with $R\sim100$ spectrophotometry to measure the geometry of cosmic expansion and the growth of structure with a single instrument. This combination of capabilities allows KST to address the three outstanding questions of dark physics in concert:

\begin{itemize}
    \item \textbf{Dynamical Dark Energy:} Testing whether the dark energy equation of state evolves ($w_0$--$w_a$) and distinguishing it from modified gravity, by combining weak-lensing cosmic shear (growth) with angular BAO, the Alcock--Paczy\'nski test in the NIR redshift desert, and a NIR-standardized Type Ia supernova Hubble diagram (geometry), plus an independent $H_0$ from lensed-SN time delays.
    \item \textbf{The Nature of Dark Matter:} Mapping dark matter from cluster cores to galactic substructure through diffraction-limited strong and weak lensing, constraining self-interaction cross-sections and the small-scale matter power spectrum to discriminate among cold, warm, fuzzy, and self-interacting dark matter.
    \item \textbf{Dark Sirens:} Delivering an independent, distance-ladder-free measurement of $H_0$ by pairing gravitational-wave standard sirens with a deep NIR photometric-redshift host catalog, while enabling rapid wide-field follow-up of kilonova counterparts.
\end{itemize}

The same survey data simultaneously yield a broad \textbf{Legacy Science} portfolio (galaxy evolution, active galactic nuclei, exoplanet microlensing, and Milky Way astronomy), at no additional observing cost apart from the dedicated bulge microlensing campaign. By providing diffraction-limited resolution ($\sim0.15''$) and NIR coverage not available on the Vera C. Rubin Observatory, while combining four-band optical imaging with contiguous $R\sim100$ NIR spectrophotometry to $1.8~\mu$m in a way that complements \textit{Euclid}, Roman, and CSST, KST fills a critical observational niche. As Korea's first meter-class space telescope dedicated to astrophysics, it will serve as both a capable standalone facility for dark-sector science and a vital complementary asset in the upcoming global era of precision cosmology.

\acknowledgments
MJJ acknowledges support for the current research from the National Research Foundation (NRF) of Korea under the programs 2022R1A2C1003130 and RS-2026-25607498.
HSH acknowledges support from the National Research
Foundation of Korea (NRF) funded by the Korea government (MSIT; RS-2026-25482692)
and the Global-LAMP Program funded by the Ministry of Education (RS-2023-00301976).
DL acknowledges support from the Basic Science Research Program through the NRF of Korea, funded by the Ministry of Education (RS-2025-25419201 and RS-2022-NR070872).

\vspace{1ex}\noindent\textbf{Data availability.} This paper presents no new observational data. The forecasts and figures are derived from the survey parameters stated in the text and the cited literature. The forecasting codes and derived data are available from the corresponding author upon reasonable request.

\vspace{1ex}\noindent\textbf{Conflict of interest.} The authors declare no conflict of interest.

\bibliography{main}

\end{document}